\documentclass[letterpaper, oneside, 12pt]{article}

\RequirePackage{amsthm,amsmath,amsfonts,amssymb}
\RequirePackage[numbers]{natbib}
\RequirePackage{graphicx}

\usepackage{amsmath}
\usepackage{graphicx}%
\usepackage{amsfonts}%
\usepackage{amssymb}
\usepackage{xcolor}
\usepackage{overpic}
\usepackage[ruled]{algorithm}
\usepackage[noend]{algpseudocode}
\usepackage{rotating}
\usepackage{epstopdf}
\usepackage{url}
\usepackage{comment}
\usepackage{enumerate}
\usepackage{mathtools}
\usepackage{multirow}
\usepackage{pgfplots}
\usepackage{natbib}
\usepackage{pst-solides3d}
\usepackage{tabularray}
\usepackage{multirow}
\usepackage{subfig}
\usepackage{array}
\usepackage{subcaption}
\usepackage[section]{placeins}
\usepackage{makecell}

\usepackage{setspace}
\usepackage[hmargin=1.0in,vmargin=1.0in]{geometry}

\usepackage[compact]{titlesec}
\titlespacing{\section}{0pt}{*0.8}{*0.8}
\titlespacing{\subsection}{0pt}{*0.8}{*0.8}
\titlespacing{\subsubsection}{0pt}{*0.8}{*0.8}

\newcommand{\bA}{ {\boldsymbol A} }

\newcommand{\bD}{ {\boldsymbol D} }

\newcommand{\bH}{ {\boldsymbol H} }

\newcommand{\bI}{ {\boldsymbol I} }

\newcommand{\bm}{ {\boldsymbol m} }

\newcommand{\bO}{ {\boldsymbol O} }

\newcommand{\bs}{ {\boldsymbol s} }
\newcommand{\bS}{ {\boldsymbol S} }

\newcommand{\bU}{ {\boldsymbol U} }

\newcommand{\bx}{ {\boldsymbol x} }

\newcommand{\by}{ {\boldsymbol y} }

\newcommand{\bZ}{ {\boldsymbol Z} }

\newcommand{\balpha}{ {\boldsymbol \alpha} }

\newcommand{\bbeta}{ {\boldsymbol \beta} }

\newcommand{\bmu}{ {\boldsymbol \mu} }

\newcommand{\bet}{ {\boldsymbol \eta} }

\newcommand{\bSigma}{ {\boldsymbol \Sigma} }

\newcommand{\bzeta}{ {\boldsymbol \zeta} }

\newcommand{\bxi}{ {\boldsymbol \xi} }

\newcommand{\bzero}{ {\boldsymbol 0} }

\title{Integrative Learning of Dynamically Evolving Multiplex Graphs and Nodal Attributes Using Neural Network Gaussian Processes with an Application to Dynamic Terrorism Graphs}

\author{Jose Rodriguez-Acosta\footnotemark[1]
    \and 
    Sharmistha Guha\footnotemark[1]
    \and
    Lekha Patel\footnotemark[2]
    \and
    Kurtis Shuler\footnotemark[2]
}

\date{}

\begin{document}

\maketitle

\footnotetext[1]{Department of Statistics, Texas A\&M University}

\footnotetext[2]{Sandia National Laboratories}




\begin{abstract}
Exploring the dynamic co-evolution of multiplex graphs and attributes observed over graph nodes or actors is a compelling scientific question in criminal and terrorism networks.
This article is motivated by the study of dynamically evolving interactions among prominent terrorist organizations, while also considering various organizational attributes such as size, ideology, leadership, and operational capacity.
Statistically principled integration of multiplex graphs with nodal attributes is significantly challenging due to the need to leverage complex shared information within and across layers, account for uncertainty in predicting unobserved links, and capture temporal evolution of node attributes. These difficulties increase when layers are only \emph{partially observed}, as in terrorism networks where connections are deliberately hidden to obscure key relationships. To address these challenges simultaneously, this article presents a principled methodological framework that integrates the multiplex graph layers and nodal attributes. The approach employs time-varying stochastic latent factor models, leveraging shared latent factors to capture the graph’s structure and its co-evolution with node attributes. These latent factors are modeled using Gaussian processes with an infinitely wide deep neural network-based covariance function, termed neural network Gaussian processes (NN-GP). 
The NN-GP framework on latent factors exploits the predictive power of Bayesian deep neural network (BNN) architecture while propagating uncertainty in inference for reliable scientific conclusions. Detailed simulation studies in the supplementary file highlight the superior performance of the proposed approach in achieving inferential objectives. The proposed approach, termed as \emph{dynamic joint learner}, enables predictive inference (with uncertainty) of diverse unobserved dynamic relationships among prominent terrorist organizations and their organization-specific attributes, as well as their clustering behavior in terms of friend-and-foe relationships, which could be informative in counter-terrorism operations.
\end{abstract}


\section{Introduction}

Covert dynamic graphs, consisting of nodes or actors (e.g., individuals or organizations) and edges representing their relationships, are deliberately obscured to evade detection. Examples include criminal networks, fraudulent financial schemes, and the historic suffragette movement \citep{ficara2021criminal, rostami2015complexity, benigni2017detection, benigni2017online}. In this article, we focus on terrorism graphs, where understanding the evolution of graph structures remains one of the most pressing challenges in global security intelligence. Terrorist graphs rely on hidden connections to shield critical relationships and protect key organizations, with ties that often shift in response to counter-terrorism pressure, geopolitical developments, or strategic realignments. Accurately predicting how these graphs evolve - including which alliances strengthen and which rivalries emerge, forming dynamically evolving multiplex graph structures - is central to effective counter-terrorism operations and the protection of civilian populations. In addition to concealed ties, multiple organizational attributes of each terrorist group evolve over time. While specific relationships may be hidden, organizations cannot simultaneously obscure all aspects of organizational change. Shifts in operational capacity, ideology, or leadership inevitably leave observable traces in attributes such as attack frequency, recruitment patterns, or public statements \citep{jackson2005aptitude}. By jointly modeling dynamic multiplex terrorist graphs with these evolving organizational attributes, it becomes possible to leverage such signals. This scientific problem can be statistically cast as the joint modeling of multiplex graph and nodal attributes over time - a central challenge addressed in this article. Such analyses are essential for identifying threats, disrupting harmful collaborations, and predicting future behaviors, thereby providing valuable insights for ongoing security and intelligence efforts.

The data consist of two parts: time-series of multiplex graphs
$\mathcal{G}(t)$ comprising $L$ layers of undirected graphs $\mathcal{G}_1(t),...,\mathcal{G}_L(t)$ with a common set of $J$ nodes $\mathcal{N}=\{\mathcal{N}_1,...,\mathcal{N}_J\}$, and a collection of $m$ time series of attributes $\bx_j(t)=(x_{j,1}(t),...,x_{j,m}(t))^T$ associated with the $j$th node $\mathcal{N}_j$, $j=1,...,J$. The $l$th layer of the multiplex graph is represented by a symmetric adjacency matrix $\bA_l(t)$ with its $(j,j')$th entry $a_{jj',l}(t)$ indicating the edge weight between nodes $\mathcal{N}_j$ and $\mathcal{N}_{j'}$ at the $l$th layer at time $t$. The edge weights can be continuous, binary or categorical depending on the scientific problem.

In some scientific problems, including in the motivating terrorism study, each layer of the multiplex graph is only partially observed. At time $t$, $a_{jj',l}(t)$ is either observed or unobserved (link status unknown). Let $\bU_l^{(o)}(t)\subseteq \{1,\dots,J\}\times\{1,\dots,J\}$ denote the set of \textbf{observed} edges and $\bU_l^{(u)}(t)\subseteq \{1,\dots,J\}\times\{1,\dots,J\}$ the set of \textbf{unobserved} edges for layer $l$ at time $t$, such that $\bU_l^{(o)}(t)\cup\bU_l^{(u)}(t) =\{1,...,J\}\times\{1,...,J\}$.
Thus, $a_{jj',l}(t)$ is an observed or unobserved link if $(j,j')\in\bU_l^{(o)}(t)$ or $(j,j')\in\bU_l^{(u)}(t)$, respectively. The goal of this article is to predict unobserved links and node-specific attributes based on the observed data, thereby estimating the temporal trajectory of both the multiplex graph and nodal attributes. Uncertainty quantification is crucial in the inference process, as the problem necessitates high-dimensional models to accurately capture the co-evolution.

The study of temporal co-evolution between graphs and nodal attributes typically follows two main approaches: modeling graph evolution conditioned on nodal attributes (selection) or modeling nodal attributes conditioned on graph structure (influence). ``Selection'' models focus on how nodal attributes influence link formation, often using Exponential Random Graph Models (ERGMs) or mixed-effects generalized linear models \citep{wasserman2005introduction, hoff2002latent, hoff2005bilinear}, or their temporal extensions (TERGMs) \citep{robins2001random, guo2007recovering, hanneke2010discrete, brandes2012visualization, lee2020model}. 
Uncertainty in TERGMs is typically assessed using resampling techniques, e.g., bootstrap \citep{desmarais2012micro, desmarais2017statistical}. Similarly, stochastic actor-oriented models (SAOMs) \citep{snijders2010introduction, steglich2010dynamic, snijders2017stochastic, koskinen2023multilevel} have been extensively studied for modeling the evolution of dynamic graphs and the co-evolution of graphs and nodal attributes. SAOMs employ graph-specific rules or effects, such as reciprocity, triadic closure, or homophily, to explain the observed evolution of graphs. Although appealing, the performance of ERGMs and SAOMs  depends heavily on the choice of sufficient statistics or graph-specific rules included in the formulation. ERGMs, in particular, often struggle to capture local features of the multiplex graph, resulting in suboptimal performance in many real-world problems \citep{handcock2003statistical}.
Alternatively, ``influence'' models use the nodal attributes as the dependent variables and estimate the effects of the graph on these attributes \citep{robins2001network, frank2004social, fujimoto2013decomposed, goldsmith2013social, sweet2020latent}.

In a selection model, a multiplex graph is modeled conditionally on the attributes, assuming the attributes are fully observed. Similarly, in a model for influence, the multiplex graph must be fully observed. However, when data consist of graphs that are only partially observed at each time point - as is often the case in covert dynamic networks studied in terrorism research - these frameworks become less suitable. Disentangling selection from influence is a key challenge in co-evolution modeling. Here we focus on frameworks that do not attempt to make a causal distinction between these two mechanisms. Instead, we adopt a joint Bayesian modeling strategy that simultaneously incorporates graph edges and nodal attributes within a unified framework. This joint specification captures their co-dependence,  avoiding the need for strong identification assumptions to separate selection and influence effects \citep{doreian2001causality, leenders2013longitudinal}. We therefore refrain from making causal claims about the direction of influence between graphs and attributes, and instead provide a principled approach to quantify associations and leverage information sharing between them for improved inference.

Extensive research exists on jointly modeling multiplex graphs using stochastic block models \citep{barbillon2017stochastic, paul2016consistent} and latent space models \citep{gollini2016joint, salter2017latent, d2023model, durante2017bayesian}. While these models leverage latent variables to capture relationships across layers of a multiplex graph, they largely overlook interactions between graphs and node-specific attributes. In the study of graph-attribute associations, \cite{fosdick2015testing} employ a multivariate normal distribution that jointly models nodal attributes and node-specific latent factors, though their method is restricted to static, single-layer graphs with continuous attributes. \cite{zhang2022joint} develop a joint latent space model in the static scenario in which shared latent variables explain both the single-layer graph structure and the multivariate nodal attributes.  \cite{wang2023joint} introduce latent variables for both graph nodes and nodal attributes, modeling their interactions jointly. However, their framework is also limited to static single-layer graphs and relies on variational inference for computational tractability.

For dynamic settings, \cite{de2010obesity} propose time-varying models for single-layer graphs with binary or categorical attributes, later extended by \cite{niezink2017co} to continuous attributes. These approaches, grounded in the ERGM and TERGM methodologies, inherit well-known limitations of these models. A flexible framework is offered by \cite{guhaniyogi2020joint}, which allows graphs and attributes of diverse types (continuous, binary, categorical) to evolve jointly over time. However, this method assumes a single-layer graph with fully observed edges at each time point.

\noindent \textbf{Outline of the proposed approach.} 
This article introduces a Bayesian framework for jointly inferring on the generative mechanism of multiplex graphs and nodal attributes, referred to as the \emph{dynamic joint learner}. Our approach leverages low-rank factor models with dynamically evolving latent factors to innovatively and flexibly capture the temporal and structural dependencies within each layer of the multiplex graph and  nodal attributes. Our model incorporates time-varying latent factors specific to each layer of the multiplex graph, along with shared latent factors that capture node interactions both specific to individual layers and common across all layers. The node-specific latent factors shared across layers also capture the time-varying associations between the graph layers. Moreover, node-specific latent factors in different layers are shared with the factor model for nodal attributes to capture dependencies in the co-evolution of multiplex graphs and nodal attributes over time.

The time-varying latent factors are modeled using Gaussian processes (GP) with deep neural network kernel functions, known as neural network GP (NN-GP) \citep{lee2017deep, garriga2018deep}. NN-GP exactly replicates the predictions of an infinitely wide deep neural network, and is expressed with only a few unknown parameters. The NN-GP priors for temporal dynamics are designed to capture the non-stationary evolution of terrorism graphs, which often undergo abrupt structural shifts due to events such as leadership decapitation, peace negotiations, or military interventions. Unlike stationary kernels (e.g., Matérn), the non-stationary covariance of the deep neural network kernel effectively models these sudden transitions. Integrating the deep neural network approach within a Bayesian framework with NN-GP priors harnesses the predictive power of deep neural networks while offering a principled approach to quantify uncertainty in predicting nodal attributes and unknown links within the multiplex graph. This framework also facilitates model regularization, promotes robust predictions, and addresses the overfitting challenges encountered in standard deep neural networks \citep{salman2019overfitting,bejani2021systematic}.

Our approach accommodates continuous, binary, and categorical links, as well as diverse nodal attributes, and effectively handles partially observed graph layers. Unlike dynamic graph neural networks (DGNNs), which model temporal graph evolution by updating nodes and edges \citep{rossi2020temporal, skarding2021foundations}, our method incorporates the joint evolution of graphs and nodal attributes while accommodating partially observed graphs at each time step. While some existing studies on graph-nodal attribute associations focus on testing global associations between the graph and all attributes jointly, as well as local associations between the graph and individual attributes \citep{guhaniyogi2020joint, fosdick2015testing}, our objective is distinct - we aim to reconstruct unobserved node relationships and predict nodal attributes over time, with uncertainty quantification. Future work will extend our framework to incorporate hypothesis testing within this context.

\noindent \textbf{Novelty of the proposed approach.} \\
\textbf{(1) Joint modeling of time-varying multiplex graph and nodal attributes.} The proposed framework uniquely captures time-varying dependencies between the multiplex graph and nodal attributes without relying on prior knowledge about the directionality of these dependencies. Since the relationship between multiplex graphs and nodal attributes may often be bi-directional, constraining it to a unidirectional framework, as in selection or influence models, may impose an overly restrictive assumption. As a result, our framework can simultaneously predict unobserved links $\{a_{jj',l}(t):(j,j')\in \bU_l^{(u)}(t)\}$ and missing attribute data, 
while also enabling prediction of the co-evolution of multiplex graphs and nodal attributes at new time points.\\
\textbf{(2) NN-GP prior on time-varying latent factors.} Unlike conventional deep neural networks, NN-GP avoids the use of a large number of parameters and is capable of providing uncertainty in inference. Although NN-GP has gained attention in deep learning for its ability to quantify uncertainty \citep{iwata2017improving, huang2015scalable, howard2025wilsonian, pang2019neural, flam2017mapping}, its potential to model dynamically evolving graphs remains unexplored. Additionally, NN-GP supports scalability for large multiplex graphs through efficient GP variants \citep{guhaniyogi2018meta, guhaniyogi2023distributed}, a direction we aim to explore in future work.\\
\textbf{(3) Uncertainty in the inference.} Our framework provides robust inference by incorporating model-based uncertainty quantification (UQ) in modeling complex data. Ordinary deep neural networks are often regarded as black boxes, highlighting the growing interest in developing UQ methods within deep learning frameworks that incorporate statistically interpretable structures.
To this end, the proposed framework is both flexible and general, introducing a \emph{novel interpretable deep neural network architecture with Bayesian inference} for joint learning of multiplex graphs and their associated nodal attributes.

The rest of the article is structured as follows. Section \ref{RealDataIntro} describes the BAAD2 (Big, Allied and Dangerous) dataset and critical security questions. Section \ref{sec:model} presents the model for dynamic co-evolution of the multiplex graph and nodal attributes with time-varying shared latent factors. It also provides a detailed explanation of the NN-GP prior construction for the latent factors. Section \ref{sec:posteriorComp} outlines the posterior computation, inferences for unobserved links and nodal attributes, and details on computational time and architecture. Section \ref{sec:realdata} demonstrates the performance of the proposed framework, along with competing models, in the analysis of covert dynamic graphs and organizational attributes for terrorism study. Finally, Section \ref{sec:conclusion} concludes the article with an eye towards future work. Detailed simulation studies, along with the details of MCMC updates are provided in the supplementary file.

\section{Dynamically Evolving Covert Terrorist Graphs}\label{RealDataIntro}
Understanding the evolution of terrorist graph structures constitutes one of the most pressing challenges in global security intelligence. These covert graphs operate through deliberately concealed relationships that shift in response to counter-terrorism pressure, geopolitical changes, and strategic realignments. The ability to predict how these graphs evolve, including which alliances strengthen, which rivalries emerge, and how organizational capabilities transform, directly impacts the effectiveness of counter-terrorism operations.

\begin{figure}[h]
    \centering
    \includegraphics[width=15cm] 
    {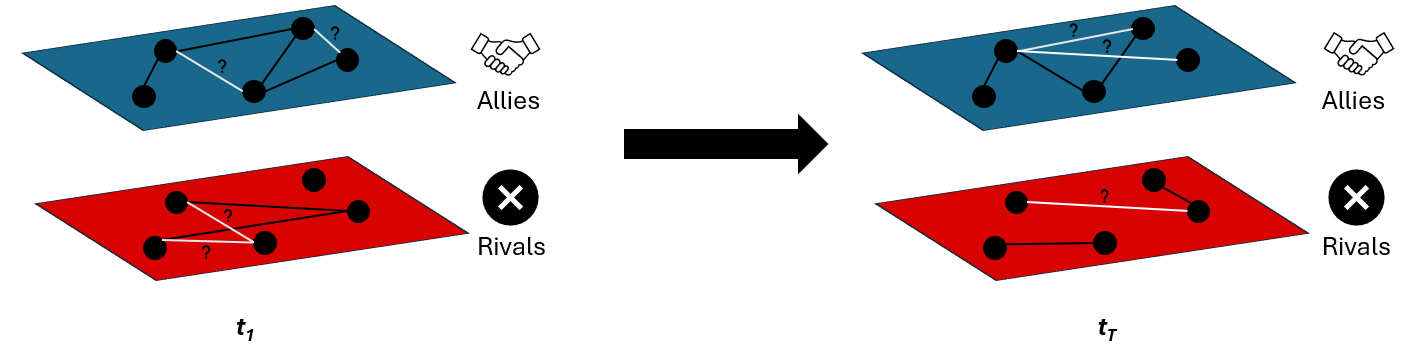}
    \caption{Diagrammatic representation of a two-layer multiplex graph with nodal attributes over $T$ time points between terrorist organizations (nodes). The two layers are represented by ``Alliance'' and ``Rivalry'' between organizations. A black edge between nodes denotes a known (observed) link between them. A white edge with a ``?'' denotes an unknown (unobserved) link between the corresponding nodes.}
    \label{model_diagram}
\end{figure}

This section introduces the BAAD2 (Big, Allied, and Dangerous) dataset \cite{baad_db}, which contains graph data detailing the structure and behavior of major terrorist and insurgent organizations from 1998 to 2012, and will be analyzed using the methodology developed in this article. This dataset captures the relationships between $15$ terrorist organizations listed in Table \ref{tab_org_list}, spanning a period of 15 years. These organizations include groups such as the Taliban and Hizballah, operating over several regions of Asia, Africa, and South America. The types of extremist ideologies represented in the groups include religious, separatist, ethnic, and leftist classifications. Two types of relationships are examined between each pair of organizations: ``Alliance'' and ``Rivalry.'' These interactions can be represented as a dynamically evolving multiplex graph, where terrorist organizations serve as nodes, as depicted in the visual illustration in Figure~\ref{model_diagram}.

This dataset also includes a variety of attributes for each organization, including organizational size, ideology, leadership, operational capacity, and other dynamically-varying factors. These evolving attributes can reflect significant shifts within the organizations, such as leadership changes, recruitment efforts, or alterations in strategic focus, thus changing the potential threat an organization can pose. Further, an organization's growth in size or change in leadership can impact both its operational capacity and relationships with other groups, leading to: new alliances, rivalries, or shifts in influence.

\noindent\underline{\textbf{Critical Security Questions.}} The evolving nature of terrorist graphs presents three fundamental challenges for security intelligence that require advanced analytical frameworks. \\
\textbf{(Q1)}. First, intelligence agencies must frequently infer complete graph structures from incomplete information. Terrorist organizations deliberately conceal critical relationships, use intermediaries to obscure connections, and maintain operational security that leaves many interactions unobserved \citep{sparrow1991application, krebs2002mapping}. A particular challenge involves detecting complex simultaneous relationships where organizations maintain both cooperative and competitive dynamics. These dual relationships, which may involve collaboration in certain operational domains while maintaining rivalry in others, represent a critical intelligence gap \citep{sageman2004understanding, everton2012disrupting}. Traditional analytical methods that treat relationships as simply present or absent in a uni-layer graph fail to capture these nuanced interactions that determine true threat landscapes. \\
\textbf{(Q2)}. Second, detecting organizational changes before they translate into operational activities is a major intelligence challenge. Shifts in attributes such as leadership, ideology, or capacity precede terrorist actions by months or years \citep{cronin2009terrorism}, yet these signals are missed when graphs and attributes are studied separately \citep{carley2006destabilizing, jackson2005aptitude}. The key difficulty is distinguishing combinations of positional and attribute changes that signal escalation from those reflecting routine organizational behavior, insights that are critical for preventive intelligence.\\
\textbf{(Q3)} Third, understanding cascading effects in terrorist ecosystems is critical for anticipating the broader impact of counterterrorism interventions. Because terrorist graphs are interconnected, actions affecting one group can trigger system-wide responses \citep{kenney2007pablo}. External pressures, such as peace talks, leadership losses, or territorial setbacks, can reshape rivalries and alliances \citep{bacon2018terrorist}. Anticipating these second-order effects requires modeling the joint evolution of graph structures and organizational attributes under such shocks.

These security challenges are compounded by the criminal dimensions of terrorist operations. As \cite{asal2019crime} shows, terrorist organizations' involvement in criminal enterprises shapes both graph structures and operational capacity. Their study of 140 insurgent groups finds that territorial control increases criminal activity. Such intersections create additional concealment, with separate political and criminal networks complicating intelligence efforts. These issues motivate our methodology, which must address partial observability, temporal dynamics, and the interdependence of graphs and organizational attributes.

\noindent\underline{\textbf{Scientific Objectives with Security Implications.}} To address these critical security questions, this paper develops a principled methodological framework with three key innovations. \\
First, we implement a fully Bayesian approach that reconstructs complete graph structures from partially observed data, providing probabilistic assessments of hidden relationships. 
Unlike traditional methods that treat unknown connections as \emph{unconditionally} missing at random (MAR) or missing completely at random (MCAR), in terrorism graphs, missingness is strategic as organizations actively conceal their critical connections. Our framework addresses this by jointly modeling graph evolution and organizational attributes through latent factors, where conditional on these latent factors, missingness can be treated as MAR. The latent factors capture the strategic patterns of concealment, allowing us to infer hidden relationships based on observable organizational changes that terrorists cannot fully conceal.
Our framework generates confidence levels for each potential link, enabling analysts to prioritize surveillance resources based on proper statistical evidence. 
Second, our joint model links changes in organizational capacity, ideology, and operational strength to evolving graph positions, identifying transformation patterns that may be missed in conventional analysis. 
Third, employing Neural Network Gaussian Processes (NN-GPs) enables us to quantify uncertainty while maintaining the predictive power of deep learning architectures. This enables the framework to distinguish high-confidence predictions from uncertain assessments, which is critical for graph operational decisions where false positives may carry significant costs. 
Here the BAAD2 dataset provides an ideal testbed for these methods. Its coverage of major terrorist organizations during periods of significant transformation (e.g., before and after 9/11) offers both the temporal depth and organizational diversity needed to validate our approach. By deliberately masking portions of this graph, we create realistic conditions that mirror the partial observability inherent in intelligence operations, demonstrating our framework's performance when critical information is concealed.

\begin{table}[H]
    \centering
    \small
    \begin{tabular}{|c|c|}
    \hline
        1 & Abu Sayyaf Group (ASG) \\
        \hline
        2 & Hamas (Ha) \\
        \hline
        3 & Hizballah (Hi) \\
        \hline
        4 & Islamic Movement of Uzbekistan (IMU) \\
        \hline
        5 & Kurdistan Workers' Party (PKK) \\
        \hline
        6 & Lord's Resistance Army (LRA) \\
        \hline
        7 & Moro Islamic Liberation Front (MILF) \\
        \hline
        8 & Mujahedin-e Khalq (MEK) \\
        \hline
        9 & National Liberation Army of Colombia (ELN) \\
        \hline
        10 & Palestinian Islamic Jihad (PIJ) \\
        \hline
        11 & Revolutionary Armed Forces Of Colombia (FARC) \\
        \hline
        12 & Taliban (Ta) \\
        \hline
        13 & United Liberation Front of Assam (ULFA) \\
        \hline
        14 & Allied Democratic Forces (ADF) \\
        \hline
        15 & Baloch Liberation Front (BLF) \\
        \hline
    \end{tabular}
    \caption{List of the 15 terrorist organizations in BAAD2 dataset.}
    \label{tab_org_list}
\end{table}

\section{Model Development and Prior Formulation}\label{sec:model}

\subsection{Model Development}
Assume $\bA_{l}^{(o)}(t)=\{a_{jj',l}(t):(j,j')\in\bU_l^{(o)}(t)\}$ denotes the set of observed edges in the partially observed $l$th layer of the multiplex graph at time $t$, and $\bx_j(t)$ is the $m$-dimensional attribute associated with node $\mathcal{N}_j$ at time $t$. The unobserved $l$th layer of the multiplex graph, $\bA_{l}^{(u)}(t)$, is defined analogously. In this article, we consider the graph at each layer to be undirected with no self relationships, such that $a_{jj',l}(t)=a_{j'j,l}(t)$ and $a_{jj,l}(t)=0$, for all $1\leq j<j'\leq J$ and $l=1,...,L$. This assumption is appropriate for terrorism networks, where organizations do not form self-relationships, and the interactions between any two groups, such as alliances or conflicts, are inherently reciprocal and thus represented as undirected. The multiplex graph and nodal attributes are observed at a finite number of time points $t_1<\cdots<t_T$ resulting in realizations of the partially observed $l$th layer of the multiplex graph $\bA_{l}^{(o)}(t_1),...,\bA_{l}^{(o)}(t_T)$, and $j$th nodal attributes $\bx_j(t_1),...,\bx_j(t_T)$.

We propose a joint time-varying latent factor model to capture the co-evolution of multiplex graphs and nodal attributes. Edges are modeled as conditionally independent given time-varying latent factors for the $J$ nodes. To capture cross-layer dependence, some factors are shared across layers while others are layer-specific. Specifically, the link between nodes $j$ and $j'$ in layer $l$ is modeled as:
\begin{align}\label{eq:network_model}
E[a_{jj',l}^{(o)}(t)]= F(\mu(t) +\bzeta_{j}(t)^T \bzeta_{j'}(t)+ \bxi_{j,l}(t)^T \bxi_{j',l}(t)),\:\:j<j',\:\:(j,j')\in\bU_l^{(o)}(t),
\end{align}
where $F(\cdot)$ is a suitable link function and $\mu(t)$ is the intercept function. In this article, we focus on $a_{jj',l}^{(o)}(t)$'s being binary with $F(\cdot)$ as the logit link function. This leads to the following specification of the model:
\begin{align}\label{eq:network_model_logit}
\text{logit}(P(a_{jj',l}^{(o)}(t)=1))= \mu(t) +\bzeta_{j}(t)^T \bzeta_{j'}(t)+ \bxi_{j,l}(t)^T \bxi_{j',l}(t),\:\:j<j',\:\:(j,j')\in\bU_l^{(o)}(t).    
\end{align}
However, more general distributions, where $a_{jj',l}^{(o)}(t)$'s are continuous or categorical, can be easily accommodated with relatively minor changes to our computational approach.

Here $\bzeta_j(t)=(\zeta_{j,1}(t),...,\zeta_{j,R_\zeta}(t))^T$ is the $R_\zeta$-dimensional $j$th node-specific latent factor shared across all graph layers at time $t$, and $\bxi_{j,l}(t)=(\xi_{j,l,1}(t),...,\xi_{j,l,R}(t))^T$ is the $R$-dimensional $j$th node-specific latent factor specific to the $l$th graph layer at time $t$.
The formation of an edge between $\mathcal{N}_j$ and $\mathcal{N}_{j'}$ depends on the angular distance of latent factors $\bzeta_j(t)$ and $\bzeta_{j'}(t)$, as well as  $\bxi_{j,l}(t)$ and $\bxi_{j',l}(t)$, in the joint latent space \citep{hoff2005bilinear,durante2017bayesian}. Specifically, the angular distance of latent factors $\bzeta_j(t)$ and $\bzeta_{j'}(t)$ characterizes the similarity between $\mathcal{N}_j$ and $\mathcal{N}_{j'}$ at time t, common to all the layers. Meanwhile, the angular distance between the layer-specific latent factors 
 $\bxi_{j,l}(t)$ and $\bxi_{j',l}(t)$ captures deviations specific to layer $l$, providing flexibility in modeling layer-specific graph structures. Smaller angular distances correspond to a higher probability of an edge existing between nodes $\mathcal{N}_j$ and $\mathcal{N}_{j'}$. 
 This framework also preserves the \emph{transitivity property} of graphs: if edges exist between $\mathcal{N}j$ and $\mathcal{N}{j'}$, and between $\mathcal{N}j$ and $\mathcal{N}{j''}$, then the probability of an edge existing between $\mathcal{N}{j'}$ and $\mathcal{N}{j''}$ is likely to be high.
 Notably, the latent factors shared between layers also model time-varying association between the multiple layers of the graph.

The dual latent factor structure in Equation \eqref{eq:network_model_logit} directly addresses the complex nature of terrorism graphs documented in the security literature. Specifically, the shared latent factors capture baseline relationships driven by common ideological positions in terms of shared friends or enemies, while the layer-specific factors model deviations representing operational competition for resources, territory, or political influence between enemies or allies. This dual structure is essential for terrorism graphs where a latent factor model with only a one-layer binary friend-or-foe classification fails to capture the nuanced reality of groups that, for example, coordinate military operations while competing for recruitment and funding  
\citep{bacon2018terrorist, asal2019crime}.

\begin{figure}[H]
    \centering
    \includegraphics[width=10cm]{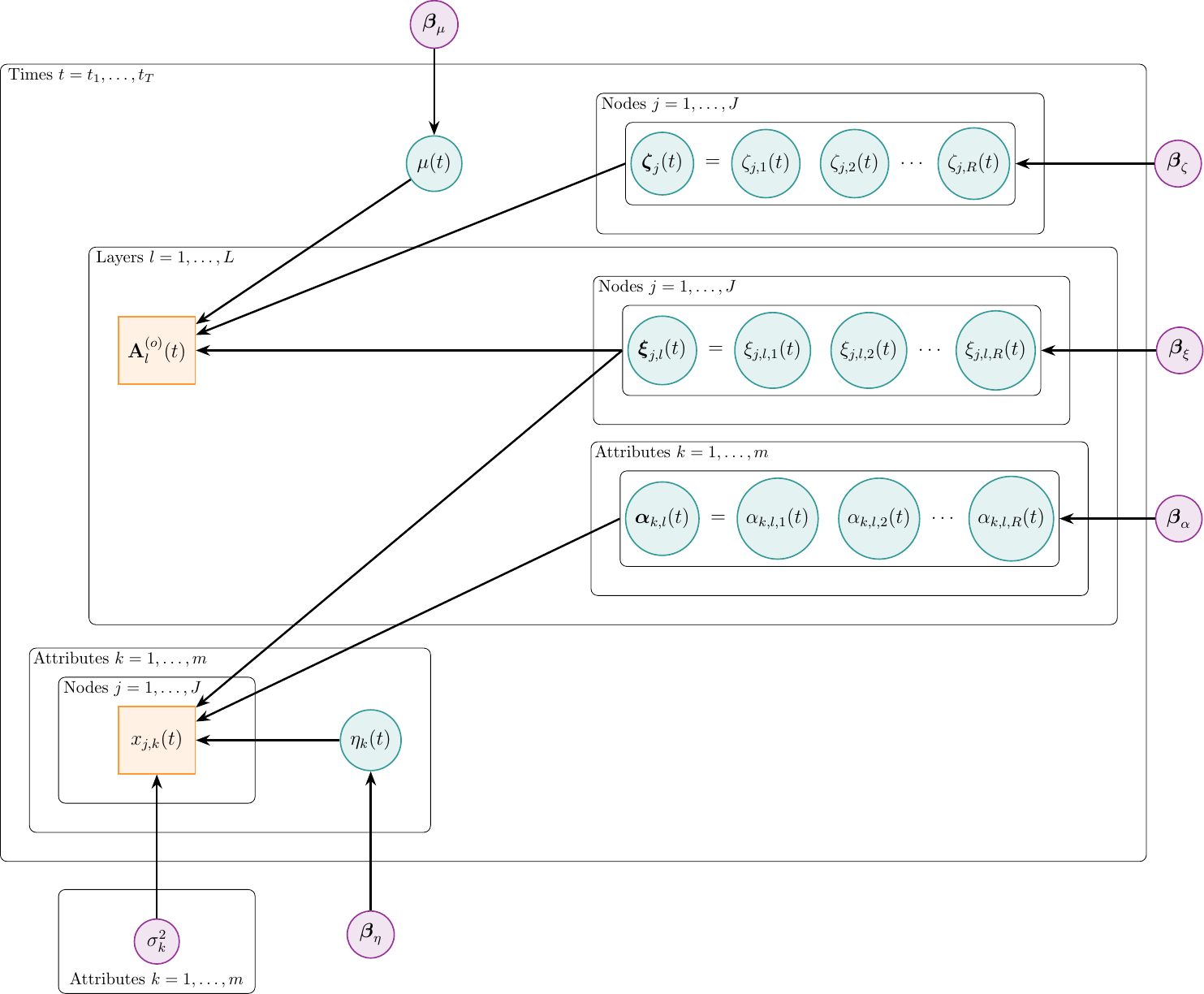} 
    \caption{Diagram of the dependencies among observations ($\bA^{(o)}_{l}(t)$), $x_{j,k}(t)$), latent variables ($\mu(t)$, $\bzeta_{j}(t)$, $\bxi_{j,l}(t)$, $\balpha_{k, l}(t)$, $\eta_{k}(t)$), and parameters ($\sigma^{2}_{k}$, $\bbeta_{\mu}$, $\bbeta_{\zeta}$, $\bbeta_{\xi}$, $\bbeta_{\alpha}$, $\bbeta_{\eta}$).}
    \label{model_param_diagram}
\end{figure}

\sloppy
To capture time-varying associations between graph layers and nodal attributes, we assume that nodes with similar latent factors in the graphs tend to exhibit similar attributes. Consequently, the latent factors driving graph connectivity also inform the nodal attributes, as described below.
\begin{align}\label{eq:nodal_attribute_model}
E[x_{j,k}(t)]= G_k(\eta_k(t)+\sum_{l=1}^L\bxi_{j,l}(t)^T\balpha_{k,l}(t)),\:\:k=1,...,m;\:\:j=1,...,J,
\end{align}
where $G_k(\cdot)$ represents the link function, $\eta_k(t)$ is an intercept function and $\balpha_{k,l}(t)=(\alpha_{k,l,1}(t),...,\alpha_{k,l,R}(t))^T$ is the coefficient for the latent factors $\bxi_{j,l}(t)$ at time $t$. This representation implies that if the probability of an edge between nodes $\mathcal{N}_j$ and $\mathcal{N}_{j'}$ is high, then $\mathcal{N}_j$ and $\mathcal{N}_{j'}$ having similar nodal attribute vectors is also more likely. When the $k$th nodal attribute is continuous, $G_k(\cdot)$  is an identity link function with (\ref{eq:nodal_attribute_model}) becoming
\begin{align}\label{eq:nodal_attribute_identity_link}
x_{j,k}(t)= \eta_k(t)+\sum_{l=1}^L\bxi_{j,l}(t)^T\balpha_{k,l}(t)+\epsilon_{j,k}(t),\:\:k=1,...,m;\:\:j=1,...,J.    
\end{align}
The $\epsilon_{j,k}(t)$'s are idiosyncratic errors following N($0,\sigma_k^2)$ i.i.d. 
Although models (\ref{eq:network_model}) and (\ref{eq:nodal_attribute_model}) do not incorporate known covariates that may influence the evolution of the graph or nodal attributes, respectively, extending them to include such effects is straightforward by adding additional regression terms to (\ref{eq:network_model}) and (\ref{eq:nodal_attribute_model}). A schematic diagram of the model with dependencies among different model components is shown in Figure~\ref{model_param_diagram}.

The joint modeling of graph structure and organizational attributes through shared latent factors exploits a fundamental vulnerability in terrorist operational security in that while organizations can conceal specific relationships, they cannot simultaneously hide all manifestations of organizational change. Growth in operational capacity, ideological shifts, or leadership changes leave observable traces in specific attributes such as attack frequency, recruitment patterns, and public statements \citep{jackson2005aptitude}. By linking evolution of the terrorism graph to these observable attributes, we leverage signals that terrorists cannot fully obscure to reconstruct concealed relationships. \\
Furthermore, the transitivity property encoded in our model addresses the fundamental characteristic of terrorism graph formation that alliance structures frequently exhibit cascading relationship patterns wherein existing bilateral connections facilitate the emergence of additional ties between previously unconnected organizations \citep{christia2012alliance}. As intelligence analysis has documented systematic evidence of such transitive closures \citep{levitt2005hamas}, the angular distance formulation in our model preserves this transitivity property, enabling the framework to capture indirect relationship formations that emerge through intermediary connections. Section~\ref{sec:transitivity_cluster} demonstrates the evidence of transitivity in the terrorism data leading to identification of clusters of terrorist organizations showing connection in terms of friend-or-foe relationships.

\subsubsection{Identifiability of Latent factors}
For any orthogonal matrix $\bO$, $\bzeta_j(t)^T\bzeta_j(t)=(\bO\bzeta_j(t))^T(\bO\bzeta_j(t))$, $\bxi_{j,l}(t)^T\bxi_{j,l}(t)=(\bO\bxi_{j,l}(t))^T(\bO\bxi_{j,l}(t))$ and $\bxi_{j,l}(t)^T\balpha_{j,l}(t)=(\bO\bxi_{j,l}(t))^T(\bO\balpha_{j,l}(t))$. To address rotation indeterminacy, we infer the equivalence class of latent factors, such as pairwise angular distances, following established approaches \citep{hoff2002latent, gollini2016joint, guhaniyogi2020joint, wang2023joint}. This is sufficient for our goal of predicting unobserved edges and attributes, which rely on  models (\ref{eq:network_model}) and (\ref{eq:nodal_attribute_model}). Latent factors can be aligned to a common orientation using a ``Procrustean'' transformation \citep{hoff2002latent, hoff2005bilinear}, which we employ in the study of transitivity property of terrorism graph data in Section~\ref{sec:transitivity_cluster}.

\subsection{Modeling Time-Varying Latent Functions Through Neural Network Gaussian Process (NN-GP) Prior}
To model time-varying association among latent factors, we assign Gaussian process priors independently on the functions 
$\{\zeta_{j,r}(t):t\in\mathcal{T}\subseteq\mathbb{R}^{+}\}$, $\{\xi_{j,l,r}(t):t\in\mathcal{T}\subseteq\mathbb{R}^{+}\}$, $\{\mu(t):t\in\mathcal{T}\subseteq\mathbb{R}^{+}\}$, $\{\eta_{k}(t):t\in\mathcal{T}\subseteq\mathbb{R}^{+}\}$ and $\{\alpha_{k,l,r}(t):t\in\mathcal{T}\subseteq\mathbb{R}^{+}\}$, for $l=1,...,L$, $k=1,..,m$, $j=1,..,J$ and $r=1,...,R$. Within the class of Gaussian processes, we utilize deep neural network kernel functions to enhance flexibility and incorporate the predictive strength of deep neural networks into our framework. We provide a brief overview of deep neural network kernels in Section 1 of the supplementary file.

\subsubsection{NN-GP Prior on Time-Varying Unknown Functions}
The unknown functions are modeled as mean-zero NN-GPs with a $F$-layer deep neural network covariance kernel, as detailed in Section 1 of the supplementary file. Specifically, we have $\mu(\cdot) \sim \text{NN-GP}(0, \kappa_F(t_i, t_j; \bbeta_\mu))$, $\eta_k(\cdot) \sim \text{NN-GP}(0, \kappa_F(t_i, t_j; \bbeta_\eta))$, $\zeta_{j,r}(\cdot) \sim \text{NN-GP}(0, \kappa_F(t_i, t_j; \bbeta_\zeta))$, $r=1,..,R_\zeta$, $\xi_{j,l,r}(\cdot) \sim \text{NN-GP}(0, \kappa_F(t_i, t_j; \bbeta_\xi))$, and $\alpha_{k,l,r}(\cdot) \sim \text{NN-GP}(0, \kappa_F(t_i, t_j; \bbeta_\alpha))$, $r=1,..,R$, where each $\bbeta$ parameter ($\bbeta_{\mu}$, $\bbeta_{\eta}$, $\bbeta_{\zeta}$, $\bbeta_{\xi}$, $\bbeta_{\alpha}$) is a two-dimensional vector involving the variance of the weights and biases, $j=1,...,J$, $l=1,...,L$, $k=1,...,m$. While we assume independent NN-GP priors on $\eta_k(t)$ for $k=1,\ldots,m$, they share information through the common parameter set $\bbeta_{\eta}$. Similarly, the sets of latent factors $\{\zeta_{j,r}(t) : j=1,\ldots,J; r=1,\ldots,R_\zeta\}$, $\{\xi_{j,l,r}(t) : j=1,\ldots,J; l=1,\ldots,L; r=1,\ldots,R\}$, and $\{\alpha_{k,l,r}(t) : k=1,\ldots,m; l=1,\ldots,L; r=1,\ldots,R\}$, share information within each set by using the same parameters for the NN-GP specifications within each set.
Each component of $\bbeta_{\eta}$, $\bbeta_{\alpha}$, $\bbeta_{\zeta}$, $\bbeta_{\xi}$, $\bbeta_{\mu}$ is assigned a discrete uniform prior on the grid $\{0.01,0.02,...,0.1\}$. For continuous 
$k$-th nodal attributes modeled via equation (\ref{eq:nodal_attribute_identity_link}), we assign an inverse-gamma prior IG($a_{\sigma},b_{\sigma}$) on the idiosyncratic variance $\sigma_k^2$.

The widely used Matérn covariance kernel suffers from unidentifiability between its length-scale and variance parameters, often requiring challenging hyperparameter tuning, especially with many unknown functions \cite{zhang2004inconsistent}.  In contrast, the deep neural network kernel avoids this issue, leveraging the representational power of deep neural networks.
Unlike over-parameterized deep learning models, NN-GP requires only two parameters, enabling effective training with fewer observations due to the structured framework of Gaussian process regression.
Critically, the choice of NN-GP priors for temporal dynamics specifically addresses the non-stationary evolution patterns observed in terrorism graphs. Unlike graphs that evolve smoothly, terrorism graphs often exhibit sudden structural shifts in response to external shocks such as leadership decapitation \citep{jordan2009heads}, peace negotiations \citep{cronin2009terrorism}, or military interventions \citep{kenney2007pablo}. The deep neural network kernel's non-stationary covariance captures these abrupt transitions that standard stationary kernels such as the Matérn cannot adequately represent, which is crucial for terrorism graphs where a single event can fundamentally restructure alliance patterns' temporal dynamics. 

\section{Posterior Computation and Inference}\label{sec:posteriorComp}
This section details Bayesian computation for the proposed model, focusing on binary edges and continuous attributes, central to the empirical studies. Posterior computation for binary or categorical attributes and categorical edge weights can follow a similar framework.

We utilize the result discussed in Theorem 1 of \cite{polson2013bayesian} to obtain
\begin{align}\label{eq:PG}
& p(a_{jj',l}^{(o)}(t)) =\frac{\exp\left\{a_{jj',l}^{(o)}(t)(\mu(t) +\bzeta_{j}(t)^T \bzeta_{j'}(t)+ \bxi_{j,l}(t)^T \bxi_{j',l}(t))\right\}}{1+\exp\left\{\mu(t) +\bzeta_{j}(t)^T \bzeta_{j'}(t)+ \bxi_{j,l}(t)^T \bxi_{j',l}(t)\right\}}\nonumber\\
&\qquad\qquad=\exp\left\{(a_{jj',l}^{(o)}(t)-0.5)\left(\mu(t) +\bzeta_{j}(t)^T \bzeta_{j'}(t)+ \bxi_{j,l}(t)^T \bxi_{j',l}(t)\right )\right\} \times \nonumber\\
& \;\int \exp\left(-\frac{\omega_{jj',l}^{(o)}(t)}{2}\left(\mu(t) +\bzeta_{j}(t)^T \bzeta_{j'}(t)+ \bxi_{j,l}(t)^T \bxi_{j',l}(t)\right )^2\right )p(\omega_{jj',l}^{(o)}(t))d\omega_{jj',l}^{(o)}(t),
\end{align}
where $p(\omega_{jj',l}^{(o)}(t))$ is the density of P\'olya-Gamma (1,0) distribution. We use equation (\ref{eq:PG}) and the data augmentation approach outlined in \cite{polson2013bayesian} for effective posterior computation. In particular, with the multiplex graph obtained at time-points $t_1,...,t_T$, the data augmented likelihood from the multiplex graph is given by
\begin{align*}
&\mathcal{L}(\{\bA_l^{(o)}(t_1),...,\bA_l^{(o)}(t_T)\}_{l=1}^L)\propto\\
&\qquad\prod_{i=1}^T\prod_{(j,j')\in\bU_l^{(o)}(t_i)}\prod_{l=1}^L\Big[
\exp\left\{(a_{jj',l}^{(o)}(t_i)-0.5)\left(\mu(t_i) +\bzeta_{j}(t_i)^T \bzeta_{j'}(t_i)+ \bxi_{j,l}(t_i)^T \bxi_{j',l}(t_i)\right )\right\}\\\
&\qquad\times \exp\left(-\frac{\omega_{jj',l}^{(o)}(t_i)}{2}\left(\mu(t_i) +\bzeta_{j}(t_i)^T \bzeta_{j'}(t_i)+ \bxi_{j,l}(t_i)^T \bxi_{j',l}(t_i)\right )^2\right )\Big].
\end{align*}
The likelihood from the nodal attributes is given by
\begin{align*}
\mathcal{L}(\{\bx_j(t_1),...,\bx_j(t_T)\}_{j=1}^J)
\propto \prod_{j=1}^J\prod_{k=1}^{m}\prod_{i=1}^T 
N(x_{j,k}(t)|\eta_k(t)+\sum_{l=1}^L\bxi_{j,l}(t)^T\balpha_{k,l}(t),\sigma_k^2).
\end{align*}
The joint posterior of parameters with the multiplex graph and nodal attributes is given by
\begin{align*}
&\mathcal{L}(\{\bA_l^{(o)}(t_1),...,\bA_l^{(o)}(t_T)\}_{l=1}^L)\times   \mathcal{L}(\{\bx_j(t_1),...,\bx_j(t_T)\}_{j=1}^J) \\ 
&\times N(\bmu|\bzero,\bSigma(\bbeta_{\mu}))\times \prod_{k=1}^m N(\bet_k|\bzero,\bSigma(\bbeta_{\eta}))\times 
\prod_{j=1}^J\prod_{r=1}^R N(\bzeta_{j,r}|\bzero,\bSigma(\bbeta_{\zeta}))\times 
\prod_{j=1}^J\prod_{l=1}^L\prod_{r=1}^R N(\bxi_{j,l,r}|\bzero,\bSigma(\bbeta_{\xi}))\\
&\times \prod_{k=1}^m\prod_{l=1}^L\prod_{r=1}^R N(\balpha_{k,l,r}|\bzero,\bSigma(\bbeta_{\alpha}))
\times \prod_{i=1}^T\prod_{(j,j')\in\bU_l^{(o)}(t_i)}\prod_{l=1}^L
PG(\omega_{jj',l}^{(o)}(t_i)|1,0)\\
&\times p(\bbeta_{\eta})\times p(\bbeta_{\zeta})\times p(\bbeta_{\xi})\times p(\bbeta_{\mu})
\times \prod_{k=1}^m IG(\sigma_k^2|a_{\sigma},b_{\sigma}),
\end{align*}
where $\bmu=(\mu(t_1),...,\mu(t_T))^T$, $\bet_k=(\eta_k(t_1),...,\eta_k(t_T))^T$, $\bzeta_{j,r}=(\zeta_{j,r}(t_1),...,\zeta_{j,r}(t_T))^T$, $\bxi_{j,l,r}=(\xi_{j,l,r}(t_1),...,\xi_{j,l,r}(t_T))^T$ and $\balpha_{k,l,r}=(\alpha_{k,l,r}(t_1),...,\alpha_{k,l,r}(t_T))^T$. Also, $\bSigma(\bbeta_{\mu}),\bSigma(\bbeta_{\eta}),\bSigma(\bbeta_{\zeta}),$ $\bSigma(\bbeta_{\alpha}),\bSigma(\bbeta_{\xi})$ are all $T\times T$ covariance matrices with $(i,i')$th entries 
$\kappa_F(t_i,t_{i'};\bbeta_{\mu}),\kappa_F(t_i,t_{i'};\bbeta_{\eta}),$  $\kappa_F(t_i,t_{i'};\bbeta_{\zeta}),\kappa_F(t_i,t_{i'};\bbeta_{\alpha})$ and $\kappa_F(t_i,t_{i'};\bbeta_{\xi})$, respectively. While the  posterior distributions for the parameters are not available in closed forms, the augmented full conditional distributions belong to standard
families. Section 3 of the supplementary file provides details of the full conditional distributions of the parameters.

\noindent \underline{\textbf{Missing edge mechanisms and nodal attribute prediction.}} As noted under the ``Scientific Objectives with Security Implications" in Section \ref{RealDataIntro}, our framework assumes that unobserved edges are missing at random (MAR) \emph{conditional on the latent factors and observed data}, rather than representing structural zeros (edges with zero probability of existence). This distinction becomes particularly relevant in terrorism graphs where certain organization pairs may have negligible interaction probability due to insurmountable geographical, ideological, or operational constraints, but intelligence about such structural constraints is uncertain. Moreover, for dynamically evolving covert terrorism graphs, the distinction between ``impossible'' and ``highly unlikely'' connections is often ambiguous and may change over time. As an alternative, our modeling approach addresses this through two mechanisms: (i) through posterior inference, the model learns latent factor positions (see Section~\ref{sec:transitivity_cluster}) where organizations with no observed interactions tend to have large angular distances in the latent space, effectively assigning near-zero probabilities to their connections without requiring explicit constraints, and (ii) the non-stationary covariance structure of the NN-GP captures temporal variations in interaction possibilities, allowing the model to adapt when structural constraints evolve (e.g., when geographical barriers are overcome or ideological positions shift). While explicit modeling of structural zeros is a potential extension, the current framework's ability to learn these patterns from data is well-suited to the uncertain nature of covert dynamic terrorism graphs.

Let $\mu^{(q)}(t_i),\bzeta_{j}^{(q)}(t_i),\bxi_{j,l}^{(q)}(t_i)$, $q=1,..,Q$, be the $Q$ post burn-in samples obtained from the MCMC chain.
To draw inference on unobserved edge $a_{jj',l}^{(u)}(t_i), (j,j')\in\bU_l^{(u)}(t_i)$, we draw samples $a_{jj',l,q}^{(u)}(t_i)$ from Ber($p_{jj',l,q}^{(u)}(t_i)$),
\begin{align*}
\text{logit}(p_{jj',l,q}^{(u)}(t_i))=\mu^{(q)}(t_i) +\bzeta_{j}^{(q)}(t_i)^T \bzeta_{j'}^{(q)}(t_i)+ \bxi_{j,l}^{(q)}(t_i)^T \bxi_{j',l}^{(q)}(t_i),\:q=1,..,Q.   
\end{align*}
We predict an edge if $\frac{1}{Q}\sum_{q=1}^Qa_{jj',l,q}^{(u)}(t_i)>\Delta_c$, $0<\Delta_c<1$. Although $\Delta_c$ is typically chosen as $0.5$, simulation studies show the performance of our approach by varying the cutoff $\Delta_c$ in a grid between $(0,1)$ and reporting the area under the ROC curve (AUC). Relying on point estimates alone can be misleading, especially when the stakes involve counter-terrorism decision-making. Uncertainty quantification (UQ) provides probabilistic confidence about whether a predicted link truly exists, allowing intelligence agencies to prioritize investigations based on the strength of evidence.  The quantity $\frac{1}{Q}\sum_{q=1}^Qa_{jj',l,q}^{(u)}(t_i)$ specifies uncertainty in estimating an edge between two nodes $\mathcal{N}_j$ and $\mathcal{N}_{j'}$ at the $l$th layer at time $t_i$. Similarly, to predict an unobserved $x_{j,k}(t_i)$, we draw MCMC samples from N($\eta_k^{(q)}(t_i)+\sum_{l=1}^L\bxi_{j,l}^{(q)}(t_i)^T\balpha_{k,l}^{(q)}(t_i),\sigma_k^{(q)2})$, for $q=1,..,Q$. The point prediction of $x_{j,k}(t_i)$ is obtained from the mean/median of these samples. In addition, these samples will allow empirical estimation of the posterior distribution of $x_{j,k}(t_i)$, from which the 95\% predictive intervals, that quantify the uncertainty in prediction, can be obtained. Attributes inferred with narrow intervals/high certainty may directly inform strategic decisions, while those with wider intervals/high uncertainty may guide further intelligence gathering.

\noindent\underline{\textbf{Out-of-sample edge and nodal attribute prediction.}} To draw inference on the edge between the $j$th and $j'$th node at a future time $t>t_1,...,t_T$, we draw posterior samples from $\mu^{(q)}(t)|\mu^{(q)}(t_1),...,\mu^{(q)}(t_T)$, $q=1,...,Q$, which follows a normal distribution determined by the NN covariance kernel of the GP. Posterior samples are also drawn from $\bzeta_j^{(q)}(t)|\bzeta_j^{(q)}(t_1),...,\bzeta_j^{(q)}(t_T)$, $\bxi_{j,l}^{(q)}(t)|\bxi_{j,l}^{(q)}(t_1),...,\bxi_{j,l}^{(q)}(t_T)$ and from $\balpha_{k,l}^{(q)}(t)|\balpha_{k,l}^{(q)}(t_1),...,\balpha_{k,l}^{(q)}(t_T)$. To draw inference on the edge $a_{jj',l}^{(u)}(t)$, we draw samples $a_{jj',l,q}(t)$ from Ber($p_{jj',l,q}(t)$), where
\begin{align*}
\text{logit}(p_{jj',l,q}(t))=\mu^{(q)}(t) +\bzeta_{j}^{(q)}(t)^T \bzeta_{j'}^{(q)}(t)+ \bxi_{j,l}^{(q)}(t)^T \bxi_{j',l}^{(q)}(t),\:q=1,..,Q.  
\end{align*}
We predict an edge if $\frac{1}{Q}\sum_{q=1}^Qa_{jj',l,q}(t)>\Delta_c$ and report AUC by varying the cut-off $\Delta_c$ in a grid between $(0,1)$. The quantity $\frac{1}{Q}\sum_{q=1}^Qa_{jj',l,q}(t)$ also specifies uncertainty in estimating an edge between two nodes $\mathcal{N}_j$ and $\mathcal{N}_{j'}$ at the $l$th layer at time $t$. To predict an unobserved nodal attribute $x_{j,k}(t)$ at time $t$, we draw MCMC samples from
N($\eta_k^{(q)}(t)+\sum_{l=1}^L\bxi_{j,l}^{(q)}(t)^T\balpha_{k,l}^{(q)}(t),\sigma_k^{(q)2}$), for $q=1,..,Q$. The point prediction and uncertainty are obtained from these MCMC samples.

\subsection{Computational Complexity}
The computational complexity of the proposed model is dominated by the complexity of
updating $\bmu=(\mu(t_i):i=1,..,T)^T,\bet_k=(\eta_k(t_i):i=1,..,T)^T,\bzeta_{j,r}=(\zeta_{j,r}(t_i):i=1,..,T)^T, \bxi_{j,l,r}=(\xi_{j,l,r}(t_i):i=1,..,T)^T, \balpha_{k,l,r}=(\alpha_{k,l,r}(t_i):i=1,..,T)^T$. The computational burden of updating these quantities is primarily governed by two parameters: the number of graph nodes $J$ and the number of time points $T$. Table \ref{table_comp_time} shows run time per MCMC iteration for different choices of $J$ and $T$. All computation times are based on model runs using a compute cluster where each of the twelve compute nodes contains two AMD EPYC 7763 processors with a combined 128 cores, 256 threads, and 1 TB of RAM.
The table shows that the computational complexity of each iteration scales $O(J^2)$ with the number of graph nodes and $O(T)$ with the number of time points. Hence, the Gibbs sampling algorithm can support inference for a moderately sized dataset. To further ease the computational burden for large $J$, we propose to employ a distributed inferential process which splits a multiplex graph into subgraphs, fits the model with data over these small subgraphs, followed by combining inferences over these different subgraphs. While this strategy has been successfully employed in Gaussian process models and their sparse approximations for correlated data \cite{guhaniyogi2018meta, guhaniyogi2023distributed}, it is yet to be investigated for graph-based stochastic process models like the one proposed here. However, exploring this strategy is beyond the scope of this article and we plan to investigate it in future work.

\begin{table}[H]
\centering
\begin{tabular}{c  c | c c c c c c c c c c}
& & \multicolumn{10}{c}{$T$} \\
& & 10 & 20 & 30 & 40 & 50 & 60 & 70 & 80 & 90 & 100 \\
\hline
\multirow{12}{*}{$J$} & 10 & 0.41  &  0.84  &  1.20  &  1.59   & 1.97  &  2.44  & 3.15 &   3.66  &  4.47  &   5.08 \\
\cline{2-12}
 & 20 & 1.27 &  2.55  & 3.62 &  4.68  & 6.05 &  6.84 &  9.60 & 11.29 & 12.84   & 14.95 \\
\cline{2-12}
 & 30 & 2.55 &  5.19  & 7.53 &  9.62 & 12.05 & 15.10 & 19.52 & 23.53 & 26.12 &  30.38\\
\cline{2-12}
 & 40 & 4.26  & 8.82  & 12.45 & 15.87  & 19.81 & 24.56 & 33.29 & 38.59 & 46.96 &  51.68 \\
\cline{2-12}
 & 50 & 6.53 & 12.75 & 18.63 & 23.79  & 30.35 & 36.37 & 53.84 & 59.95 & 66.61 &  79.39 \\
\cline{2-12}
& 60 & 9.72 & 17.88 & 27.20 & 35.18 & 45.80 & 50.11 & 71.92 & 86.28 & 95.61  & 108.60\\
\cline{2-12}
& 80 & 16.42 & 31.04 &  45.18 & 59.05 & 72.84 & 87.97 & 130.07 & 146.47 & 166.11 &  200.71  \\
\cline{2-12}
& 100 & 25.75 & 46.25 & 71.06 & 91.27 & 112.53 & 148.13 & 191.87 & 220.88 & 262.17 & 288.00 \\
\cline{2-12}
& 120 & 37.58 & 69.31 & 99.61 & 131.76 & 168.30 & 199.87 & 261.34 & 315.22 & 371.86 & 410.20 \\
\cline{2-12}
& 150 & 58.04 & 105.34 & 154.23 & 195.52 & 248.95 & 302.01 & 425.40 & 496.83 & 586.24 & 663.90 \\
\cline{2-12}
& 180 & 87.27 & 148.08 & 212.02 & 290.04 & 343.62 & 427.67 & 603.49 & 737.42 & 806.63 &  958.84 \\
\cline{2-12}
& 200 & 94.92 & 176.45 & 265.47 & 349.97 & 434.93 & 569.09 & 740.89 & 855.75 & 984.33 & 1164.02 \\
\hline
\end{tabular}
\caption{Computation time in seconds per iteration for data generated with different values of $J$ and $T$. In each case, the data is generated with $L = 2$, $m = 5$, and DJL is fit with $R = 4$, $F = 1$.}
\label{table_comp_time}
\end{table}

\section{Analysis of Covert Terrorist Graph Data}\label{sec:realdata}

In the terrorism graph data described in Section \ref{RealDataIntro}, there are $J=15$ organizations, $L=2$ relationship types, and $T=14$ time points, with the final time point being held out. Additionally, after standardizing the continuous organizational attributes and one-hot encoding the categorical organizational attributes, the number of nodal attributes in the analysis is $m=14.$ These rich, time-varying multiplex graphs with nodal attributes allow a comprehensive view of both the evolving relationships and internal dynamics of terrorist organizations.

Our analyses directly address the three critical security questions \textbf{(Q1)-(Q3)} outlined in Section \ref{RealDataIntro}, demonstrating how the proposed methodology can yield actionable intelligence insights that conventional approaches may fail to uncover. Specifically, we evaluate our framework's ability to reconstruct hidden relationships, detect organizational transformations, and predict cascading effects within the terrorist ecosystem.

We assess the performance of some competing methods relative to our approach by comparing their ability to predict edges and attributes across three settings: \emph{in-sample}, \emph{missing}, and \emph{out-of-sample}, following the evaluation framework used in Section 2 of the supplementary file (Simulation Studies), where the competing methods are described in detail.
The \emph{in-sample} scenario refers to predicting edges and attributes that are part of the dataset used to train the model. The \emph{missing} scenario involves predicting graph edges within the time points $t_1,...,t_T$, which were masked during the training of all models. The results for predicting nodal attributes are also presented for the \emph{missing} scenario. The \emph{out-of-sample} scenario evaluates performance on edges and attributes at time points beyond $t_1,...,t_T$. For the out-of-sample assessment, we simulate the multiplex graphs and nodal attributes at additional future time points.

To assess performance in the \emph{missing} scenario, edges are deliberately masked at different time points for each layer to construct \emph{partially observed} graphs. The time points with missing edges are selected with a $0.1$ probability, and edges are removed with a $0.25$ probability at each selected time point. To assess \emph{out-of-sample} performance, data from the held out, final time point is used to predict edges and attributes.

We fit the proposed joint model for multiplex graphs and nodal attributes, referred to as the \emph{dynamic joint learner} (DJL), by setting the dimensions of the shared and layer-specific latent factors equal, i.e., $R=R_\zeta=4$. This assumption is not required, as we demonstrate in Section 2 of the supplementary file. Further $F=1$ is assumed based on the simulation results in Section 2 of the supplementary file. The expression for NN-GP kernel with $F=1$ is available in Section 1 of the supplementary file. To evaluate the benefits of jointly modeling nodal attributes and multiplex graphs, we compare our approach with two competitors. The first, \emph{dynamic marginal learner (DML)}, fits separate models for multiplex graphs (Model (\ref{eq:network_model_logit})) and nodal attributes (Model (\ref{eq:nodal_attribute_identity_link})), allowing us to assess the advantages of joint modeling. The second competitor, \emph{joint latent factor model (JLAFAC)} \citep{guhaniyogi2020joint}, integrates a single time-varying graph layer with nodal attributes using AR(1) processes for latent functions. For the terrorism study with $L=2$ graph layers, we fit JLAFAC separately to each layer and the nodal attributes, referring to these models as \emph{JLAFAC(first)} and \emph{JLAFAC(second)}.

\subsection{Graph Reconstruction, Prediction of Organizational Attributes and their Operational Intelligence Implications}

\begin{figure}[h!]
    \centering
    \subfloat[Edge Prediction]{\includegraphics[width=.3\linewidth]{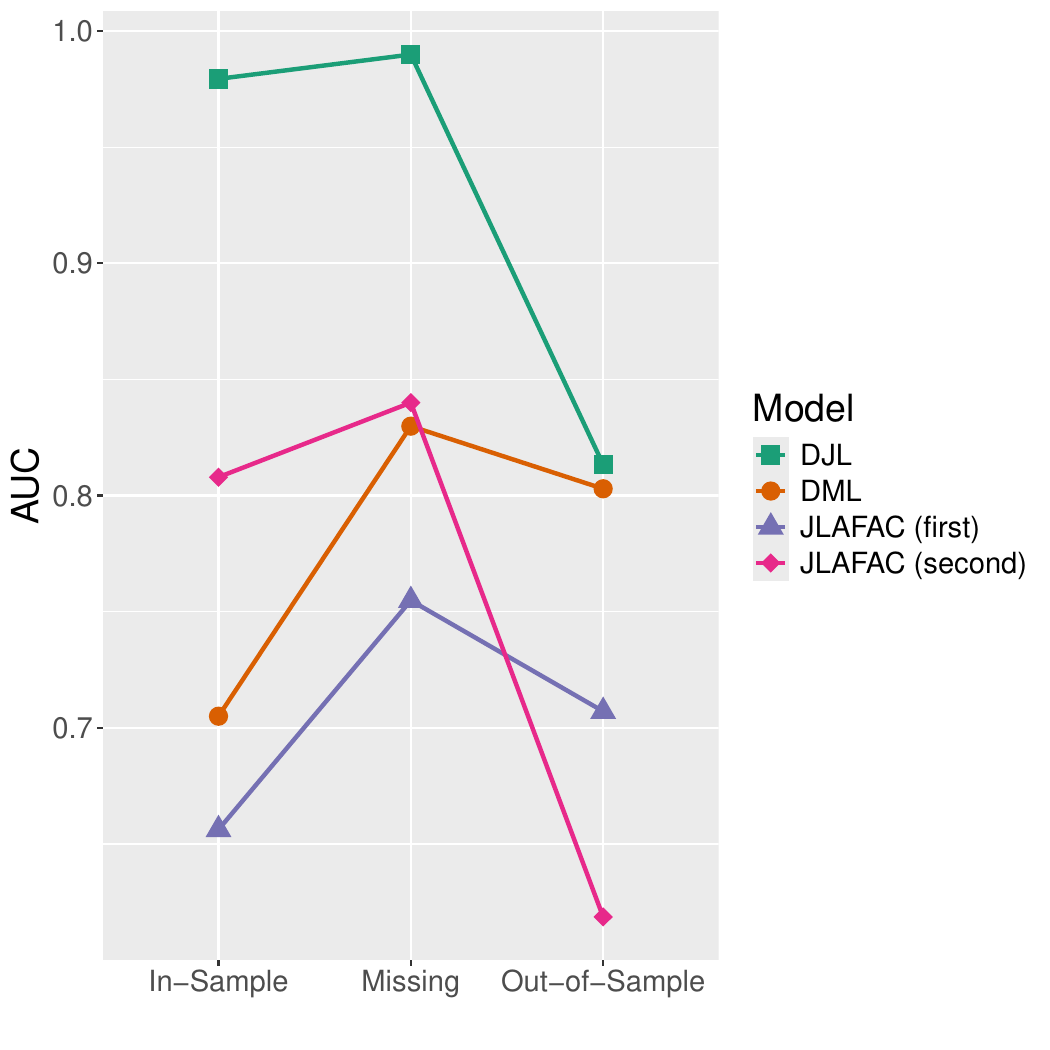}}
    \subfloat[Nodal Attribute Prediction] {\includegraphics[width=.3\linewidth]{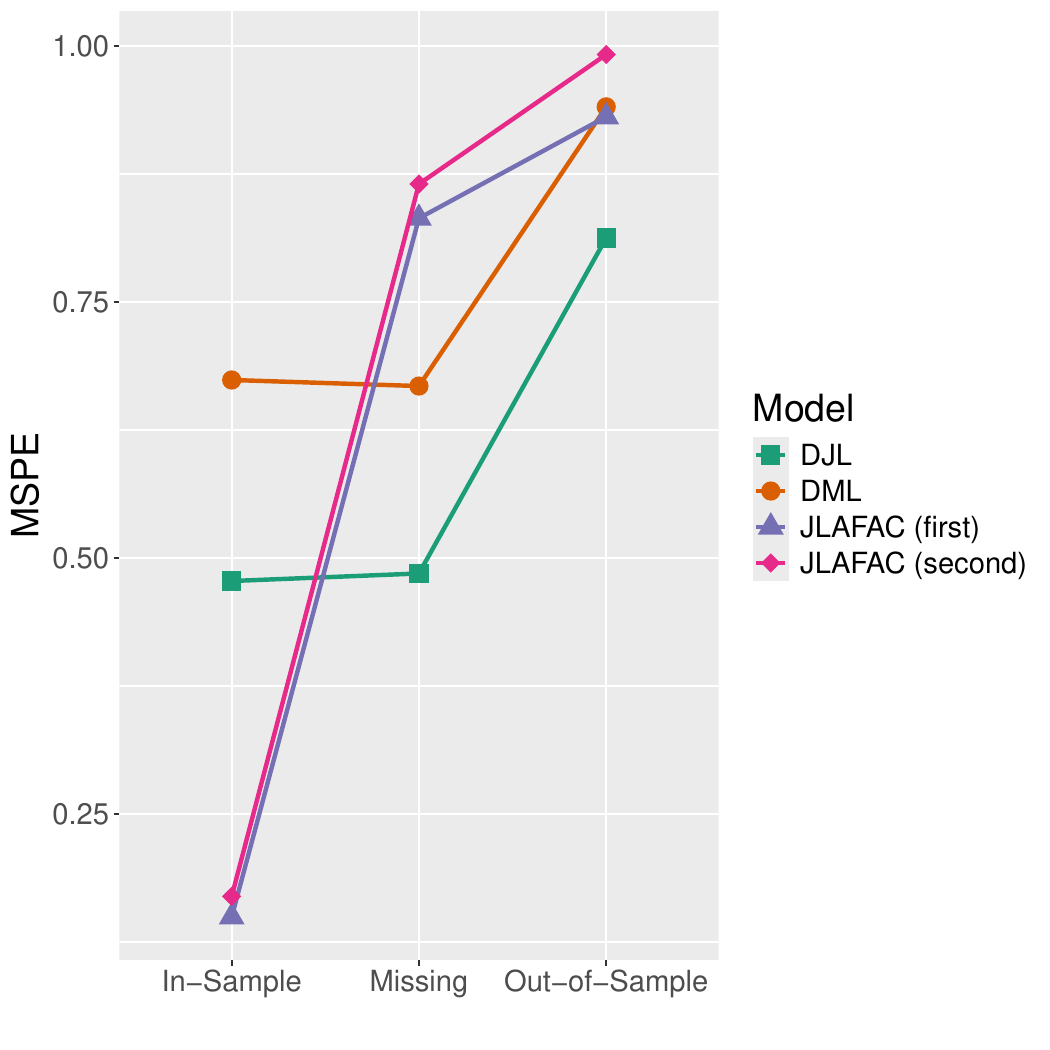}} \\
    \subfloat[95\% PI Coverage]{\includegraphics[width=.3\linewidth]{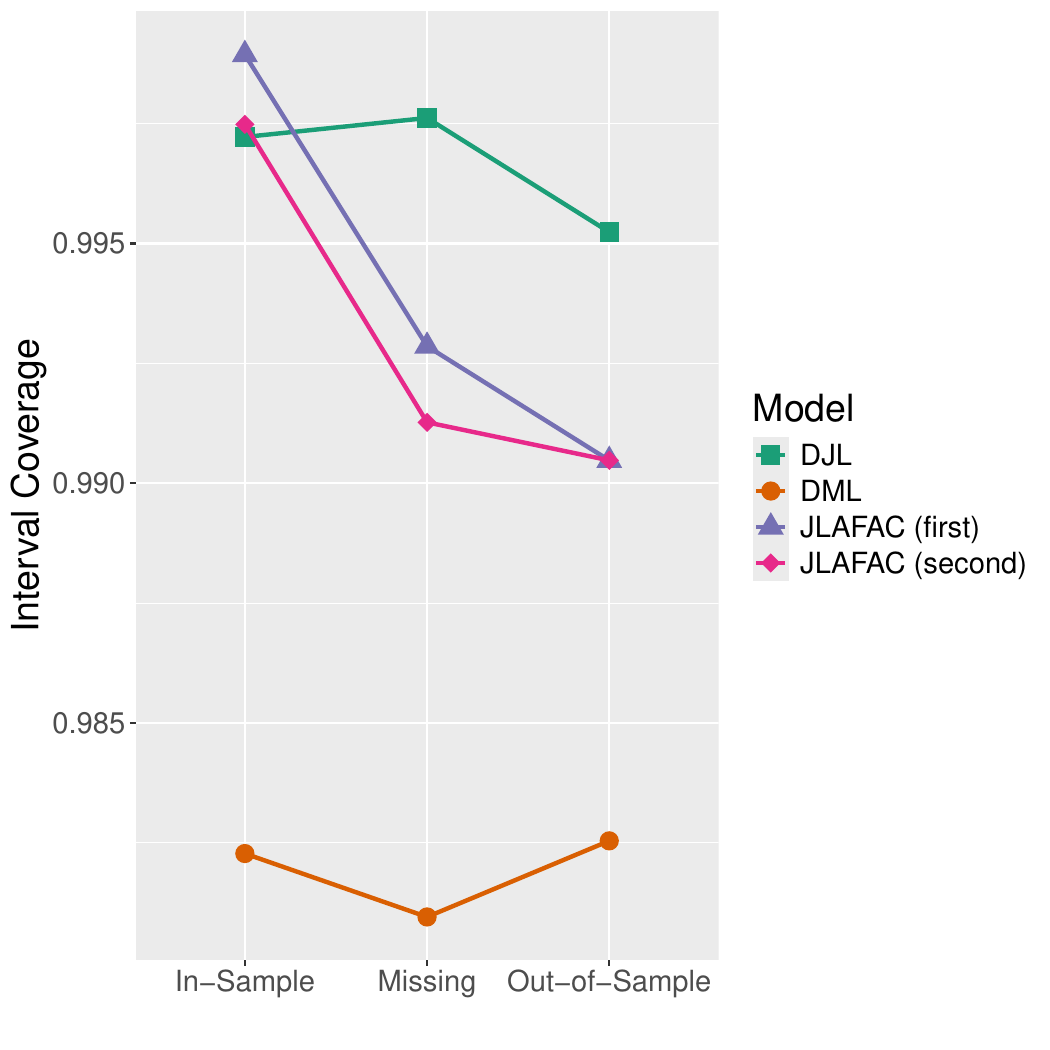}}
    \subfloat[95\% PI Length] {\includegraphics[width=.3\linewidth]{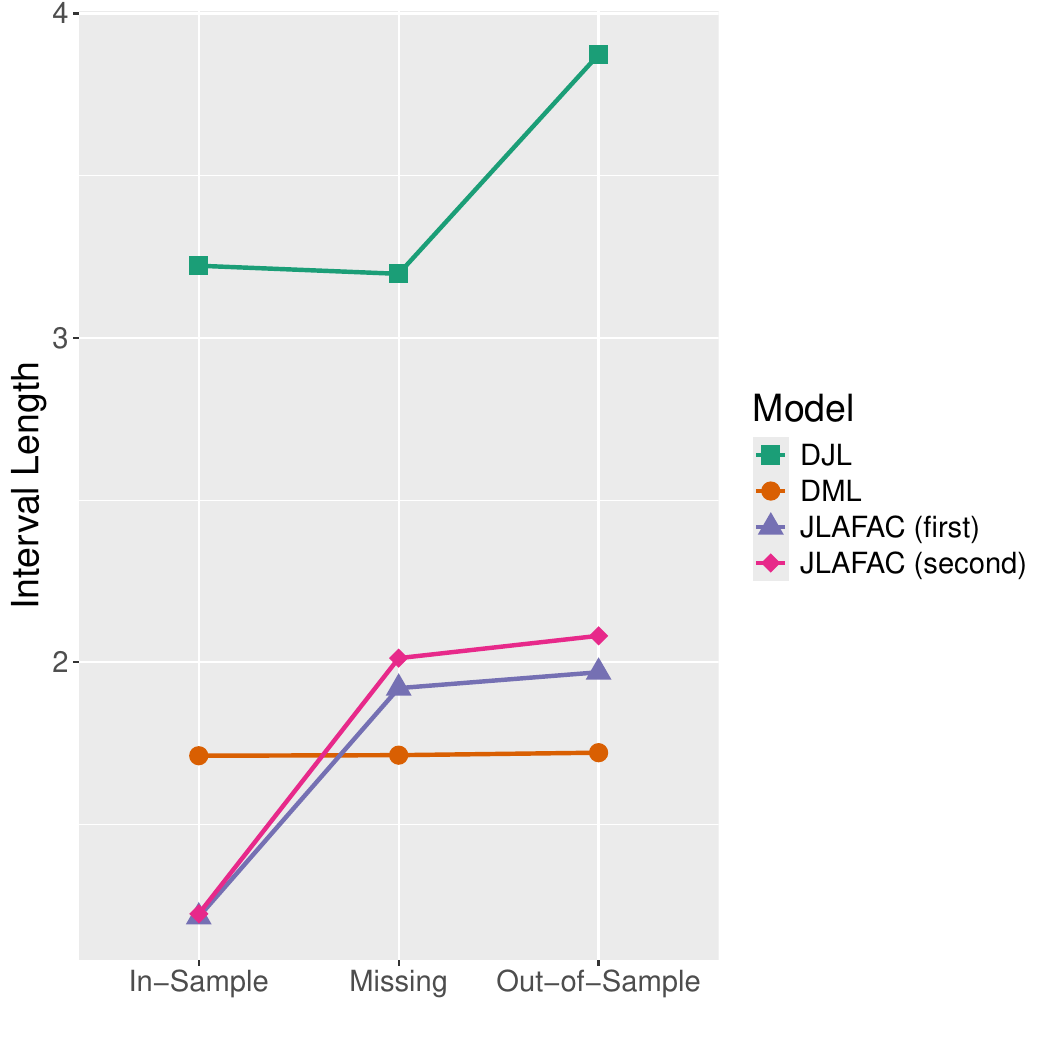}} 
    \caption{First row shows results for edge prediction and point prediction of nodal attributes for the proposed DJL, along with DML, JLAFAC(first) and JLAFAC(second) at \emph{in-sample}, \emph{missing} and \emph{out-of-sample}
    scenarios. The second row shows coverage and length of 95\% prediction intervals (PIs) for nodal attributes for all competitors at \emph{in-sample}, \emph{missing} and \emph{out-of-sample} scenarios.}\label{real_data_fig}
\end{figure}

Figure \ref{real_data_fig} demonstrates DJL's superior edge prediction performance, achieving near-perfect AUC scores in the \emph{in-sample} and \emph{missing} scenarios and around 0.8 in \emph{out-of-sample} predictions. These performance metrics have direct operational significance: with 25\% of relationships deliberately masked to simulate realistic intelligence conditions, the model's ability to achieve AUC scores exceeding 0.93 for missing link prediction means that intelligence agencies operating with incomplete information may identify high-probability unobserved connections with sufficient confidence to guide intelligence and resource allocation. Competing models perform significantly worse in the \emph{in-sample} and \emph{missing} cases, with only DML showing competitive \emph{out-of-sample} results with DJL. DJL also excels in predicting nodal attributes, particularly in \emph{missing} and \emph{out-of-sample} scenarios, offering reliable forecasts of organizational changes. While all models slightly over-cover 95\% predictive intervals, DJL produces marginally wider intervals, providing conservative uncertainty estimates. 

The superior performance of DJL compared to traditional approaches validates the importance of joint modeling for intelligence applications. The separate modeling approaches (DML) and single-layer joint models (JLAFAC) consistently show under-performance, particularly in out-of-sample scenarios where DJL maintains approximately 80\% accuracy, while competitors show significant degradation. This performance gap may have direct operational relevance: intelligence assessments that fail to capture graph-attribute interdependencies may miss critical signals that precede escalations in terrorist activity.

Furthermore, the 95\% prediction intervals in DJL enable risk-calibrated decision-making. When predicting unobserved relationships, uncertainty is critical information in presence of false positives that may misdirect limited surveillance resources, or false negatives that could leave threats undetected. The slightly wider prediction intervals produced by DJL compared to competitors reflect a more honest assessment of uncertainty, a factor that is essential for operational decisions where overconfidence can have catastrophic consequences \citep{betts1978analysis}.

\subsection{Detection of Complex Alliance-Rivalry Dual Relationships}
\begin{figure}[h]
    \centering
    \includegraphics[width=0.55\linewidth]{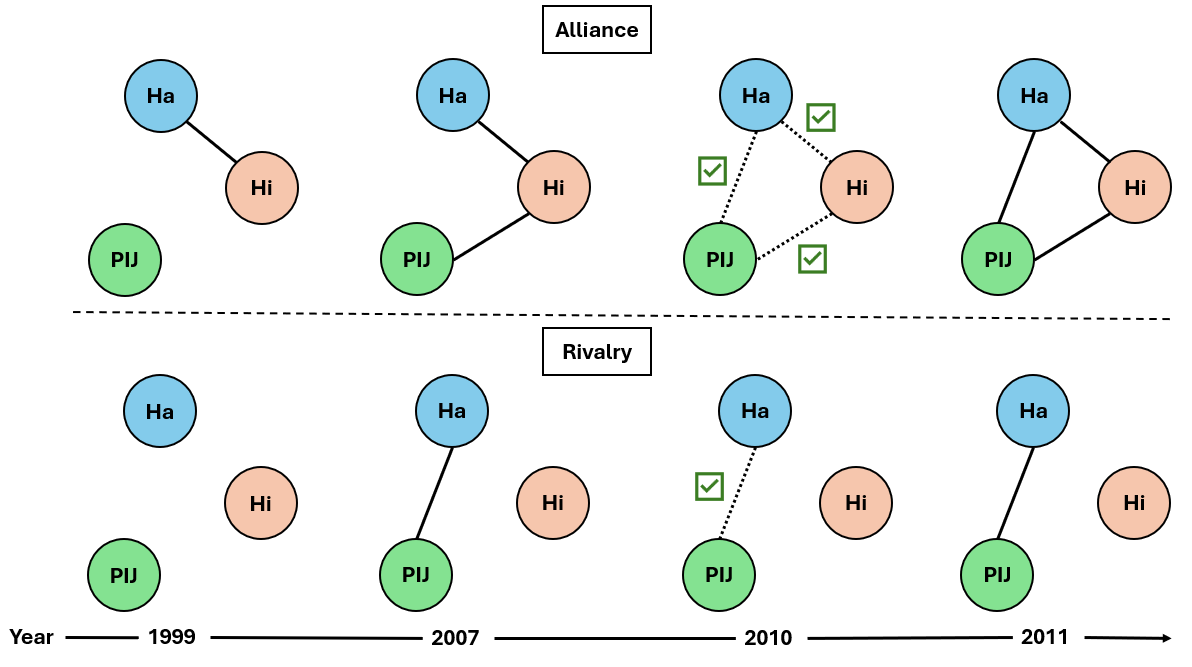}
    \caption{Visual represention of the dynamic relationships between between Hizballah (Hi), Hamas (Ha), and the Palestinian Islamic Jihad (PIJ) over time. The edges in 2010 are dotted and have a check mark to represent that they were unobserved and correctly predicted by DJL as being present.}
    \label{real_example_diagram}
\end{figure}

The evolving relationships among Hizballah, Hamas, and the Palestinian Islamic Jihad (PIJ), portrayed in Figure~\ref{real_example_diagram}, highlight key dynamics over time. Initially, Hizballah and the PIJ had no reported alliance or rivalry in 1998-1999, but DJL correctly predicted their alliance in subsequent years. Similarly, DJL accurately forecasted the persistent alliance between Hizballah and Hamas in 2010. DJL also identified the complex simultaneous alliance and rivalry between Hamas and the PIJ in 2010, reflecting their shifting relations: no ties in 1998-1999, rivalry in 2006-2007, and revived alliances during and after the Gaza War (2008-2009). These predictions align with historical events such as the Second Intifada \cite{baad_db, Hamas_Narrative, Hizballah_Narrative}, Gaza Civil War \cite{Levitt_2005, NCTC_2023}, and conflicts with Israel \cite{Deutsche_Welle_2024}, emphasizing the nuanced interplay of alliances and rivalries among these groups.

The model's ability to detect such dual relationships addresses a critical intelligence gap identified by our first security question \textbf{(Q1)} in Section \ref{RealDataIntro}. Traditional intelligence assessments often categorize relationships as either cooperative or adversarial, and may miss nuanced realities where organizations collaborate in military operations against common enemies competing for political control and resources \citep{bacon2018terrorist}. The simultaneous alliance-rivalry between Hamas and PIJ reflects this complexity via the joint operations against Israel combined with competition for Palestinian political leadership. In particular, this intelligence may have immediate operational relevance for diplomatic strategies that must engage multiple stakeholders with overlapping but competing interests.

\subsection{Early Warning Through Organizational Transformation 
}
DJL also offer insights into the dynamic relationship between the Revolutionary Armed Forces of Colombia (FARC) and the National Liberation Army (ELN). In 1998, they were allied without rivalry, but by 1999, their relationship evolved into a simultaneous alliance and rivalry, which persisted in subsequent years. DJL accurately predicted this pattern for the missing year of 2005, by which the dual nature of their relationship had solidified \cite{baad_db}. Their alliance stemmed from joint efforts to combat the Colombian government and paramilitaries, maintaining territorial control and operational sustainability \cite{Farah_1998}. However, rivalry emerged in 1999 as peace talks began between the government and FARC, initially excluding the ELN \cite{state_col_1999}. This rivalry involved violence as both groups sought leverage in negotiations and to elevate their status \cite{BBC_News_2005}. By 2005, this complex mix of cooperation and opposition continued, reflecting the intricate dynamics DJL predicted.

This case directly addresses our second security question \textbf{(Q2)} regarding early detection of organizational transformations. The model's joint analysis of graph evolution and organizational attributes has captured how external intervention has triggered measurable changes in both organizations' operational characteristics while simultaneously altering their relationship dynamics. The framework detected these transformation patterns years before they fully manifested, demonstrating the potential for providing intelligence agencies with months or years of lead time before organizational changes result in operational escalation. The MSPE values below 0.5 for attribute prediction in missing scenarios confirm that such organizational changes can be reliably detected even when direct observation is limited.

\subsection{Inferences on Transitivity of the Terrorism Graph}\label{sec:transitivity_cluster}
Transitivity in a graph fosters tightly knit structures that naturally lead to clustering, reflecting cohesive communities where information, resources, and behaviors can spread rapidly. Transitivity-induced clustering is especially important for terrorism graphs, as it amplifies cascading effects, allowing strategies, ideologies, or operational tactics to diffuse quickly within and across organizations. This directly relates to the third critical question (\textbf{Q3}), which concerns how changes in one or a few organizations may cascade to others. To investigate this, we draw inference on the latent positions of each organization over time.

\begin{figure}[H]
    \centering
    \subfloat[1998]{\includegraphics[width=0.25\textwidth]{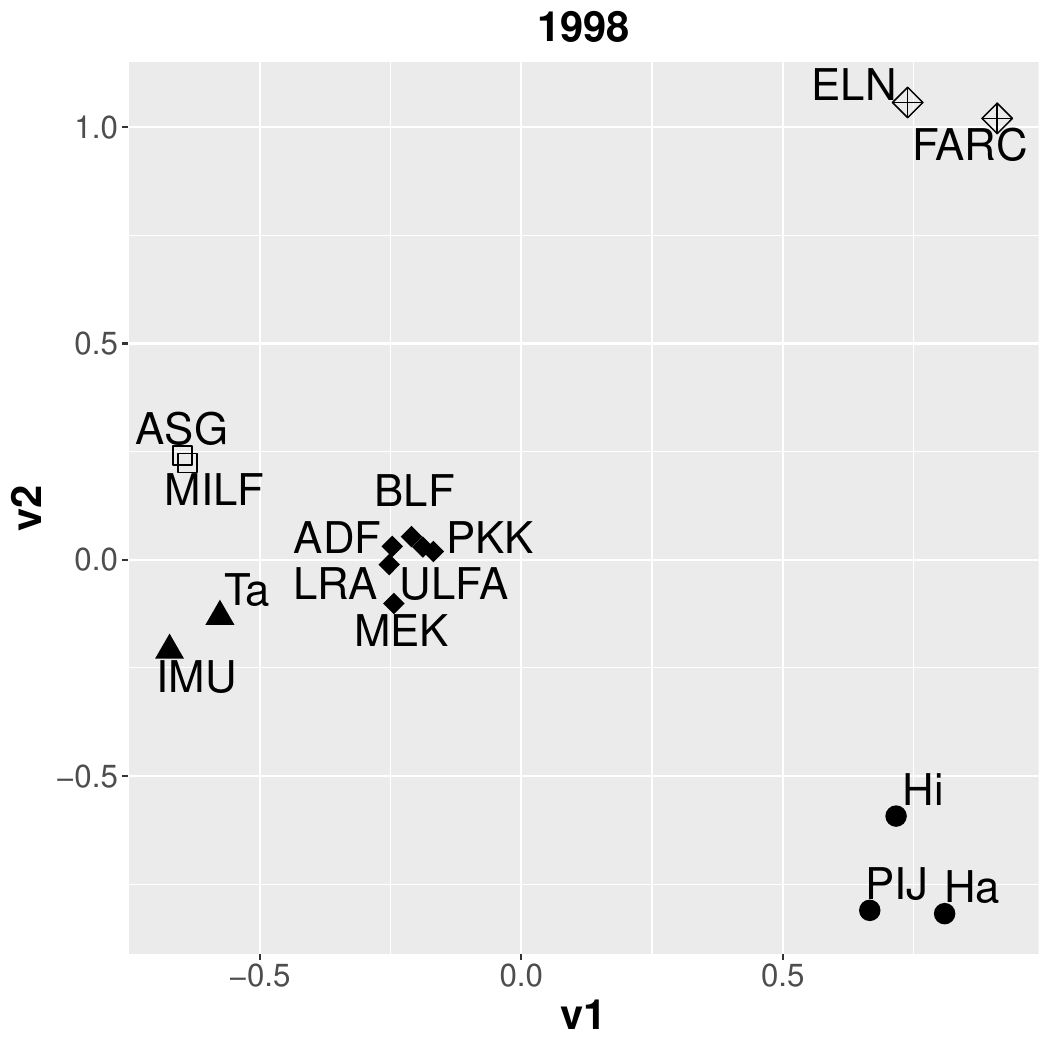}}
    \hfill
    \subfloat[2002]{\includegraphics[width=0.25\textwidth]{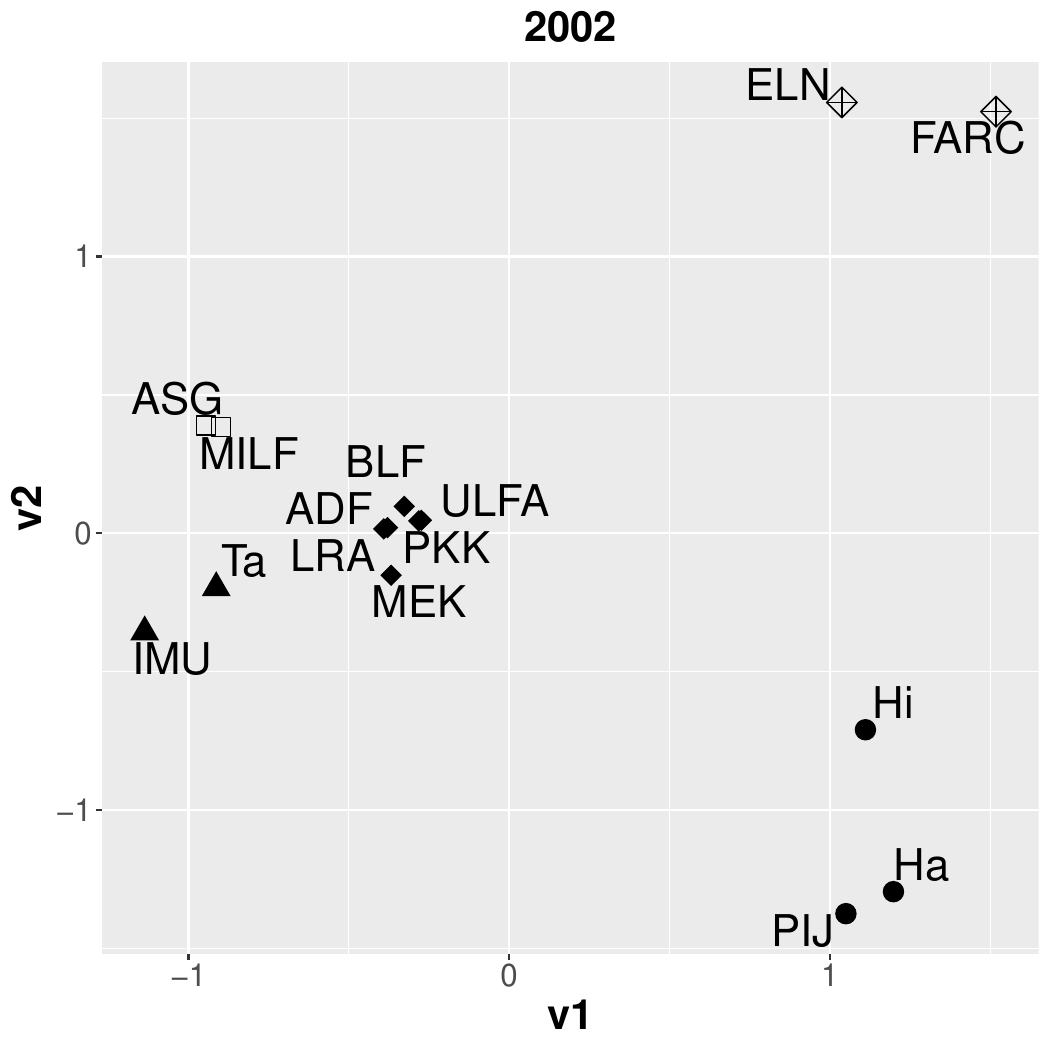}}
    \hfill
    \subfloat[2005]{\includegraphics[width=0.25\textwidth]{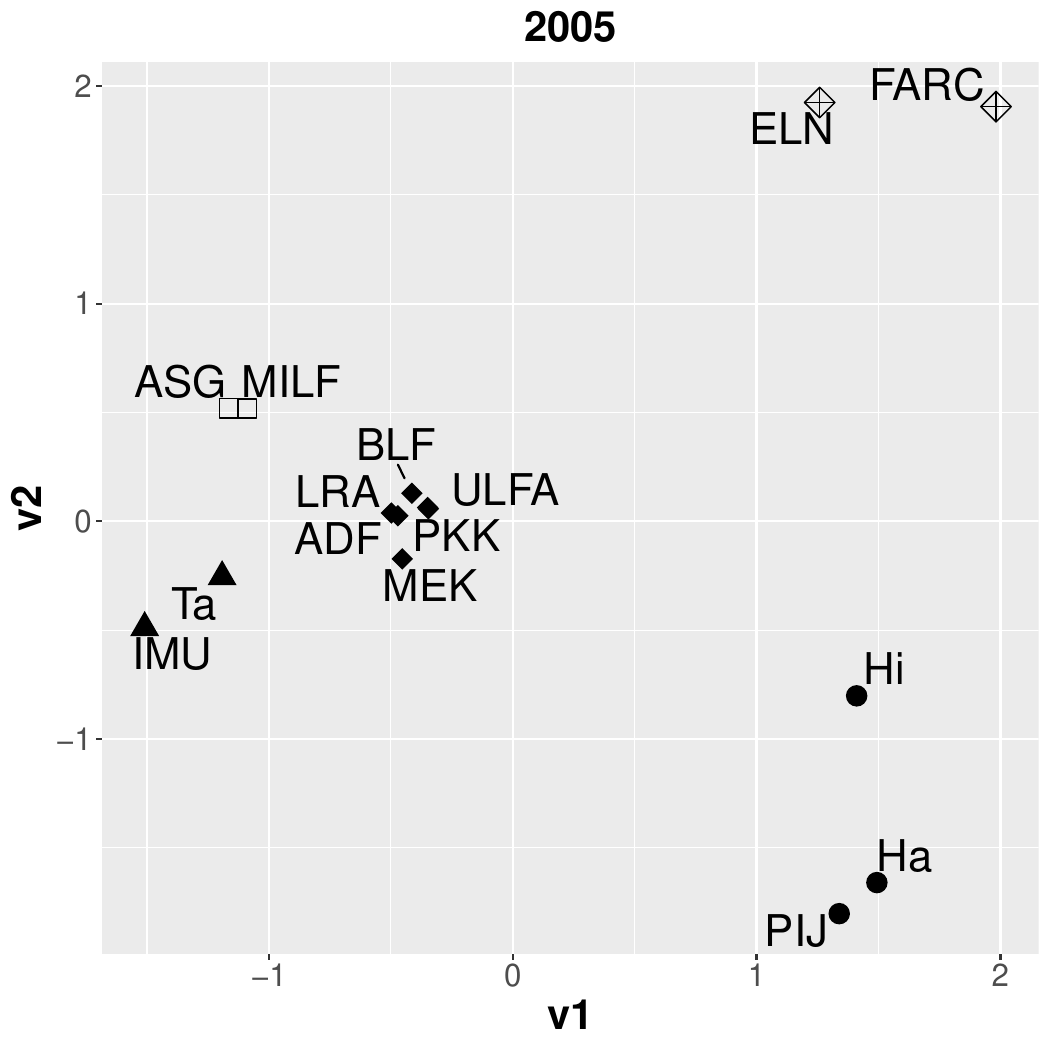}}\\
    \subfloat[2007]{\includegraphics[width=0.25\textwidth]{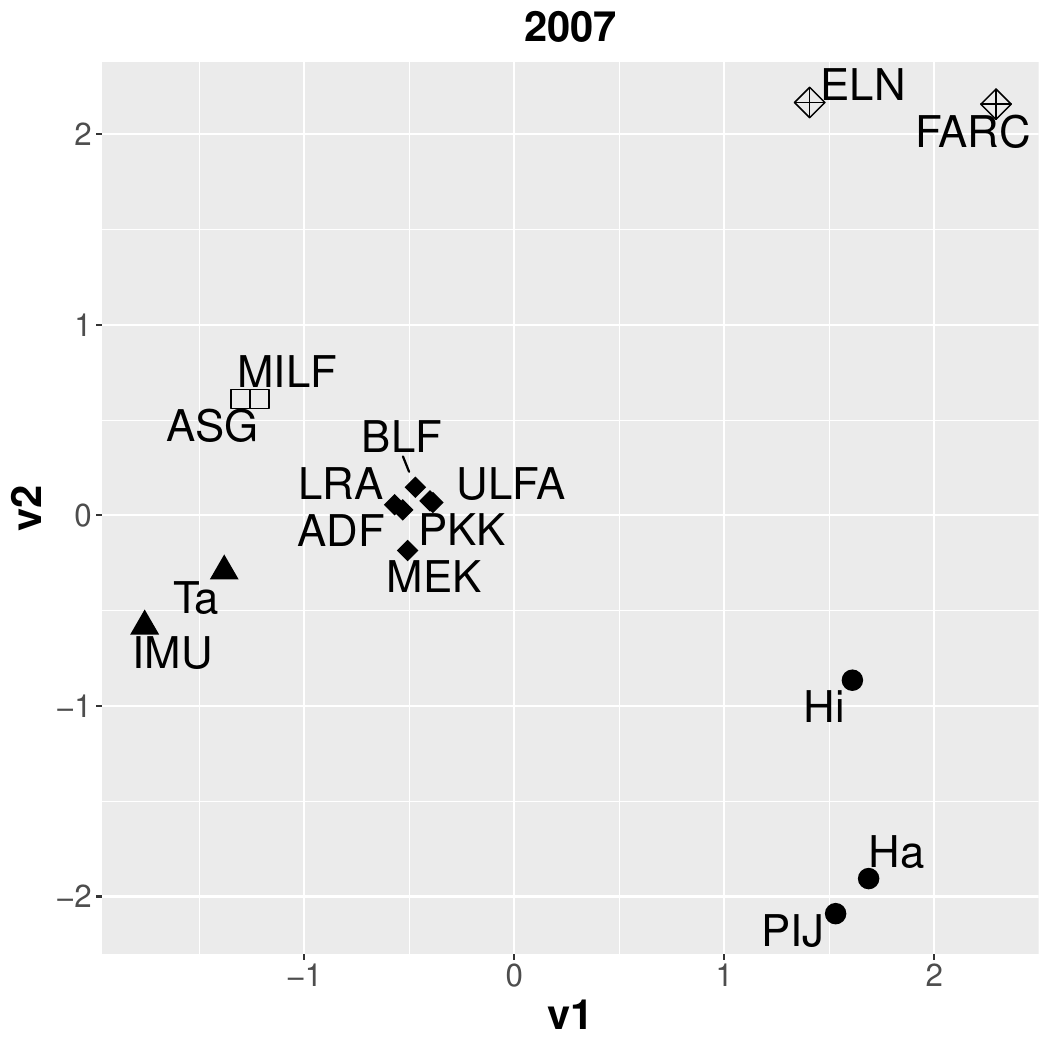}}
    \hfill
    \subfloat[2010]{\includegraphics[width=0.25\textwidth]{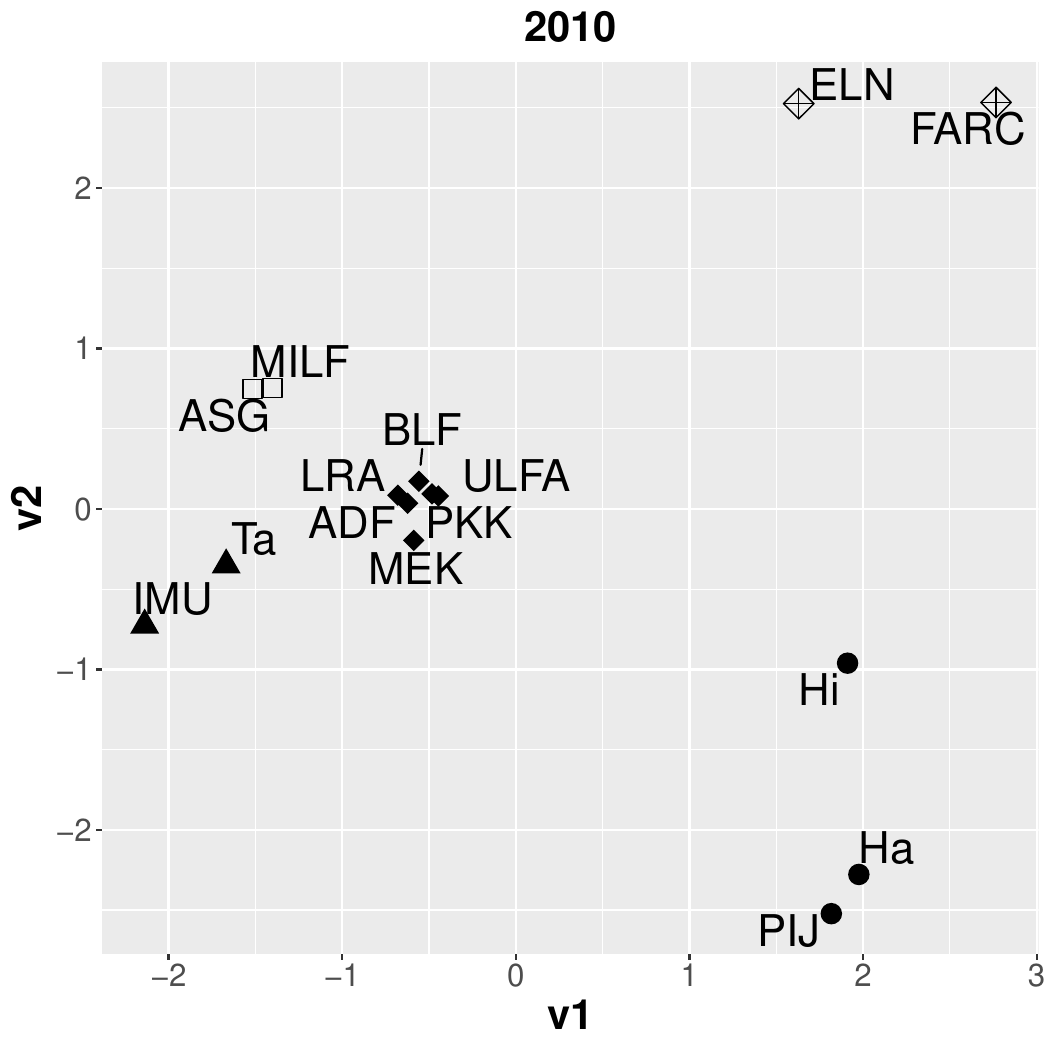}}
    \hfill
    \subfloat[2012]{\includegraphics[width=0.25\textwidth]{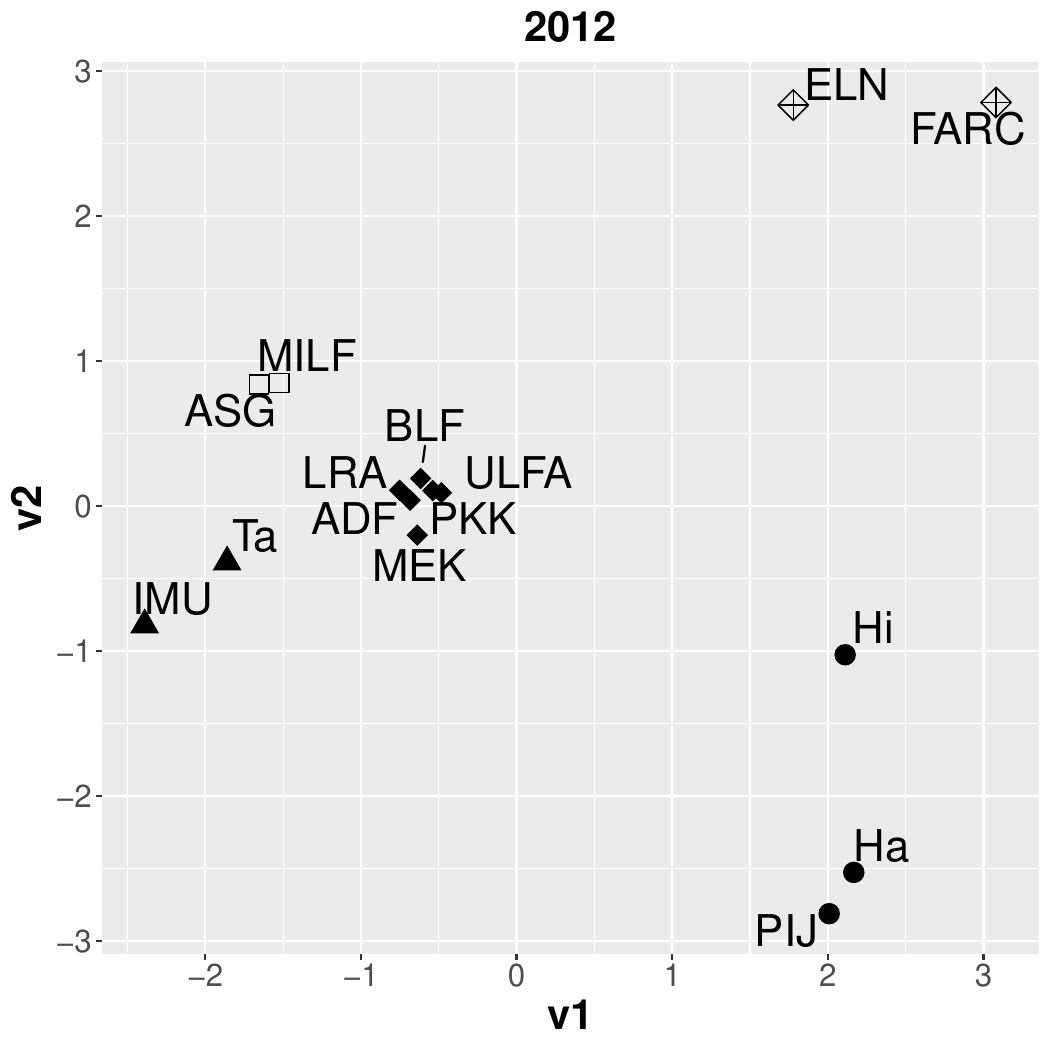}}
    \caption{Plots over time of the first two principal components of the latent factors, $\bzeta_{j}(t)$, for each organization after performing Procrustes transformations.}\label{latent_plots}
\end{figure}

Because the shared latent functions $\bzeta_j(t)$ are identifiable up to rotation, we align posterior samples to a common orientation using Procrustean transformations \citep{Sibson1978, hoff2005bilinear}. Specifically, for each $t=t_1, ..., t_T$, we obtain $\bZ(t) =[\bzeta_1(t),\ldots,\bzeta_J(t)]^T \in \mathbb{R}^{J \times R}$. Letting $\bZ^{(q)}(t)$ denote the $q$-th MCMC iterate and $\bZ_0(t)$ the reference matrix (taken as the first post–burn-in sample), we compute the rotated version $\widetilde{\bZ}^{(q)}(t)$ that minimizes its squared distance from $\bZ_0(t)$. The posterior mean $\widehat{\bZ}(t)=\dfrac{1}{Q}\sum_{q=1}^Q \widetilde{\bZ}^{(q)}(t)$ serves as our estimate of $\bZ(t)$. We then apply PCA to $\widehat{\bZ}(t)$, using the first two principal components ($v1, v2$) to visualize latent positions. The results, shown in Figure~\ref{latent_plots} for a few time points, illustrate dynamic community structures, including both observed and out-of-sample predictions (e.g., for 2012).

The analysis reveals interpretable clustering patterns aligned with known geopolitical contexts: Hizballah, Hamas, and Palestinian Islamic Jihad forming a Palestinian conflict cluster \citep{hamas_2024}; ELN and FARC in Colombia \citep{eln_2022}; the Taliban and the Islamic Movement of Uzbekistan (IMU) in Central Asia \citep{IMU_2006}; Moro Islamic Liberation Front (MILF) and Abu Sayyaf (ASG) in the Philippines \citep{abu_sayyaf_2006}; and separatist groups such as Kurdistan Workers' Party (PKK),  Mujahedin-e Khalq (MEK), and Baloch Liberation Front (BLF) forming another community \citep{pkk_2025, mek_2025, blf_2025}. Finally, the clusters of MILF and ASG appear close in latent space to the cluster of the Taliban and IMU, suggesting greater cooperation or rivalry between these two clusters compared to their ties with the ELN–FARC cluster, which is likely weakened by geographic distance. These findings demonstrate how transitivity and clustering help uncover cascading effects across terrorist organizations.

\section{Conclusion and Future Work}\label{sec:conclusion}
This article introduces a novel modeling framework to address the dynamic co-evolution of multiplex graphs and nodal attributes. The proposed framework utilizes time-varying factor models to represent each layer of the graph and nodal attributes, with some of the time-varying latent factors shared across different layers and nodal attributes. The latent factors are assigned NN-GP priors to uncover associations between the graph and nodal attributes, as well as their temporal dynamics. The NN-GP prior leverages the predictive power of a DNN, while enabling uncertainty quantification in inference and significantly reducing the number of fitted parameters, compared to a standard DNN. The framework supports inference even when each layer of the multiplex graph, as well as nodal attributes, are partially observed at different time points. Empirical evaluations using both simulated data and covert terrorist network graphs demonstrate the strong predictive performance of the proposed approach in estimating nodal attributes and unobserved relationships between nodes over time. The joint modeling framework significantly outperforms methods that model multiplex graphs and nodal attributes separately, as well as approaches that fit $L$ separate joint models for each graph layer and nodal attributes.

Given that the motivating terrorism study focuses on predicting nodal attributes and unobserved links between nodes, statistical hypothesis testing on the association between nodal attributes and the graph is beyond the scope of this article. We plan to address this in future work with applications specifically targeting this scientific question. Additionally, the NN-GP is computationally feasible when the number of graph nodes $J$ and number of time points $T$ are moderately large, as in the motivating application. For cases where $J$ is large, we intend to use distributed Bayesian inference following the strategy outlined in \citep{guhaniyogi2018meta, guhaniyogi2023distributed} and adapting it to our graph-based framework.

\section*{Acknowledgments}
Portions of this research were conducted with the advanced computing resources provided by Texas A\&M Department of Statistics Arseven Computing Cluster.

\begin{singlespace}
\bibliographystyle{plain}
\bibliography{refs}
\end{singlespace}

\end{document}



\maketitle




\begin{abstract}
This file is organized into four sections. Section 1 provides a brief overview of the Gaussian process prior with the deep neural network kernel, referred to as the neural network Gaussian process (NN-GP) in the main article. Section 2 presents a detailed simulation study comparing the performance of our proposed approach with competing methods across various data-generating schemes and true parameter settings, including a detailed sensitivity analysis to assess robustness of modeling assumptions. Section 3 outlines the full conditional distributions required to implement the Gibbs sampler for our method. Finally, Section 4 demonstrates satisfactory convergence of the proposed Dynamic Joint Learning (DJL) framework by reporting effective sample sizes for representative model parameters.
\end{abstract}

\newpage
\section{Gaussian Process Prior with Deep Neural Network Kernel}\label{sec:nn-gp}
This section provides a brief overview of the Neural Network Gaussian Processes (NN-GP) used in modeling the unknown latent functions in Section 3.2 of the main article.
\cite{lee2017deep} show that deep Bayesian neural networks (BNNs) \citep{neal2012bayesian, wilson2020bayesian} induce non-stationary covariance functions and that untrained, infinitely wide BNNs are equivalent to Gaussian processes with recursively defined kernels, referred to as Neural Network Gaussian Processes (NN-GP). We briefly review the NN-GP covariance kernel and its link to deep neural networks, focusing on one-dimensional input since our analysis uses time as input.

Consider a fully connected Deep Neural Network (DNN) $\mathcal{H}_N$ with $F$ layers, where each neuron in one layer is connected to all neurons in the next layer \citep{goodfellow2016deep}. Let $u_{f,s}(t)$ represent the $s$th neuron in the $f$th layer at time $t$, with the $f$th layer having a width of $W_f$. A fully connected DNN with a time input $t_i$ can be described by the following set of recursive equations,
\begin{align*}
 u_{f,s}(t_i)=c_{f,s}+\sum_{w=1}^{W_f}G_{f,s,w}h(u_{f-1,w}(t_i)),\:\:
 u_{0,s}(t_i)=c_{0,s}+G_{0,s,w}t_i,
\end{align*}
where $h(\cdot)$ denotes an activation function such as ReLU \citep{goodfellow2016deep}, and $\{c_{f,s}\}$ and $\{G_{f,s,w}\}$ are the bias and weight parameters for different layers, respectively. These weight and bias parameters are independently assigned zero-mean prior distributions with variances $\text{Var}(c_{f,s}) = \sigma_c^2$ and $\text{Var}(G_{f,s,w}) = \sigma_g^2 / W_f$.

When each layer of the DNN has infinite width, i.e., $W_f\rightarrow\infty$, $u_{f,s}(t_i)$ converges to a Gaussian distribution with mean zero and a finite variance. Further, $(u_{f,s}(t_1),...,u_{f,s}(t_T))^T$ is a finite dimensional realization from a GP with mean zero and a valid covariance kernel function $\kappa_f(t_i,t_j;\bbeta)$, where $\bbeta=\{\sigma_c^2,\sigma_g^2\}$, $\sigma_c^2,\sigma_g^2>0$. This covariance function for different layers assumes a recursive structure \citep{lee2017deep}
given by
\begin{align*}
&\kappa_f(t_i,t_j;\bbeta)=\sigma_c^2+\sigma_g^2 K^h(\kappa_{f-1}(t_i,t_i;\bbeta),\kappa_{f-1}(t_i,t_j;\bbeta), \kappa_{f-1}(t_j,t_j;\bbeta)),\:\:f=1,...,F\\
&\kappa_0(t_i,t_j;\bbeta)=\sigma_c^2+\sigma_g^2 t_i t_j,
\end{align*}
\sloppy
where $K^h(\cdot,\cdot,\cdot)$ is a non-linear function depending on the activation function $h(\cdot)$. We focus on the ReLU activation function that admits a closed form for this function with\\
 $K^h(\kappa_{f-1}(t_i,t_i;\bbeta),\kappa_{f-1}(t_i,t_j;\bbeta), \kappa_{f-1}(t_j,t_j;\bbeta))=\sqrt{\kappa_{f-1}(t_i,t_i;\bbeta)\kappa_{f-1}(t_j,t_j;\bbeta)} (\sin(\gamma_{f}(t_i,t_j))+\\(\gamma_{f}(t_i,t_j)-\pi)\cos(\gamma_{f}(t_i,t_j)))/2\pi$, with
 $\gamma_{f}(t_i,t_j)=\arccos\left(\kappa_{f-1}(t_i,t_j;\bbeta)[\kappa_{f-1}(t_i,t_i;\bbeta)\kappa_{f-1}(t_j,t_j;\bbeta)]^{-1/2}\right)$.
Integrating out the posterior distribution of weights and biases,
the prediction from Bayesian DNN will be identical to the prediction from a GP with the covariance kernel $\kappa_F$. When $W_f$ is finite, the NN-GP will approximate prediction from the DNN.

In the multiplex terrorism graph data analysis in Section 5 of the main article, we use $F=1$, which leads to the following covariance kernel:
\begin{align}
&\kappa_1(t_i,t_j;\bbeta)=  \sigma_c^2+\sigma_g^2 K^h(\kappa_{0}(t_i,t_i;\bbeta),\kappa_{0}(t_i,t_j;\bbeta), \kappa_{0}(t_j,t_j;\bbeta)) \nonumber\\
&=\sigma_c^2+\sigma_g^2\sqrt{\kappa_0(t_i,t_i;\bbeta)\kappa_0(t_j,t_j;\bbeta)}\Big(\sin(\gamma_{1}(t_i,t_j))+(\gamma_{1}(t_i,t_j)-\pi)\cos(\gamma_{1}(t_i,t_j))\Big)/2\pi,
\end{align}
with $\gamma_{1}(t_i,t_j)=\arccos\left(\kappa_{0}(t_i,t_j;\bbeta)[\kappa_{0}(t_i,t_i;\bbeta)\kappa_{0}(t_j,t_j;\bbeta)]^{-1/2}\right)$. The covariance function is highly non-stationary in time and ideal for the terrorism study, as opposed to the standard Matérn class of stationary covariance functions.

\section{Simulation Studies}\label{sec:simulations}
In this section, we investigate the performance of our method and compare it with some competitors in various simulation scenarios. The simulation studies have five fold objectives: \textbf{(O1)} to assess the accuracy of our method in predicting unobserved edges in time-varying graphs as well as the prediction of nodal attributes over time; \textbf{(O2)} to examine the benefits of jointly modeling time-varying multiplex graphs and nodal attributes compared to modeling the time-varying layers and nodal attributes separately; \textbf{(O3)} to demonstrate the flexibility of the NN-GP prior for time-varying latent functions, showing its capability to offer accurate inference without relying on restrictive stationarity assumptions, which are common with popular covariance kernels for GPs; \textbf{(O4)} to assess the impact of model parameters on the inference; and \textbf{(O5)} to investigate model performance under mis-specification in the data generation scheme.

The simulation study considers three different simulation schemes. The first two schemes simulate data from the fitted model, while the third scheme investigates the performance of the proposed joint model under mis-specification.
More specifically, for the first two simulation schemes, we generate the time-varying multiplex graph with binary edges and time-varying nodal attributes following equations (2) and (4) of the main article, respectively. In the simulation settings considered, we assume that the true latent dimension for the layer-specific latent factors, $R^*$, and that for the shared latent factors, $R_\zeta^*$, are the same for data generation, i.e., $R^*=R_\zeta^*$. For Simulation Schemes 1 and 3, the true number of NN-GP layers used in data generation is denoted by $F^*$. Both 
$R^*$ and $F^*$ are varied across simulations, as discussed later. The true time-varying functions $\mu(t),\{\eta_k(t):k=1,..,m\},\{\zeta_{j,r}(t):j=1,..,J;\:r=1,..,R_\zeta^*\}, \{\xi_{j,l,r}(t):j=1,..,J;\:l=1,..,L;\:r=1,..,R^*\},\{\alpha_{k,l,r}(t):k=1,..,m;\:l=1,..,L;\:r=1,...,R^*\}$ are simulated following two different strategies in the two schemes. \\
\underline{\emph{Simulation Scheme 1:}} In the first simulation scheme, the true time-varying functions are simulated from NN-GP models as follows:
\begin{align*}
  & \mu(\cdot) \sim \text{NN-GP}(0, \kappa_{F^*}(t_i, t_j; \bbeta_\mu^*)), \:\:
    \eta_k(\cdot) \sim \text{NN-GP}(0, \kappa_{F^*}(t_i, t_j; \bbeta_\eta^*)), \\
  & \zeta_{j,r}(\cdot) \sim \text{NN-GP}(0, \kappa_{F^*}(t_i, t_j; \bbeta_\zeta^*)), \\
  & \xi_{j,l,r}(\cdot) \sim \text{NN-GP}(0, \kappa_{F^*}(t_i, t_j; \bbeta_\xi^*)), \:\:
    \alpha_{k,l,r}(\cdot) \sim \text{NN-GP}(0, \kappa_{F^*}(t_i, t_j; \bbeta_\alpha^*)).
\end{align*}
where $\bbeta_{\mu}^*,\bbeta_\eta^*,\bbeta_\zeta^*,\bbeta_\xi^*,\bbeta_\alpha^*$ are the true parameters to simulate these latent functions. The true parameter values are set at $\bbeta_{\mu}^* = \bbeta_\eta^* = \bbeta_\zeta^* = \bbeta_\xi^* = \bbeta_\alpha^* = \{\sigma^{*2}_{w}, \sigma^{*2}_{b}\} =  \{0.4, 0.01\}$.\\
\underline{\emph{Simulation Scheme 2:}} In the second simulation scheme, the true time-varying functions are simulated from a first-order autoregressive (AR) process:
\begin{align*}
\mu(t_s)\sim N(\rho_{\mu}^*\mu(t_{s-1}),\sigma_\mu^{*2}),\:\: \mu(t_0)\sim N(0,\sigma_\mu^{*2}/(1-\rho_\mu^{*2})),\:|\rho_{\mu}^*|<1,\:\sigma_\mu^{*2}>0\nonumber\\   
\eta_k(t_s)\sim N(\rho_{\eta}^*\eta(t_{s-1}),\sigma_\eta^{*2}),\:\:\eta_k(t_0)\sim N(0,\sigma_\eta^{*2}/(1-\rho_\eta^{*2})),\:|\rho_{\mu}^*|<1,\:\sigma_\eta^{*2}>0\nonumber\\
\zeta_{j,r}(t_s)\sim N(\rho_{\zeta}^*\zeta_{j,r}(t_{s-1}),\sigma_\zeta^{*2}),\:\:
\zeta_{j,r}(t_0)\sim N(0,\sigma_\zeta^{*2}/(1-\rho_\zeta^{*2})),\:|\rho_{\zeta}^*|<1,\:\sigma_\zeta^{*2}>0\nonumber\\
\xi_{j,l,r}(t_s)\sim N(\rho_{\xi}^*\xi_{j,l,r}(t_{s-1}),\sigma_\xi^{*2},\:\:\xi_{j,l,r}(t_0)\sim N(0,\sigma_\xi^{*2}/(1-\rho_\xi^{*2})),\:\:|\rho_{\xi}^*|<1,\:\sigma_\xi^{*2}>0\nonumber\\
\alpha_{k,l,r}(t_s)\sim N(\rho_{\alpha}^*\alpha_{k,l,r}(t_{s-1}),\sigma_\alpha^{*2}),\:\:\alpha_{k,l,r}(t_0)\sim N(0,\sigma_\alpha^{*2}/(1-\rho_\alpha^{*2})),\:\:|\rho_{\alpha}^*|<1,\:\sigma_\alpha^{*2}>0.
\end{align*}
 The functional values at time $t_0$ allow marginal prior distributions of each of these time-varying functions at each time point to have $0$ mean and constant variance. The true parameter values are given by $\rho_{\mu}^* = \rho_{\eta}^* = \rho_{\zeta}^* = \rho_{\xi}^*= \rho_{\alpha}^* = 0.5$ and $\sigma_\mu^{*2} = \sigma_\eta^{*2} = \sigma_\zeta^{*2} = \sigma_\xi^{*2} = \sigma_\alpha^{*2} = 4$. The true parameter values indicate moderate autocorrelation and marginal variance for time-varying functions.\\
 \underline{\emph{Simulation Scheme 3:}} This simulation scheme is designed to investigate the performance of the proposed approach under model mis-specification. Specifically, each layer of the multiplex graph is simulated conditionally on the nodal attributes using a temporal exponential random graph model (TERGM) that includes two sufficient statistics: $\mathcal{S}_{1,l}(t)=\sum_{j=1}^J\sum_{j'=1}^J\sum_{k=1}^m a_{jj',l}^{(o)}(t)x_{j,k}(t)x_{j',k}(t)$ and $\mathcal{S}_{2,l}(t)=\sum_{j=1}^J\sum_{j'=1}^J a_{jj',l}^{(o)}(t)$, with corresponding coefficients $\theta_1=t/T$ and $\theta_2=0.5$, respectively. Here, the nodal attributes, $x_{j,k}(t)$, are generated as in Simulation Scheme 1. The multiplex graph data is thus simulated from a completely mis-specified TERGM, incorporating complex higher-order dependencies between the multiplex graph and nodal attributes.

All simulations use $T=20$ time points. For each simulation scheme, the number of graph layers varies between $L \in \{2, 3, 4\}$, the number of graph nodes varies between $J\in\{20,40,100\}$, the number of nodal attributes varies between $m\in\{2,5,8,10\}$, the dimensions of the true shared and layer-specific latent functions varies between $R^*\in\{3,4,5\}$ and the true number of NN-GP layers for Simulation schemes 1 and 3 varies between $F^*\in\{1,2,3\}$. Different combinations of $J$, $m$, $L$, $R^{*}$, $F^{*}$ for the three simulation schemes are shown in Table~\ref{tab:simulation_combinations}. In each simulation scheme, time points are randomly masked per layer with a 
$0.1$ probability, and 25\% of edges are removed at each masked time point. The multiplex graphs and nodal attributes are then treated as \emph{partially observed} for fitting all competing models. For the simulations, results are averaged across 10 replications.

\noindent\underline{\emph{Competitors:}} For all simulated datasets, we fit the proposed joint model for multiplex graphs and nodal attributes, referred to as the dynamic joint learner (DJL). The dimensions of the shared and layer-specific latent factors are set equal, i.e., $R=R_\zeta=4$. This assumption is not critical, as we demonstrate in Section~\ref{sec:dim_layer_shared}. All posterior inferences are based on $10000$ MCMC samples, collected after a burn-in period of $5000$ iterations, without thinning. Convergence is evaluated by monitoring the behavior of the model likelihood over iterations. Additionally, trace plots for some model parameters under Simulation Scheme 1 are provided in Section~\ref{sec:convergence_analysis} of this supplementary file, demonstrating good mixing performance of the algorithm. When fitting the DJL, we set $F=1$, corresponding to a two-layer nonstationary NN-GP covariance kernel. Since $R$ and $F$ are key hyperparameters, Section 2.2 of this supplementary file presents a detailed sensitivity analysis in inference with different choices of $(R,F)$. Overall, the results remain largely stable as $R$ increases. The inference seems to be very similar for $F=1$ or $F=2$, either of which delivers the best performance depending on the simulation scenarios, while higher values of $F$ lead to less accurate out-of-sample inference. This justifies our choice of $F=1$ in all simulation studies.

To evaluate the benefits of jointly modeling nodal attributes and multiplex graphs, we compare our approach with two competitors. The first, \emph{dynamic marginal learner (DML)}, fits separate models for multiplex graphs (Equation (2) of the main article) and nodal attributes (Equation (4) of the main article), allowing us to assess the advantages of joint modeling. The second competitor, \emph{joint latent factor model (JLAFAC)} \citep{guhaniyogi2020joint}, integrates a single time-varying graph layer with nodal attributes using AR(1) processes for latent functions. For simulated data with $L=2$ graph layers, we fit JLAFAC separately to each layer and the nodal attributes, referring to these models as \emph{JLAFAC(first)} and \emph{JLAFAC(second)}. For simulation settings with 
$L>2$, we likewise fit JLAFAC separately to each of the 
$L$ layers and the nodal attributes; however, for clarity of presentation, the results are averaged across the 
$L$ graph layers and reported under \emph{JLAFAC(first)}. Comparisons with JLAFAC(first) and JLAFAC(second) highlight the benefits of jointly modeling all graph layers and attributes, as well as the advantages of using the flexible NN-GP kernel over simpler AR(1) latent function modeling.

\subsection{Prediction of Edges and Attributes}
We evaluate the performance of the competing models based on their ability to predict edges and attributes in three scenarios: \emph{in-sample}, \emph{missing}, and \emph{out-of-sample}. The \emph{in-sample} scenario refers to predicting edges and attributes that are part of the dataset used to train the model. The \emph{missing} scenario involves predicting graph edges within the time points $t_1,...,t_T$, which were masked during the training of all models. The results for predicting nodal attributes are also presented for the \emph{missing} scenario. The \emph{out-of-sample} scenario evaluates performance on edges and attributes at time points beyond $t_1,...,t_T$. For the out-of-sample assessment, we simulate the multiplex graphs and nodal attributes at additional future time points. Details of the predictive inference for edges and attributes across the \emph{in-sample}, \emph{missing}, and \emph{out-of-sample} scenarios are provided in Section 4 of the main article. 

Table~\ref{Tab1:Prediction} shows that DJL significantly outperforms all competing methods in both edge and attribute point prediction under Simulation Scheme 1 for both the \emph{missing} and \emph{out-of-sample} scenarios. In the \emph{in-sample} scenario, JLAFAC(first) and JLAFAC(second) exhibit signs of overfitting when estimating attributes, while performing worse than DJL in edge prediction. 
While increasing the number of nodal attributes has only a minor effect on edge prediction across all models, it leads to a noticeable decline in overall performance in terms of attribute prediction for all competitors. This decline is likely due to the increasing complexity introduced by a larger number of time-varying latent functions (e.g., compare Combinations 1, 2 and 3; or Combinations 4, 5 and 6) . Increasing both $L$ and $F^{*}$ also results in performance declines (e.g., compare Combination 2 with 11, or Combination 5 with 12). This is likely because a larger number of time-varying latent functions must be estimated with increasing $L$, and the true data generating latent functions become more complex when modeled with additional NN-GP layers for higher values of $F^*$. On the other hand, increasing the number of graph nodes does not appear to have a significant impact on model performance.


\begin{table}[h]
    \centering
    \begin{tabular}{c|c c c c c c c c c c c c }
        Combination & 1 & 2 & 3 & 4 & 5 & 6 & 7 & 8 & 9 & 10 & 11 & 12 \\
        \hline
        $J$ & 20 & 20 & 20 & 40 & 40 & 40 & 100 &  20 & 40 &  100 &  20 & 40\\
        $m$ &  2 & 5 & 8 & 2 & 5 & 8 & 2 & 10 & 10 & 10 & 5 & 5\\
        $L$ &  2 & 2 & 2 & 2 & 2 & 2 & 2 & 3 & 3 & 3 & 4 & 4\\
        $R^{*}$ &  4 & 4 & 4 & 4 & 4 & 4 & 4 & 3 & 3 & 3 & 5 & 5 \\
        $F^{*}$ &  1 & 1 & 1 & 1 & 1 & 1 & 1 & 2 & 2 & 2 & 3 & 3 \\
        \hline       
    \end{tabular}
    \caption{Simulation combinations with different choices of $J$, $m$, $L$, $R^{*}$, and $F^{*}$. (Note: $F^{*}$ is only used for data generation in Simulation Schemes 1 and 3).}
    \label{tab:simulation_combinations}
\end{table}


\begin{table}[h]
\centering
\subfloat[AUC for Edge Prediction] 
{\resizebox{\textwidth}{!}{%
\begin{tabular}{|c|c|c|c|c|c|c|c|c|c|c|c|c|c|}
\hline
& & \multicolumn{12}{|c|}{AUC}  \\ \hline
& Combination & \makecell{1 \\ $J = 20$ \\ $m = 2$ \\ $L = 2$ \\ $R^{*} = 4$ \\ $F^{*} = 1$} & \makecell{2 \\ $J = 20$ \\ $m = 5$ \\ $L = 2$ \\ $R^{*} = 4$ \\ $F^{*} = 1$} & \makecell{3 \\ $J = 20$ \\ $m = 8$ \\ $L = 2$ \\ $R^{*} = 4$ \\ $F^{*} = 1$} & \makecell{4 \\ $J = 40$ \\ $m = 2$ \\ $L = 2$ \\ $R^{*} = 4$ \\ $F^{*} = 1$} & \makecell{5 \\ $J = 40$ \\ $m = 5$ \\ $L = 2$ \\ $R^{*} = 4$ \\ $F^{*} = 1$} & \makecell{6 \\ $J = 40$ \\ $m = 8$ \\ $L = 2$ \\ $R^{*} = 4$ \\ $F^{*} = 1$} & \makecell{7 \\ $J = 100$ \\ $m = 5$ \\ $L = 2$ \\ $R^{*} = 4$ \\ $F^{*} = 1$} & \makecell{8 \\ $J = 20$ \\ $m = 10$ \\ $L = 3$ \\ $R^{*} = 3$ \\ $F^{*} = 2$} & \makecell{9 \\ $J = 40$ \\ $m = 10$ \\ $L = 3$ \\ $R^{*} = 3$ \\ $F^{*} = 2$} & \makecell{10 \\ $J = 100$ \\ $m = 10$ \\ $L = 3$ \\ $R^{*} = 3$ \\ $F^{*} = 2$} & \makecell{11 \\ $J = 20$ \\ $m = 5$ \\ $L = 4$ \\ $R^{*} = 5$ \\ $F^{*} = 3$} & \makecell{12 \\ $J = 40$ \\ $m = 5$ \\ $L = 4$ \\ $R^{*} = 5$ \\ $F^{*} = 3$} \\ \hline
&  DJL & \textbf{0.9814} & \textbf{0.9779} & \textbf{0.9732} & \textbf{0.9763} & \textbf{0.9644} & \textbf{0.9503} & \textbf{0.9828} & \textbf{0.8844} & \textbf{0.8791} & \textbf{0.8961} & \textbf{0.7750} & \textbf{0.7670} \\
In & DML & 0.8778 & 0.8776 & 0.8776 & 0.8419 & 0.8433 & 0.8434 & 0.8311 & 0.7875 & 0.7660 & 0.7788 & 0.7030 & 0.6933\\
&  JLAFAC(first) & 0.7463 & 0.7259 & 0.7548 & 0.9238 & 0.9024 & 0.8724 & 0.9200 & 0.7453 & 0.8117 & 0.8790 & 0.5725 & 0.5877  \\ 
&  JLAFAC(second) & 0.7424 & 0.8472 & 0.6717 & 0.8696 & 0.8406 & 0.8365 & 0.9296   &   ---    &  ---  &    ---   &   ---   &   --- \\
\hline
&  DJL & \textbf{0.9953} & \textbf{0.9944} & \textbf{0.9814} & \textbf{0.9931} & \textbf{0.9835} & \textbf{0.9666} & \textbf{0.9614} & \textbf{0.8998} & \textbf{0.9252} & \textbf{0.9149} & \textbf{0.7538} & \textbf{0.7095} \\
Mis. & DML & 0.9053 & 0.9110 & 0.9045 & 0.8609 & 0.8539 & 0.8483 & 0.8133 & 0.8284 & 0.8261 & 0.8044 & 0.6947 & 0.6664 \\
&  JLAFAC(first) & 0.7090 & 0.6195 & 0.6970 & 0.8286 & 0.8580 & 0.7930 & 0.8881 & 0.6934 & 0.7350 & 0.8025 & 0.5521 & 0.5290 \\
& JLAFAC(second) & 0.6527 & 0.8194 & 0.6533 & 0.8050 & 0.7878 & 0.7817 & 0.7633 &   ---    &  ---  &    ---   &   ---   &   --- \\
 \hline
&  DJL & \textbf{0.9998} & \textbf{0.9978} & \textbf{0.9979} & \textbf{0.9978} & \textbf{0.9926} & \textbf{0.9827} & \textbf{0.9994} & \textbf{0.9706} & \textbf{0.9732} & \textbf{0.9821} & \textbf{0.9169} & \textbf{0.9088} \\
Out & DML & 0.9125 & 0.9139 & 0.9120 & 0.8802 & 0.8792 & 0.8779 & 0.8583 & 0.8678 & 0.8546 & 0.8563 & 0.8340 & 0.8153 \\
& JLAFAC(first) & 0.5653 & 0.5788 & 0.5510 & 0.8855 & 0.7618 & 0.6412 & 0.7983 & 0.6721 & 0.5671 & 0.7587 & 0.5429 & 0.5800 \\
& JLAFAC(second) & 0.5967 & 0.8292 & 0.5960 & 0.7634 & 0.8129 & 0.6684 & 0.8331  &   ---    &  ---  &    ---   &   ---   &   ---  \\
\hline
\end{tabular}
}}
\hfill
\subfloat[MSPE for Nodal Attribute Prediction] 
{\resizebox{\textwidth}{!}{%
\begin{tabular}{|c|c|c|c|c|c|c|c|c|c|c|c|c|c|}
\hline
& & \multicolumn{12}{|c|}{MSPE}  \\ \hline
& Combination & \makecell{1 \\ $J = 20$ \\ $m = 2$ \\ $L = 2$ \\ $R^{*} = 4$ \\ $F^{*} = 1$} & \makecell{2 \\ $J = 20$ \\ $m = 5$ \\ $L = 2$ \\ $R^{*} = 4$ \\ $F^{*} = 1$} & \makecell{3 \\ $J = 20$ \\ $m = 8$ \\ $L = 2$ \\ $R^{*} = 4$ \\ $F^{*} = 1$} & \makecell{4 \\ $J = 40$ \\ $m = 2$ \\ $L = 2$ \\ $R^{*} = 4$ \\ $F^{*} = 1$} & \makecell{5 \\ $J = 40$ \\ $m = 5$ \\ $L = 2$ \\ $R^{*} = 4$ \\ $F^{*} = 1$} & \makecell{6 \\ $J = 40$ \\ $m = 8$ \\ $L = 2$ \\ $R^{*} = 4$ \\ $F^{*} = 1$} & \makecell{7 \\ $J = 100$ \\ $m = 5$ \\ $L = 2$ \\ $R^{*} = 4$ \\ $F^{*} = 1$} & \makecell{8 \\ $J = 20$ \\ $m = 10$ \\ $L = 3$ \\ $R^{*} = 3$ \\ $F^{*} = 2$} & \makecell{9 \\ $J = 40$ \\ $m = 10$ \\ $L = 3$ \\ $R^{*} = 3$ \\ $F^{*} = 2$} & \makecell{10 \\ $J = 100$ \\ $m = 10$ \\ $L = 3$ \\ $R^{*} = 3$ \\ $F^{*} = 2$} & \makecell{11 \\ $J = 20$ \\ $m = 5$ \\ $L = 4$ \\ $R^{*} = 5$ \\ $F^{*} = 3$} & \makecell{12 \\ $J = 40$ \\ $m = 5$ \\ $L = 4$ \\ $R^{*} = 5$ \\ $F^{*} = 3$} \\ \hline
&  DJL & 0.3741 & 0.4269 & 0.4640 & 0.2318 & 0.4413 & 0.4521 & 0.1806 & 0.5536 & 0.3911 & 0.4115 & 0.4659 & 0.5531 \\
In & DML & 0.9856 & 0.9952 & 0.9747 & 0.9790 & 0.9891 & 0.9964 & 0.9169 & 0.9743 & 0.9837 & 0.9856 & 0.9837 & 0.9870 \\
&  JLAFAC(first) & 0.0083 & 0.0094 & \textbf{0.0082} & \textbf{0.0033} & 0.0049 & 0.0042 & \textbf{0.0203} & \textbf{0.0476} & \textbf{0.0131} & \textbf{0.1231} & \textbf{0.1090} & \textbf{0.1112}  \\ 
&  JLAFAC(second) & \textbf{0.0064} & \textbf{0.0045} & 0.0085 & 0.0045 & \textbf{0.0032} & \textbf{0.0041} & 0.0233  &    ---    &  ---  &    ---   &   ---   &   ---  \\
\hline
&  DJL & \textbf{0.3683} & \textbf{0.4337} & \textbf{0.4672} & \textbf{0.2311} & \textbf{0.4179} & \textbf{0.4578} & \textbf{0.1818} & \textbf{0.5820} & \textbf{0.4050} & \textbf{0.3997} & \textbf{0.5784} & \textbf{0.6782} \\
Mis. & DML & 0.9927 & 1.0081 & 0.9743 & 0.9645 & 0.9944 & 1.0062 & 0.9147 & 0.9819 & 0.9914 & 0.9880 & 0.9869 & 0.9854 \\
&  JLAFAC(first) & 0.9856 & 0.9828 & 0.9652 & 0.8536 & 0.8269 & 1.0397 & 0.9177 & 0.9857 & 1.0581 & 1.0112 & 0.9470 & 0.9484\\
& JLAFAC(second) &  1.0202 & 0.8035 & 1.0489 & 0.9296 & 0.9259 & 2.7128 & 0.8146  &    ---    &  ---  &    ---   &   ---   &   ---  \\
 \hline
&  DJL & \textbf{0.3209} & \textbf{0.4163} & \textbf{0.4614} & \textbf{0.2344} & \textbf{0.4191} & \textbf{0.4546} & \textbf{0.1838} & \textbf{0.5216} & \textbf{0.3852} & \textbf{0.3567} & \textbf{0.3205} & \textbf{0.4623} \\
Out & DML & 0.9922 & 0.9963 & 0.9769 & 0.9787 & 0.9876 & 0.9938 & 0.9308 & 0.9723 & 0.9854 & 0.9868 & 0.9870 & 0.9849\\
& JLAFAC(first) & 0.9818 & 0.9768 & 0.9608 & 0.8271 & 0.8279 & 1.0058 & 0.8742 & 0.9760 & 0.9766 & 0.9533 & 0.9089 & 0.8990\\
& JLAFAC(second) & 1.0166 & 0.7780 & 1.0596 & 0.9307 & 0.8925 & 0.9614 & 0.7954   &    ---    &  ---  &    ---   &   ---   &   ---  \\
\hline
\end{tabular}
}}
\caption{Simulation results for our proposed approach DJL with competitors DML, JLAFAC(first) and JLAFAC(second) under 12 different simulation combinations for Simulation Scheme 1. Notably, for simulation scenarios with $L>2$, we present the average performance over different layers and present it under JLAFAC(first), hence cells under JLAFAC(second) are kept blank for these scenarios. We report area under the curve (AUC) in predicting an edge/no edge for the multiplex graphs, as well as mean squared prediction error (MSPE) in predicting nodal attributes. Here ``In'' , ``Mis.'' and ``Out'' refer to in-sample, missing and out-of-sample predictive scenarios, respectively. The best performing model under each combination and each scenario is boldfaced.}\label{Tab1:Prediction}
\end{table}

In JLAFAC (first) and JLAFAC (second), the time-varying latent effects are modeled using an AR(1) process, which aligns with the method used to simulate latent effects in Scheme 2. This, combined with their tendency to overfit attributes in the \emph{in-sample} scenario, results in higher \emph{in-sample} accuracy for JLAFAC compared to DJL and DML in most cases. However, as shown in Table~\ref{Tab2:Prediction}, DJL remains competitive in terms of edge prediction. DJL, however, significantly outperforms JLAFAC when predicting \emph{missing} edges and attributes when number of graph nodes is smaller, as seen in Combinations 1, 2, 3, 4, 8, 11 and 12. As the number of graph nodes increases, the performance gap between DJL and JLAFAC narrows. For the graph and attribute prediction in \emph{out-of-sample} scenario, all models experience a noticeable drop in performance, though DJL maintains a significant advantage over its competitors. DML performs similarly to JLAFAC in both edge and attribute prediction. Similar to Scheme 1, increasing the number of attributes or the number of layers degrades attribute prediction performance, whereas edge prediction remains largely unaffected.

\begin{table}[h]
\centering
\subfloat[AUC for Edge Prediction] 
{\resizebox{\textwidth}{!}{%
\begin{tabular}{|c|c|c|c|c|c|c|c|c|c|c|c|c|c|}
\hline
& & \multicolumn{12}{|c|}{AUC}  \\ \hline
& Combination & \makecell{1 \\ $J = 20$ \\ $m = 2$ \\ $L = 2$ \\ $R^{*} = 4$ } & \makecell{2 \\ $J = 20$ \\ $m = 5$ \\ $L = 2$ \\ $R^{*} = 4$} & \makecell{3 \\ $J = 20$ \\ $m = 8$ \\ $L = 2$ \\ $R^{*} = 4$ } & \makecell{4 \\ $J = 40$ \\ $m = 2$ \\ $L = 2$ \\ $R^{*} = 4$ } & \makecell{5 \\ $J = 40$ \\ $m = 5$ \\ $L = 2$ \\ $R^{*} = 4$ } & \makecell{6 \\ $J = 40$ \\ $m = 8$ \\ $L = 2$ \\ $R^{*} = 4$ } & \makecell{7 \\ $J = 100$ \\ $m = 5$ \\ $L = 2$ \\ $R^{*} = 4$ } & \makecell{8 \\ $J = 20$ \\ $m = 10$ \\ $L = 3$ \\ $R^{*} = 3$} & \makecell{9 \\ $J = 40$ \\ $m = 10$ \\ $L = 3$ \\ $R^{*} = 3$} & \makecell{10 \\ $J = 100$ \\ $m = 10$ \\ $L = 3$ \\ $R^{*} = 3$ } & \makecell{11 \\ $J = 20$ \\ $m = 5$ \\ $L = 4$ \\ $R^{*} = 5$} & \makecell{12 \\ $J = 40$ \\ $m = 5$ \\ $L = 4$ \\ $R^{*} = 5$} \\ \hline
&  DJL & \textbf{0.8522} & 0.8316 & 0.8185 & 0.8555 & 0.8339 & 0.8200 & 0.8593 & 0.8048 & 0.8210 & 0.8470 & \textbf{0.8290} & 0.8443 \\
In & DML & 0.7798 & 0.7798 & 0.7817 & 0.7531 & 0.7533 & 0.7531 & 0.7550 & 0.7759 & 0.7483 & 0.7587 & 0.7541 & 0.7569 \\
&  JLAFAC(first) & 0.6360 & 0.7351 & 0.8054 & 0.8684 & \textbf{0.9098} & \textbf{0.9035} & 0.9517 & \textbf{0.8492} & \textbf{0.9211} & \textbf{0.9709} & 0.6983 & \textbf{0.8605}  \\ 
&  JLAFAC(second) & 0.8077 & \textbf{0.8418} & \textbf{0.8234} & \textbf{0.8858} & 0.8884 & 0.8980 & \textbf{0.9530}    &    ---    &  ---  &    ---   &   ---   &   ---   \\
\hline
&  DJL & \textbf{0.8324} & \textbf{0.8061} & \textbf{0.7852} & \textbf{0.8264} & 0.8047 & 0.7939 & 0.8566 & \textbf{0.7783} & 0.7938 & 0.8296 & \textbf{0.8357} & \textbf{0.8040} \\
Mis. & DML &  0.7730 & 0.7826 & 0.7830 & 0.7400 & 0.7333 & 0.7367 & 0.7575 & 0.7655 & 0.7501 & 0.7479 & 0.7599 & 0.7226 \\
&  JLAFAC(first) & 0.6197 & 0.6145 & 0.6928 & 0.7879 & \textbf{0.8164} & \textbf{0.8292} & 0.8969 & 0.6951 & \textbf{0.8387} & \textbf{0.9249} & 0.6574 & 0.7641 \\
& JLAFAC(second) & 0.6908 & 0.6872 & 0.6474 & 0.8045 & 0.7840 & 0.8059 & \textbf{0.9070}   &    ---    &  ---  &    ---   &   ---   &   ---   \\
 \hline
&  DJL &  \textbf{0.7932} &  \textbf{0.7759} &  \textbf{0.7581} &  \textbf{0.7675} &  \textbf{0.7468} &  \textbf{0.7375} &  \textbf{0.7891} &  \textbf{0.7273} &  \textbf{0.7224} &  \textbf{0.7775} &  \textbf{0.7285} &  \textbf{0.7838} \\
Out & DML & 0.7346 &  0.7336 &  0.7319 &  0.6938 &  0.6958 &  0.6953 &  0.7142 &  0.6949 &  0.6686 &  0.7070 &  0.6592 &  0.7185 \\
& JLAFAC(first) & 0.5914 & 0.5777 & 0.5319 & 0.5757 & 0.6294 & 0.6682 & 0.7007 & 0.5720 & 0.7150 & 0.7240 & 0.5981 & 0.6587 \\
& JLAFAC(second) & 0.6352 &  0.6958 & 0.6035 & 0.6175 & 0.6568 & 0.6936 & 0.6902   &    ---    &  ---  &    ---   &   ---   &   ---   \\
\hline
\end{tabular}
}}
\hfill
\subfloat[MSPE for Nodal Attribute Prediction] 
{\resizebox{\textwidth}{!}{%
\begin{tabular}{|c|c|c|c|c|c|c|c|c|c|c|c|c|c|}
\hline
& & \multicolumn{12}{|c|}{MSPE}  \\ \hline
& Combination & \makecell{1 \\ $J = 20$ \\ $m = 2$ \\ $L = 2$ \\ $R^{*} = 4$ } & \makecell{2 \\ $J = 20$ \\ $m = 5$ \\ $L = 2$ \\ $R^{*} = 4$ } & \makecell{3 \\ $J = 20$ \\ $m = 8$ \\ $L = 2$ \\ $R^{*} = 4$ } & \makecell{4 \\ $J = 40$ \\ $m = 2$ \\ $L = 2$ \\ $R^{*} = 4$ } & \makecell{5 \\ $J = 40$ \\ $m = 5$ \\ $L = 2$ \\ $R^{*} = 4$ } & \makecell{6 \\ $J = 40$ \\ $m = 8$ \\ $L = 2$ \\ $R^{*} = 4$ } & \makecell{7 \\ $J = 100$ \\ $m = 5$ \\ $L = 2$ \\ $R^{*} = 4$ } & \makecell{8 \\ $J = 20$ \\ $m = 10$ \\ $L = 3$ \\ $R^{*} = 3$ } & \makecell{9 \\ $J = 40$ \\ $m = 10$ \\ $L = 3$ \\ $R^{*} = 3$ } & \makecell{10 \\ $J = 100$ \\ $m = 10$ \\ $L = 3$ \\ $R^{*} = 3$ } & \makecell{11 \\ $J = 20$ \\ $m = 5$ \\ $L = 4$ \\ $R^{*} = 5$ } & \makecell{12 \\ $J = 40$ \\ $m = 5$ \\ $L = 4$ \\ $R^{*} = 5$} \\ \hline
&  DJL & 0.5764 & 0.7034 & 0.7253 & 0.5770 & 0.7141 & 0.6659 & 0.5569 & 0.7129 & 0.6796 & 0.6949 & 0.6559 & 0.6345 \\
In & DML & 0.9751 & 0.9882 & 0.9892 & 0.9933 & 0.9924 & 0.9952 & 0.9889 & 0.9932 & 0.9955 & 0.9973 & 0.9938 & 0.9971 \\
&  JLAFAC(first) & 0.2006 & 0.1461 & \textbf{0.0969} & 0.1122 & 0.0523 & 0.0291 & \textbf{0.3650} & \textbf{0.0638} & \textbf{0.0260} & \textbf{0.1356} & \textbf{0.1887} & \textbf{0.1654} \\ 
&  JLAFAC(second) & \textbf{0.1348} & \textbf{0.0630} & 0.1007 & \textbf{0.1058} & \textbf{0.0472} & \textbf{0.0270} & 0.4749  &    ---    &  ---  &    ---   &   ---   &   ---  \\
\hline
&  DJL &  \textbf{0.6222} & \textbf{0.7170} & \textbf{0.7422} & \textbf{0.5862} & \textbf{0.7111} & \textbf{0.6584} & \textbf{0.5323} & \textbf{0.7365} & \textbf{0.7156} & \textbf{0.7039} & \textbf{0.6999} & \textbf{0.6850} \\
Mis. & DML & 0.9742 & 0.9983 & 1.0019 & 0.9981 & 0.9958 & 0.9957 & 0.9830 & 0.9954 & 0.9982 & 0.9980 & 0.9921 & 1.0053\\
&  JLAFAC(first) & 1.0065 &  1.0658  & 1.0209 &  0.9652 &  0.9603  & 0.8951 &  0.9353 &  0.9576  & 0.8699 & 0.8245 &  0.9073
& 0.9676\\
& JLAFAC(second) & 0.9316 & 0.8602&  1.0409 & 0.9609 & 0.9315 & 0.9032 & 0.9632  &    ---    &  ---  &    ---   &   ---   &   ---   \\
 \hline
&  DJL & 0.9906 & \textbf{0.7720} & \textbf{0.7450} & 1.2347 & \textbf{0.7887} & \textbf{0.8604} & \textbf{0.6646} & \textbf{0.7554} & \textbf{0.8142} & \textbf{0.9399} & \textbf{0.7449} & \textbf{0.7382} \\
Out & DML & 0.9640 & 1.0071 & 0.9959 & 1.0126 & 0.9959 & 1.0102 & 0.9834 & 0.9996 & 1.0147 & 0.9916 & 0.9943 & 1.0062\\
& JLAFAC(first) & 1.0200 &  1.0490  & 1.0172 &  \textbf{0.9656}  & 0.9550 &  0.9100 &  0.9359  & 0.9530 &  0.8774 & 41.4615  & 0.9146
 &  0.9672 \\
& JLAFAC(second) & \textbf{0.9258} & 0.8717 & 1.0364 & 0.9701 & 0.9251 & 0.9090 & 0.9644  &  ---    &  ---  &    ---   &   ---   &   ---  \\
\hline
\end{tabular}
}}
\caption{Simulation results for our proposed approach DJL with competitors DML, JLAFAC(first) and JLAFAC(second) under 12 different simulation combinations for Simulation Scheme 2. Since Simulation Scheme 2 generates latent functions from AR(1) processes, the simulation combinations do not involve $F^*$. Notably, for simulation scenarios with $L>2$, we present the average performance over different layers and present it under JLAFAC(first), hence cells under JLAFAC(second) are kept blank for these scenarios. We report area under the curve (AUC) in predicting an edge/no edge for the multiplex graphs, as well as mean squared prediction error (MSPE) in predicting nodal attributes. Here ``In'' , ``Mis.'' and ``Out'' refer to in-sample, missing and out-of-sample predictive scenarios, respectively. The best performing model under each combination and each scenario is boldfaced.}\label{Tab2:Prediction}
\end{table}

Under Simulation Scheme 3, which introduces model mis-specification, the accuracy of both edge and attribute prediction declines across all competing methods, as shown in Table~\ref{Tab3:Prediction}. Despite this drop in performance, DJL maintains an edge prediction accuracy of approximately 60\% and an MSPE of around 0.5 for attribute prediction in the \emph{in-sample} scenario, reflecting moderate predictive performance. JLAFAC performs better in attribute prediction for the \emph{in-sample} scenario, yet DJL significantly outperforms all competitors in attribute prediction for \emph{missing} and \emph{out-of-sample} scenarios.
For edge prediction, JLAFAC remains competitive with DJL in the \emph{in-sample} and \emph{missing} scenarios but falls behind DJL in the out-of-sample scenario with $L=2$ and $F^*=1$. As $L, F^*$ increases, the performance of both competitors decline substantially.


\begin{table}[h]
\centering
\subfloat[AUC for Edge Prediction] 
{\resizebox{\textwidth}{!}{%
\begin{tabular}{|c|c|c|c|c|c|c|c|c|c|c|c|c|c|}
\hline
& & \multicolumn{12}{|c|}{AUC}  \\ \hline
& Combination & \makecell{1 \\ $J = 20$ \\ $m = 2$ \\ $L = 2$ \\ $R^{*} = 4$ \\ $F^{*} = 1$} & \makecell{2 \\ $J = 20$ \\ $m = 5$ \\ $L = 2$ \\ $R^{*} = 4$ \\ $F^{*} = 1$} & \makecell{3 \\ $J = 20$ \\ $m = 8$ \\ $L = 2$ \\ $R^{*} = 4$ \\ $F^{*} = 1$} & \makecell{4 \\ $J = 40$ \\ $m = 2$ \\ $L = 2$ \\ $R^{*} = 4$ \\ $F^{*} = 1$} & \makecell{5 \\ $J = 40$ \\ $m = 5$ \\ $L = 2$ \\ $R^{*} = 4$ \\ $F^{*} = 1$} & \makecell{6 \\ $J = 40$ \\ $m = 8$ \\ $L = 2$ \\ $R^{*} = 4$ \\ $F^{*} = 1$} & \makecell{7 \\ $J = 100$ \\ $m = 5$ \\ $L = 2$ \\ $R^{*} = 4$ \\ $F^{*} = 1$} & \makecell{8 \\ $J = 20$ \\ $m = 10$ \\ $L = 3$ \\ $R^{*} = 3$ \\ $F^{*} = 2$} & \makecell{9 \\ $J = 40$ \\ $m = 10$ \\ $L = 3$ \\ $R^{*} = 3$ \\ $F^{*} = 2$} & \makecell{10 \\ $J = 100$ \\ $m = 10$ \\ $L = 3$ \\ $R^{*} = 3$ \\ $F^{*} = 2$} & \makecell{11 \\ $J = 20$ \\ $m = 5$ \\ $L = 4$ \\ $R^{*} = 5$ \\ $F^{*} = 3$} & \makecell{12 \\ $J = 40$ \\ $m = 5$ \\ $L = 4$ \\ $R^{*} = 5$ \\ $F^{*} = 3$} \\ \hline
&  DJL & \textbf{0.7266} & \textbf{0.7204} & \textbf{0.6999} & \textbf{0.5846} & \textbf{0.5777} & 0.5646 & \textbf{0.5539} & \textbf{0.6081} & \textbf{0.5610} & 0.5351 & \textbf{0.6115} & \textbf{0.5712} \\
In & DML & 0.7036 & 0.7134 & 0.6983 & 0.5724 & 0.5743 & \textbf{0.5696} & 0.5446 & 0.5973 & 0.5594 & \textbf{0.5368} & 0.5879 & 0.5487 \\
&  JLAFAC(first) & 0.6731 & 0.6034 & 0.5884 & 0.5289 & 0.5252 & 0.5239 & 0.5100 & 0.5760 & 0.5253 & 0.5109 & 0.5755 & 0.5261  \\ 
&  JLAFAC(second) & 0.6731 & 0.5947 & 0.5981 & 0.5314 & 0.5257 & 0.5248 & 0.5109  &  ---    &  ---  &    ---   &   ---   &   ---   \\
\hline
&  DJL & 0.6459 &  \textbf{0.7623} & \textbf{0.6452} & 0.5255 & \textbf{0.5477} & 0.5123 & 0.5051 & \textbf{0.6058} & 0.5422 & 0.5048 & 0.5742 & 0.4936 \\
Mis. & DML & \textbf{0.6576} & 0.7311 & 0.6249 & 0.5211 & 0.5238 & 0.5488 & \textbf{0.5117} & 0.5013 & 0.5118 & 0.5037 & 0.5684 & 0.5155\\
&  JLAFAC(first) & 0.6050 & 0.5744 & 0.5524 & 0.4944 & 0.5117 & \textbf{0.5594} & 0.5076 & 0.5783 & \textbf{0.5449} & \textbf{0.5057} & \textbf{0.6308} & \textbf{0.5399}\\
& JLAFAC(second) & 0.6408 & 0.6754 & 0.5625 & \textbf{0.5278} & 0.5215 & 0.5312 & 0.4952   &  ---    &  ---  &    ---   &   ---   &   ---    \\
 \hline
&  DJL & 0.7957 & 0.7643 & 0.7770 & \textbf{0.5648} & \textbf{0.5395} & \textbf{0.5405} & 0.5088 & \textbf{0.5527} & 0.5162 & 0.5038 & 0.5388 & 0.5133 \\
Out & DML & \textbf{0.8031} & \textbf{0.7846} & \textbf{0.7986} & 0.5631 & 0.5326 & 0.5386 & 0.5012 & 0.5436 & 0.5035 & 0.5012 & \textbf{0.5485} & 0.5120 \\
& JLAFAC(first) & 0.6676 & 0.5318 & 0.5859 & 0.5366 & 0.5055 & 0.5188 & 0.5025 & 0.5442 & \textbf{0.5171} & \textbf{0.5073} & 0.5455 & \textbf{0.5147} \\
& JLAFAC(second) & 0.6689 & 0.5436 & 0.5232 & 0.5459 & 0.5184 & 0.5077 & \textbf{0.5111}  &  ---    &  ---  &    ---   &   ---   &   ---    \\
\hline
\end{tabular}
}}
\hfill
\subfloat[MSPE for Nodal Attribute Prediction] 
{\resizebox{\textwidth}{!}{%
\begin{tabular}{|c|c|c|c|c|c|c|c|c|c|c|c|c|c|}
\hline
& & \multicolumn{12}{|c|}{MSPE}  \\ \hline
& Combination & \makecell{1 \\ $J = 20$ \\ $m = 2$ \\ $L = 2$ \\ $R^{*} = 4$ \\ $F^{*} = 1$} & \makecell{2 \\ $J = 20$ \\ $m = 5$ \\ $L = 2$ \\ $R^{*} = 4$ \\ $F^{*} = 1$} & \makecell{3 \\ $J = 20$ \\ $m = 8$ \\ $L = 2$ \\ $R^{*} = 4$ \\ $F^{*} = 1$} & \makecell{4 \\ $J = 40$ \\ $m = 2$ \\ $L = 2$ \\ $R^{*} = 4$ \\ $F^{*} = 1$} & \makecell{5 \\ $J = 40$ \\ $m = 5$ \\ $L = 2$ \\ $R^{*} = 4$ \\ $F^{*} = 1$} & \makecell{6 \\ $J = 40$ \\ $m = 8$ \\ $L = 2$ \\ $R^{*} = 4$ \\ $F^{*} = 1$} & \makecell{7 \\ $J = 100$ \\ $m = 5$ \\ $L = 2$ \\ $R^{*} = 4$ \\ $F^{*} = 1$} & \makecell{8 \\ $J = 20$ \\ $m = 10$ \\ $L = 3$ \\ $R^{*} = 3$ \\ $F^{*} = 2$} & \makecell{9 \\ $J = 40$ \\ $m = 10$ \\ $L = 3$ \\ $R^{*} = 3$ \\ $F^{*} = 2$} & \makecell{10 \\ $J = 100$ \\ $m = 10$ \\ $L = 3$ \\ $R^{*} = 3$ \\ $F^{*} = 2$} & \makecell{11 \\ $J = 20$ \\ $m = 5$ \\ $L = 4$ \\ $R^{*} = 5$ \\ $F^{*} = 3$} & \makecell{12 \\ $J = 40$ \\ $m = 5$ \\ $L = 4$ \\ $R^{*} = 5$ \\ $F^{*} = 3$} \\ \hline
&  DJL & 0.2742 & 0.5105 & 0.5938 & 0.3354 & 0.6345 & 0.6773 & 0.4411 & 0.5393 & 0.6438 & 0.7510 & 0.5106 & 0.6388 \\
In & DML & 0.9754 & 0.9881 & 0.9796 & 1.0014 & 0.9822 & 0.9868 & 0.9945 & 0.9811 & 0.9821 & 1.0064  & 0.9843 & 0.9943\\
&  JLAFAC(first) & \textbf{0.0053} & 0.0075 & 0.0129 & 0.0035 & \textbf{0.0117} & \textbf{0.0049} & 0.0030 & \textbf{0.0316} & \textbf{0.0249} & \textbf{0.0250} & \textbf{0.1452} & \textbf{0.1412} \\ 
&  JLAFAC(second) & 0.0054 & \textbf{0.0074} & \textbf{0.0105} & \textbf{0.0026} & 0.0134 & 0.0050 & \textbf{0.0014}  &   ---    &  ---  &    ---   &   ---   &   ---    \\
\hline
&  DJL & \textbf{0.2769} & \textbf{0.5161} & \textbf{0.6134} & \textbf{0.3476} & \textbf{0.6441} & \textbf{0.6857} & \textbf{0.4482} & \textbf{0.5562} & \textbf{0.6634} & \textbf{0.7695} & \textbf{0.5352} & \textbf{0.7284} \\
Mis. & DML & 0.9930 & 0.9898 & 0.9886 & 0.9972 & 0.9909 & 0.9898 & 1.0019 & 0.9868 & 0.9953 & 0.9950 & 0.9896 & 0.9922\\
&  JLAFAC(first) & 0.9295 & 1.0520 & 1.1028 & 0.9305 & 0.8884 & 0.9837 & 2.3284 & 0.9285 & 0.9582 & 5.0687 & 0.9553 & 0.9752 \\
& JLAFAC(second) & 0.8642 & 0.9422 & 0.9790 & 0.7403 & 1.0145 & 2.9219 & 1.2648  &  ---    &  ---  &    ---   &   ---   &   ---   \\
 \hline
&  DJL & \textbf{0.2643} & \textbf{0.5097} & \textbf{0.5924} & \textbf{0.3291} & \textbf{0.6239} & \textbf{0.6720} & \textbf{0.4385} & \textbf{0.5163} & \textbf{0.6315} & \textbf{0.7444} & \textbf{0.3857} & \textbf{0.5212}\\
Out & DML & 0.9795 & 0.9894 & 0.9791 & 0.9952 & 0.9852 & 0.9859 & 0.9952 & 0.9766 & 0.9872 & 1.0046 & 0.9840 & 0.9930 \\
& JLAFAC(first) & 0.9406 & 1.0711 & 1.1173 & 0.8959 & 0.8529 & 0.9780 & 0.9782 & 0.9227 & 0.9599 & 0.9441 & 0.9327 & 0.9557   \\
& JLAFAC(second) & 0.8564 & 0.9439 & 0.9781 & 0.7057 & 1.0163 & 0.8874 & 2.8784  &  ---    &  ---  &    ---   &   ---   &   ---  \\
\hline
\end{tabular}
}}
\caption{Simulation results for our proposed approach DJL with competitors DML, JLAFAC(first) and JLAFAC(second) under 12 different simulation combinations for Simulation Scheme 3. Notably, for simulation scenarios with $L>2$, we present the average performance over different layers and present it under JLAFAC(first), hence cells under JLAFAC(second) are kept blank for these scenarios. We report area under the curve (AUC) in predicting an edge/no edge for the multiplex graphs, as well as mean squared prediction error (MSPE) in predicting nodal attributes. Here ``In'' , ``Mis.'' and ``Out'' refer to in-sample, missing and out-of-sample predictive scenarios, respectively. The best performing model under each combination and each scenario is boldfaced.}\label{Tab3:Prediction}
\end{table}

\begin{figure}[h]
    \centering
    \subfloat[Scheme 1: In-Sample]{\includegraphics[width=.33\linewidth]{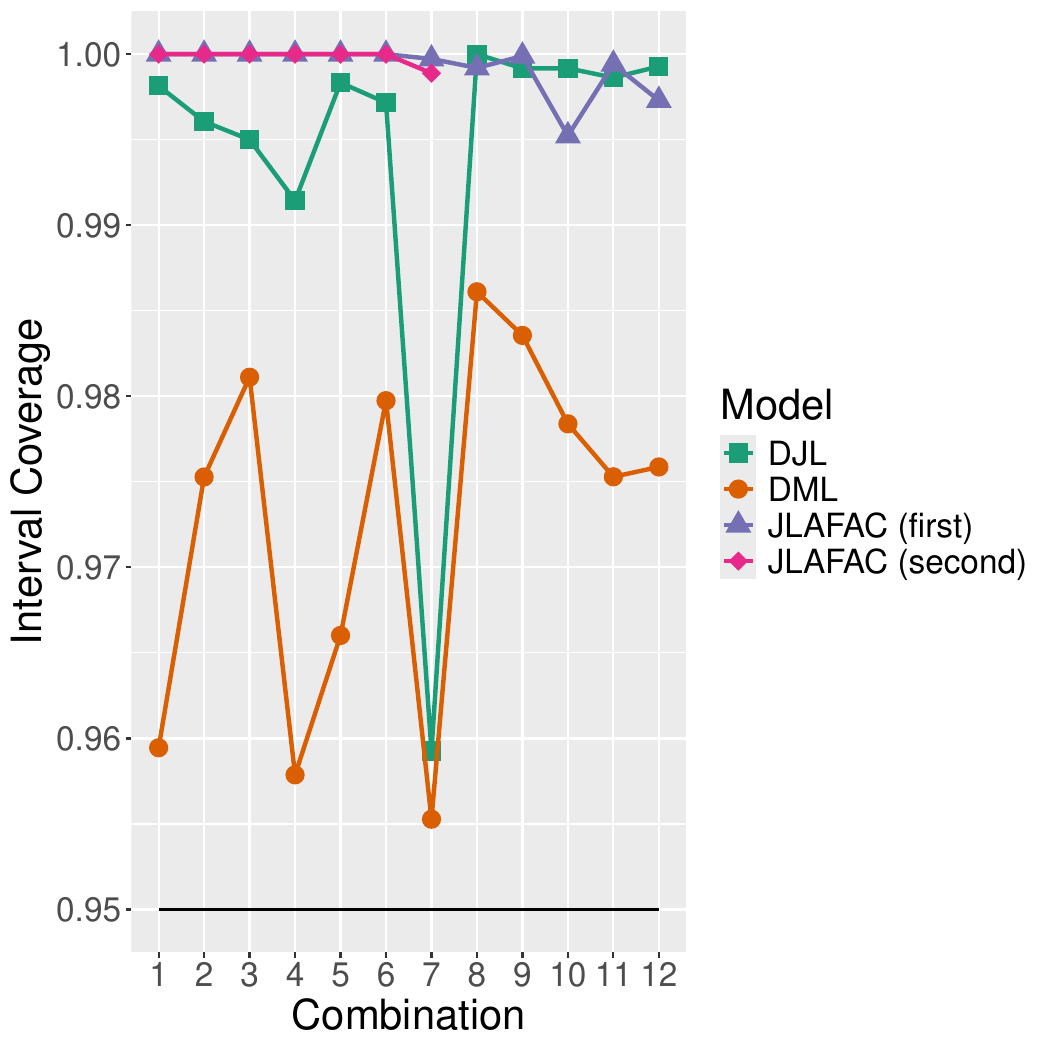}}
    \subfloat[Scheme 1: Missing] {\includegraphics[width=.33\linewidth]{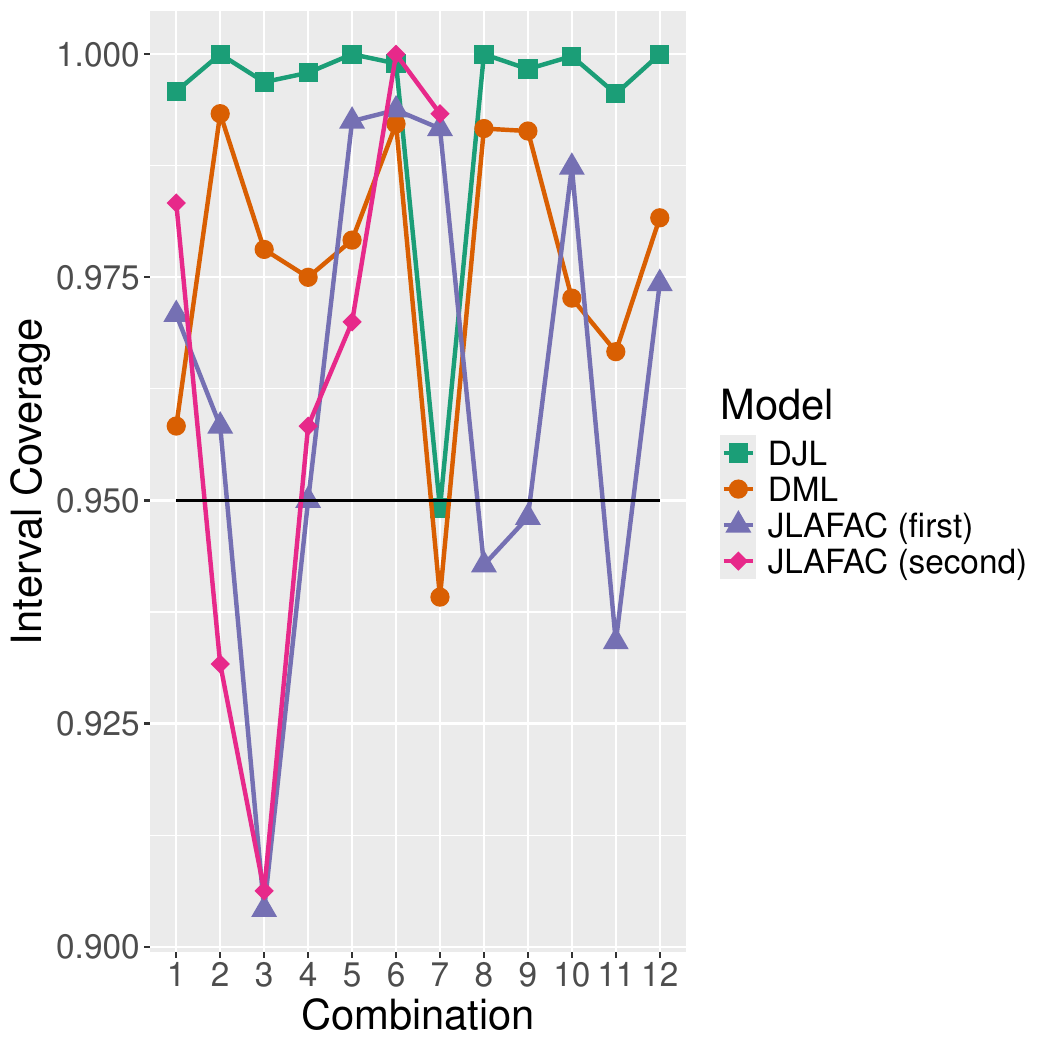}} 
    \subfloat[Scheme 1: Out-of-Sample]{\includegraphics[width=.33\linewidth]{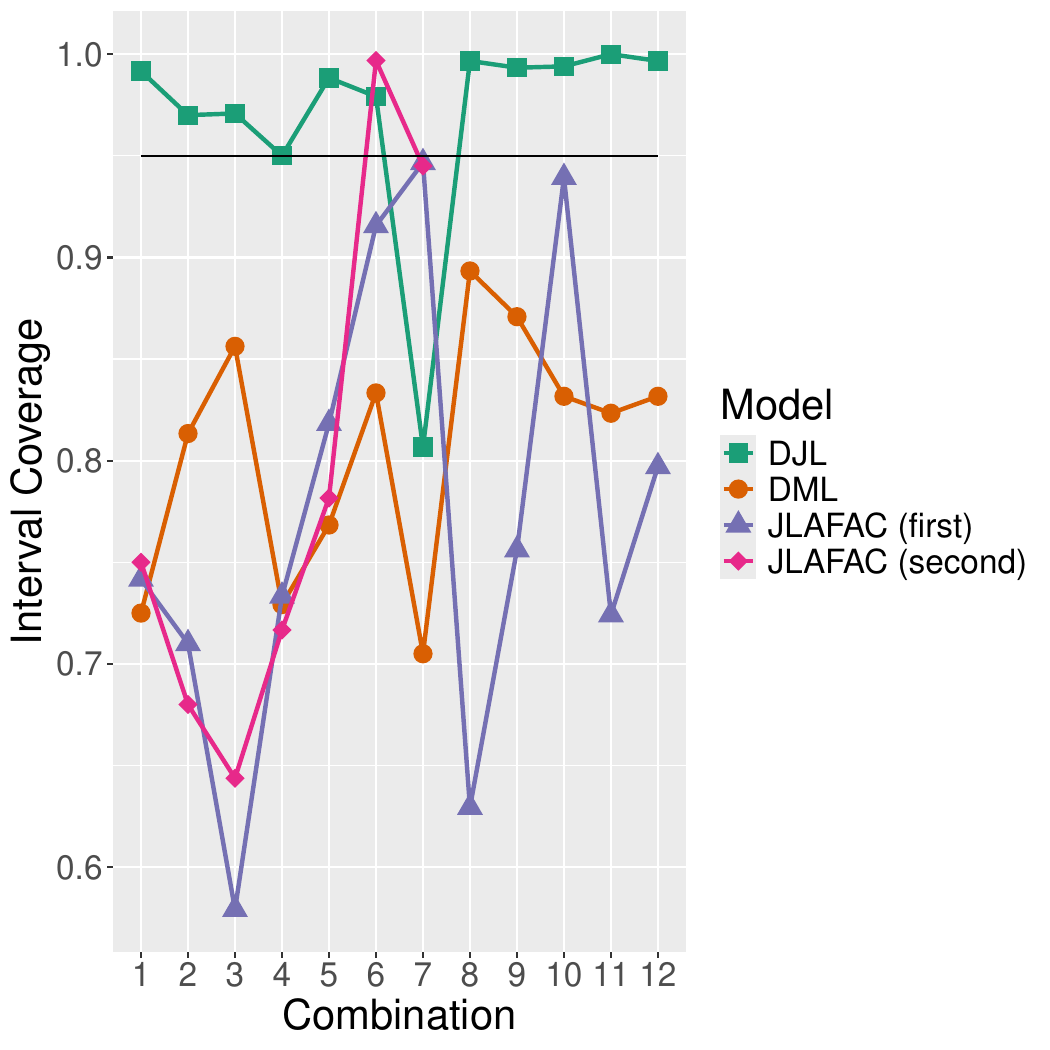}} \\
    \subfloat[Scheme 2: In-Sample]{\includegraphics[width=.33\linewidth]{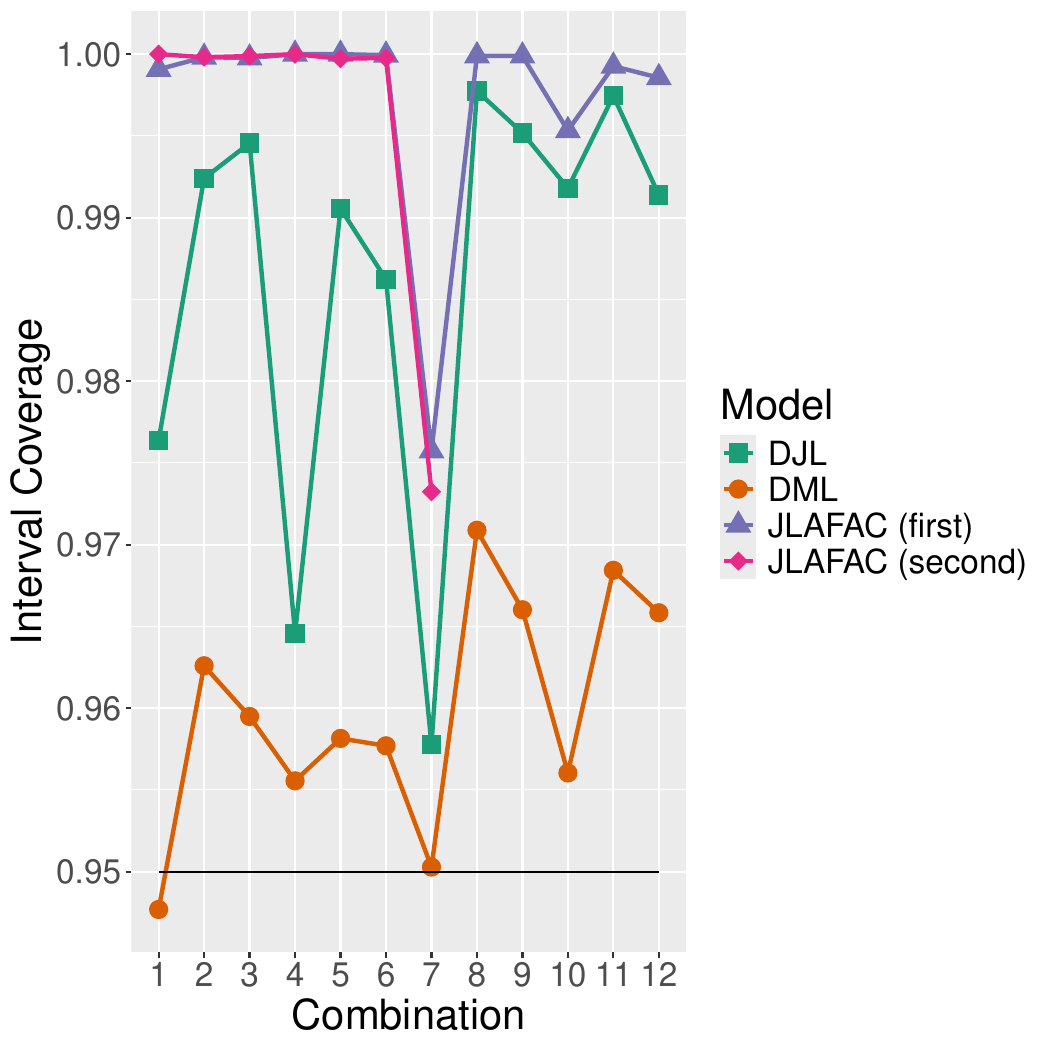}}
    \subfloat[Scheme 2: Missing] {\includegraphics[width=.33\linewidth]{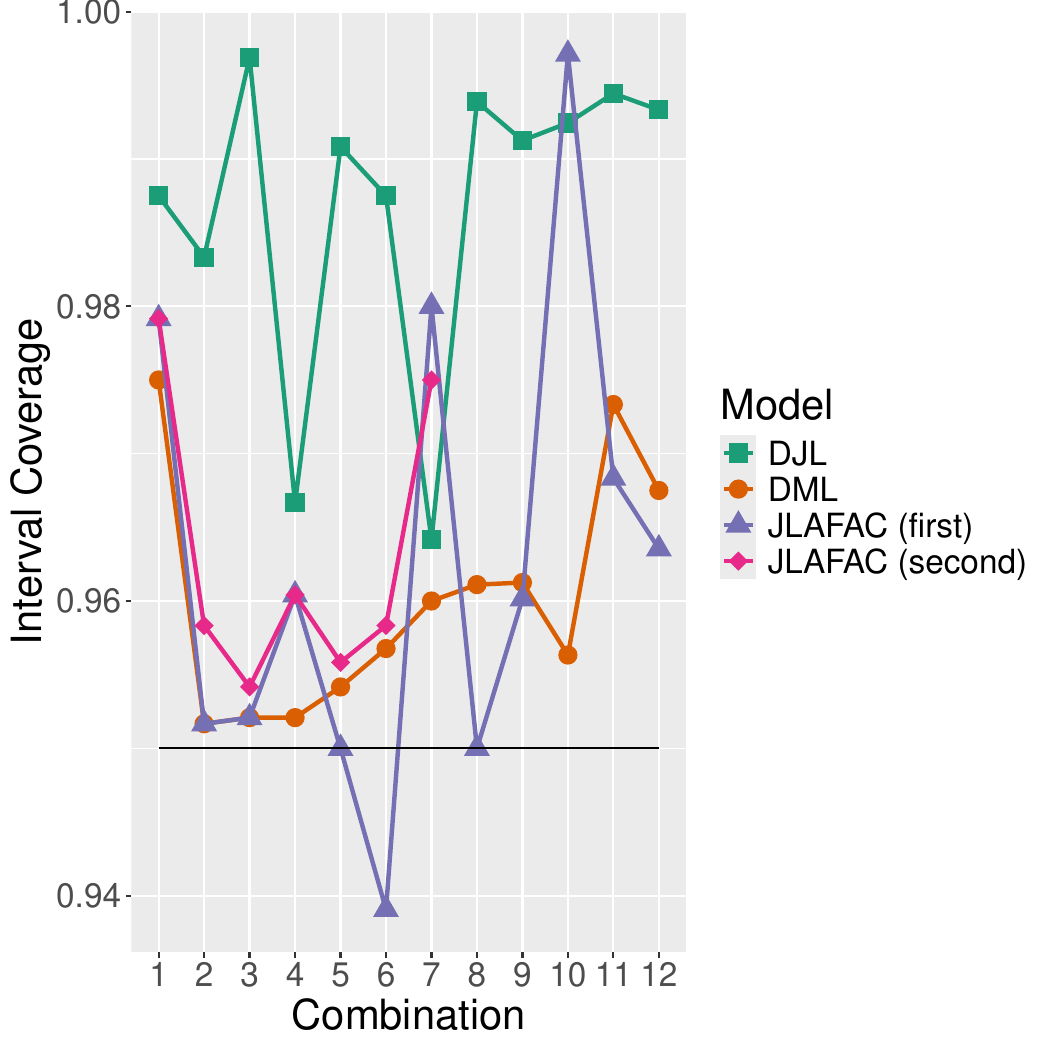}} 
    \subfloat[Scheme 2: Out-of-Sample]{\includegraphics[width=.33\linewidth]{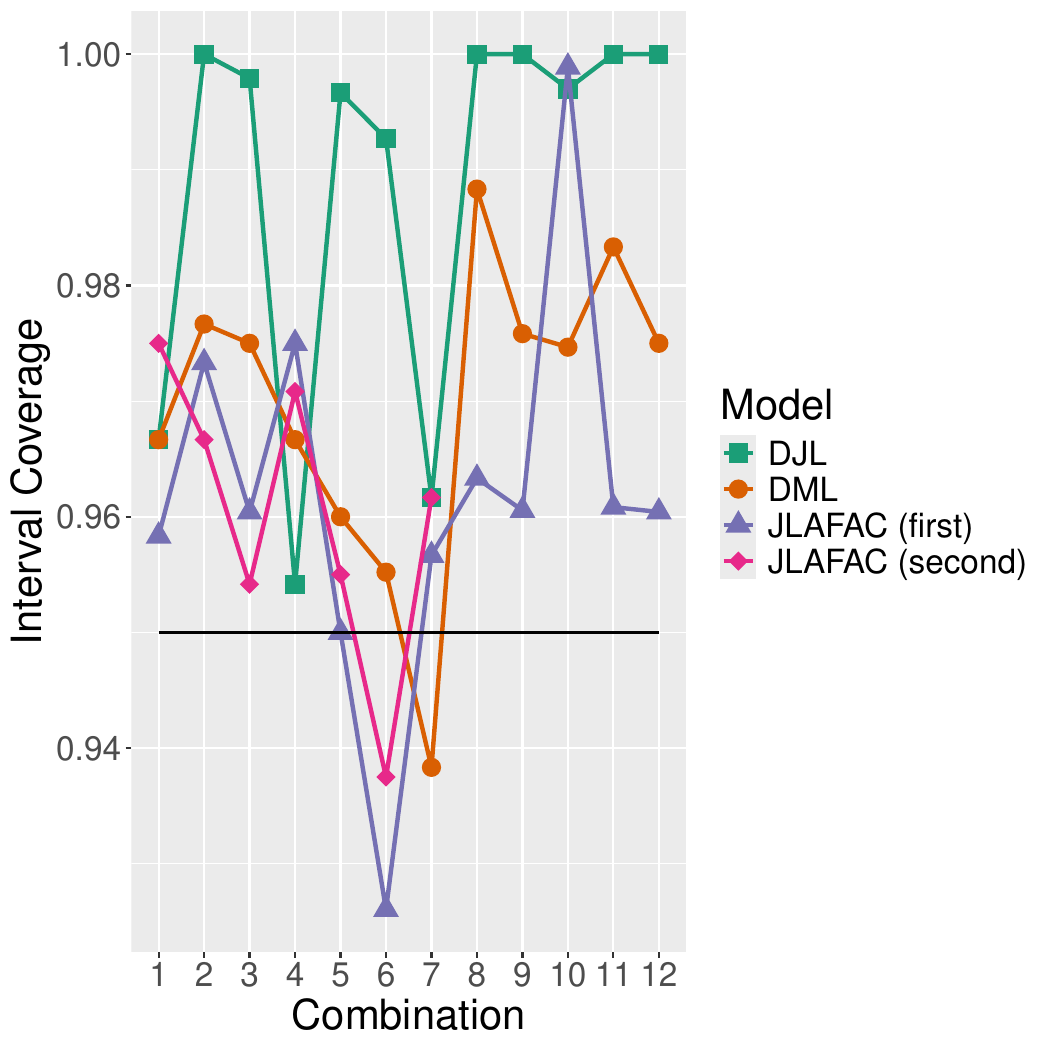}} \\
    \subfloat[Scheme 3: In-Sample]{\includegraphics[width=.33\linewidth]{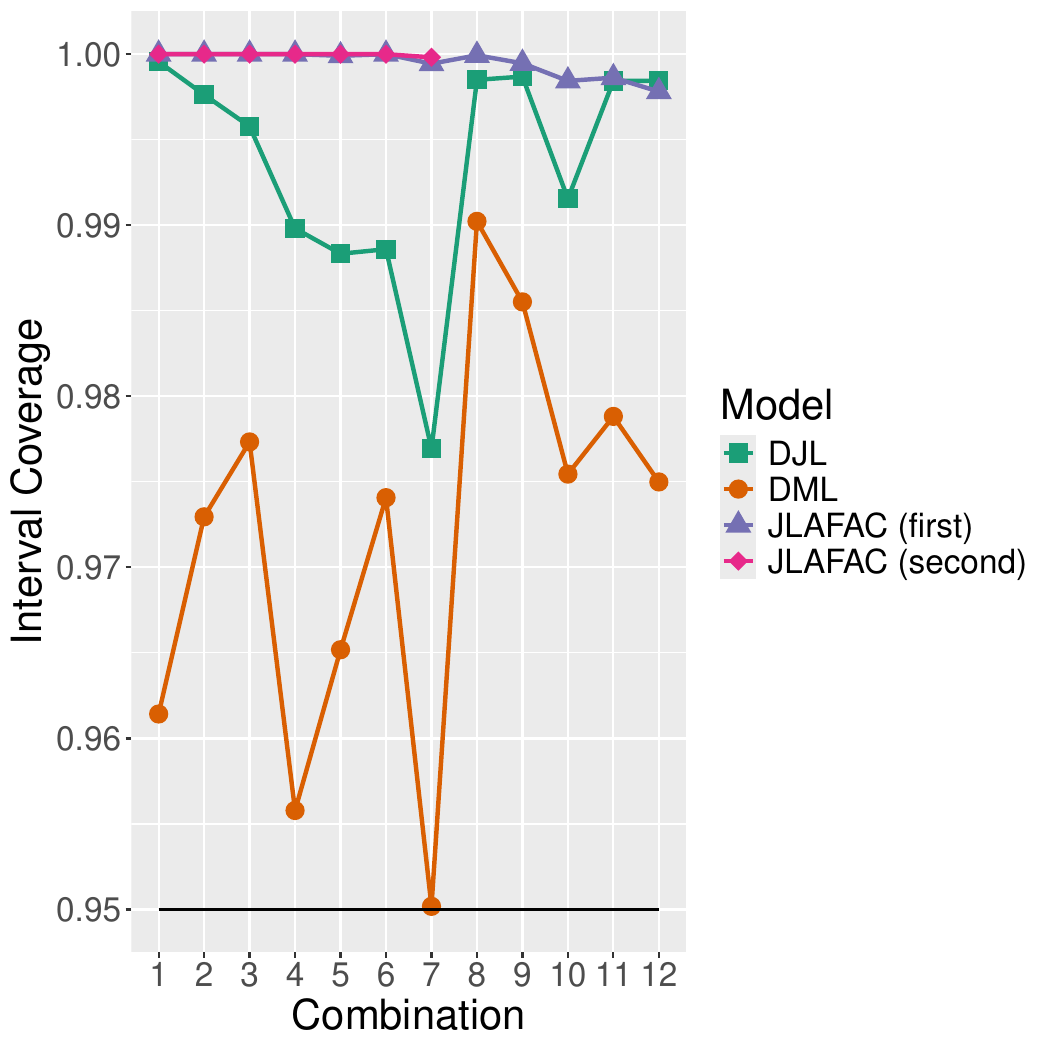}}
    \subfloat[Scheme 3: Missing] {\includegraphics[width=.33\linewidth]{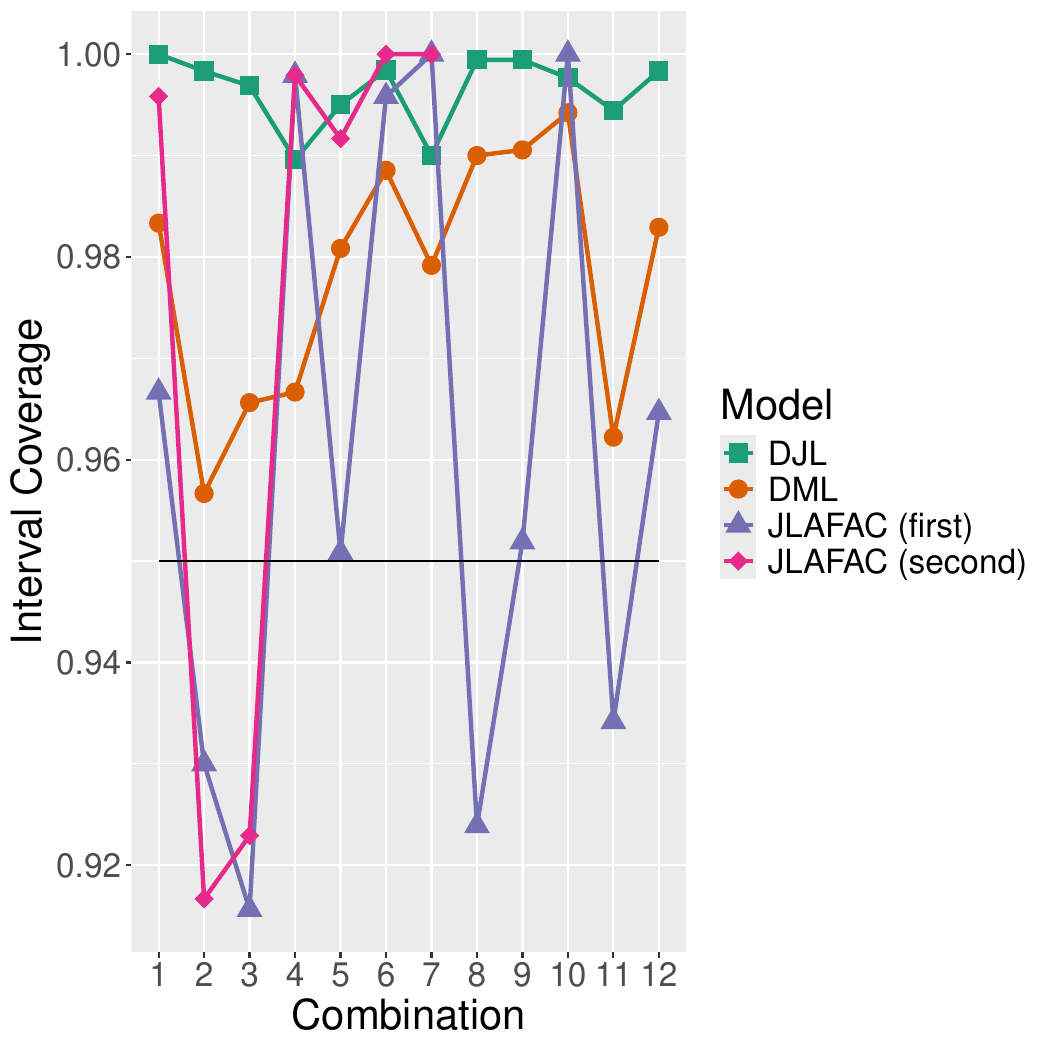}} 
    \subfloat[Scheme 3: Out-of-Sample]{\includegraphics[width=.33\linewidth]{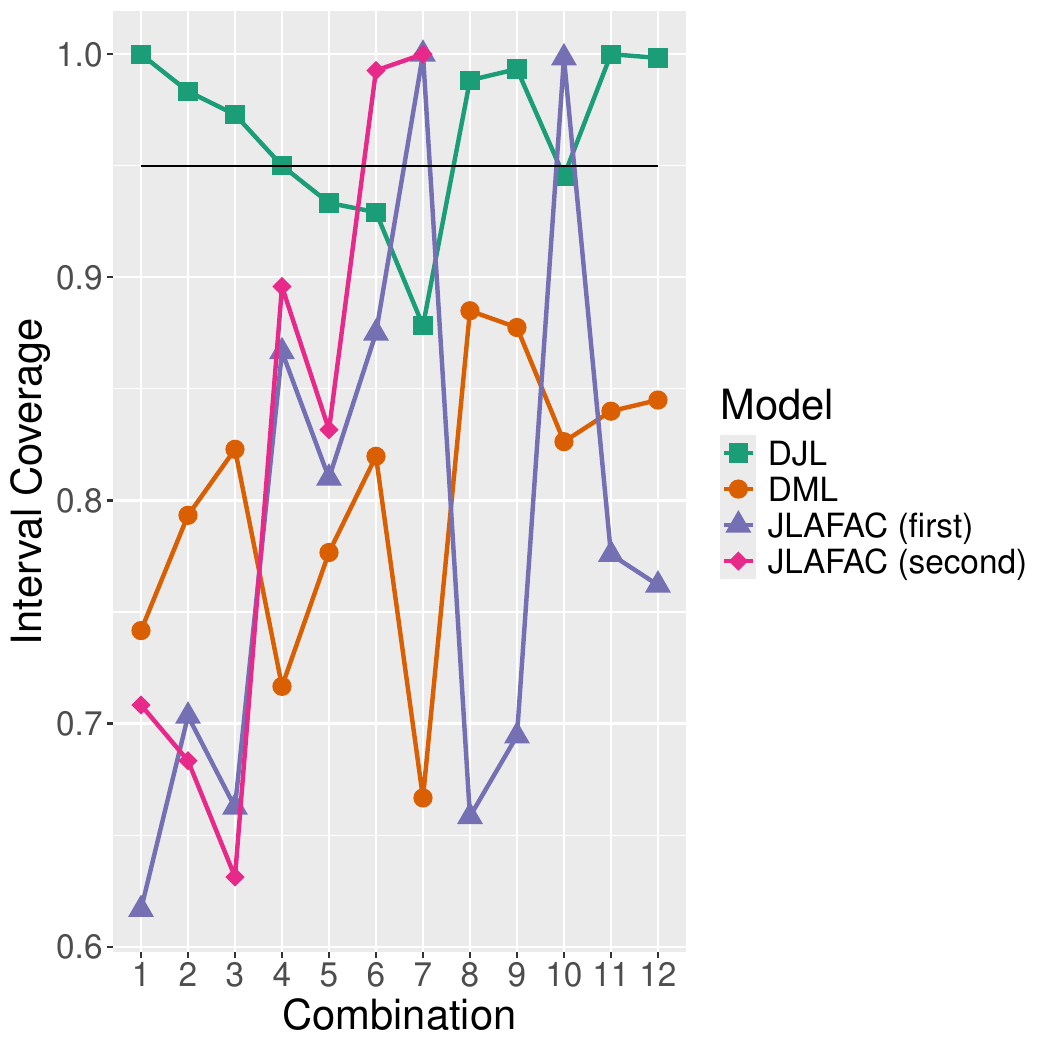}}
    \caption{Coverages of 95\% predictive intervals for attributes \emph{in-sample}, \emph{missing} and \emph{out-of-sample} under Simulation Schemes 1, 2, and 3.}\label{node_cov_sim}
\end{figure}

\begin{figure}[h]
    \centering
    \subfloat[Scheme 1: In-Sample]{\includegraphics[width=.33\linewidth]{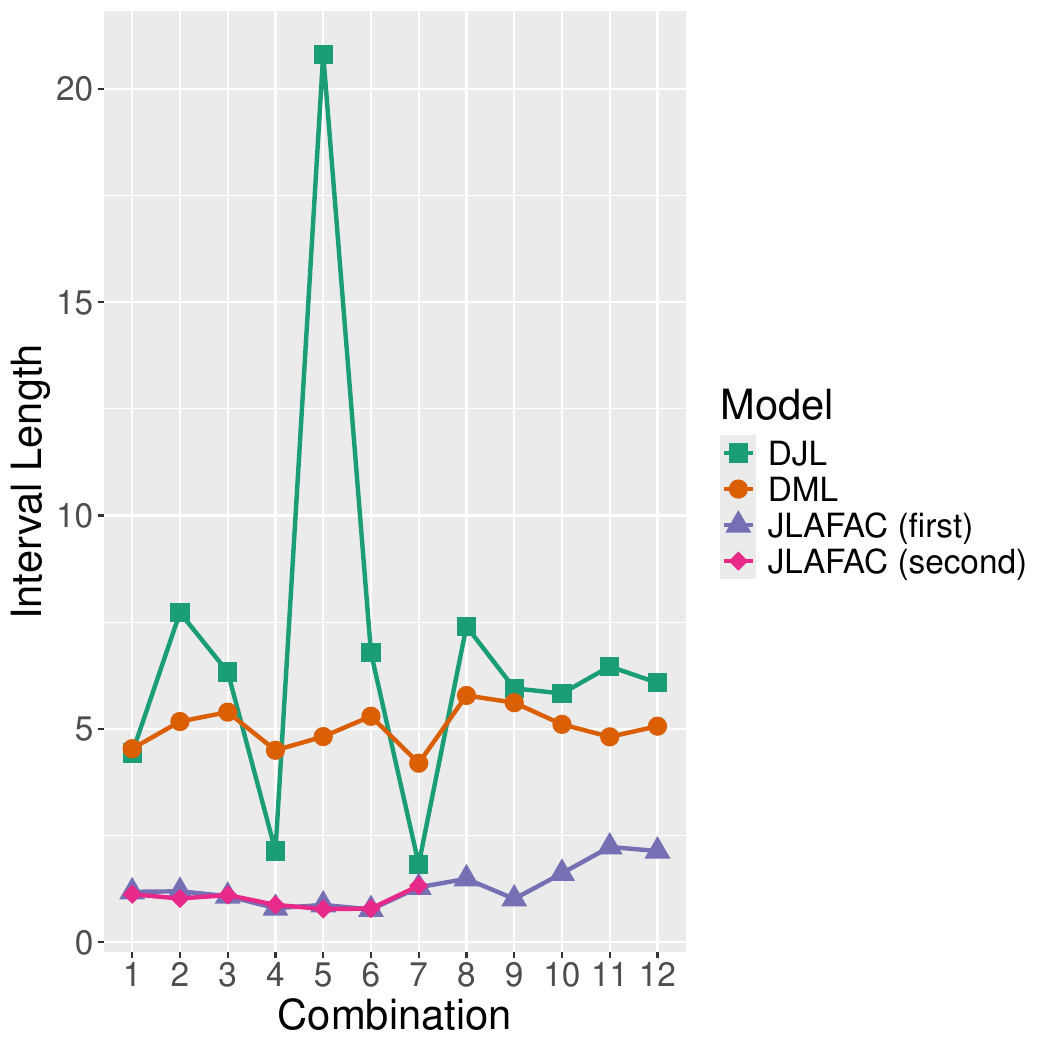}}
    \subfloat[Scheme 1: Missing] {\includegraphics[width=.33\linewidth]{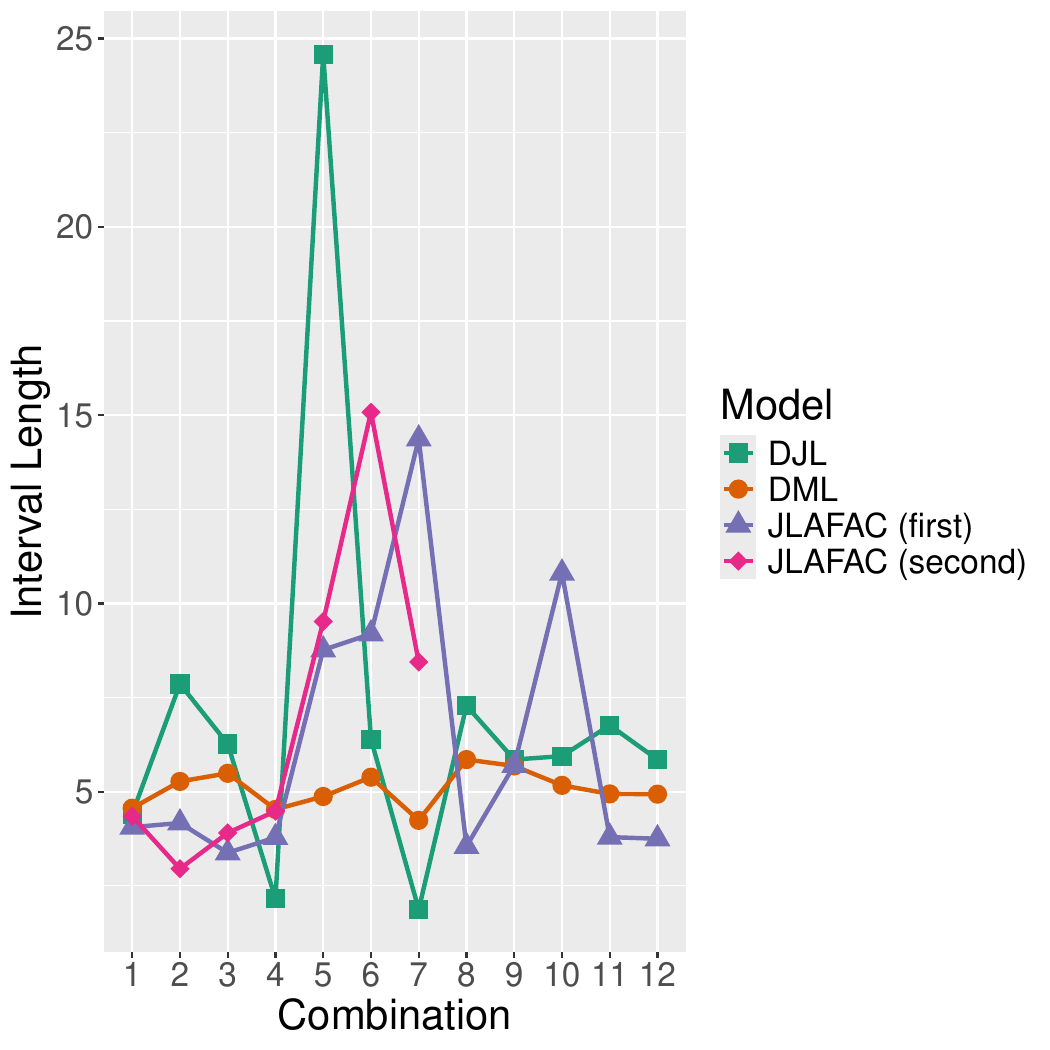}} 
    \subfloat[Scheme 1: Out-of-Sample]{\includegraphics[width=.33\linewidth]{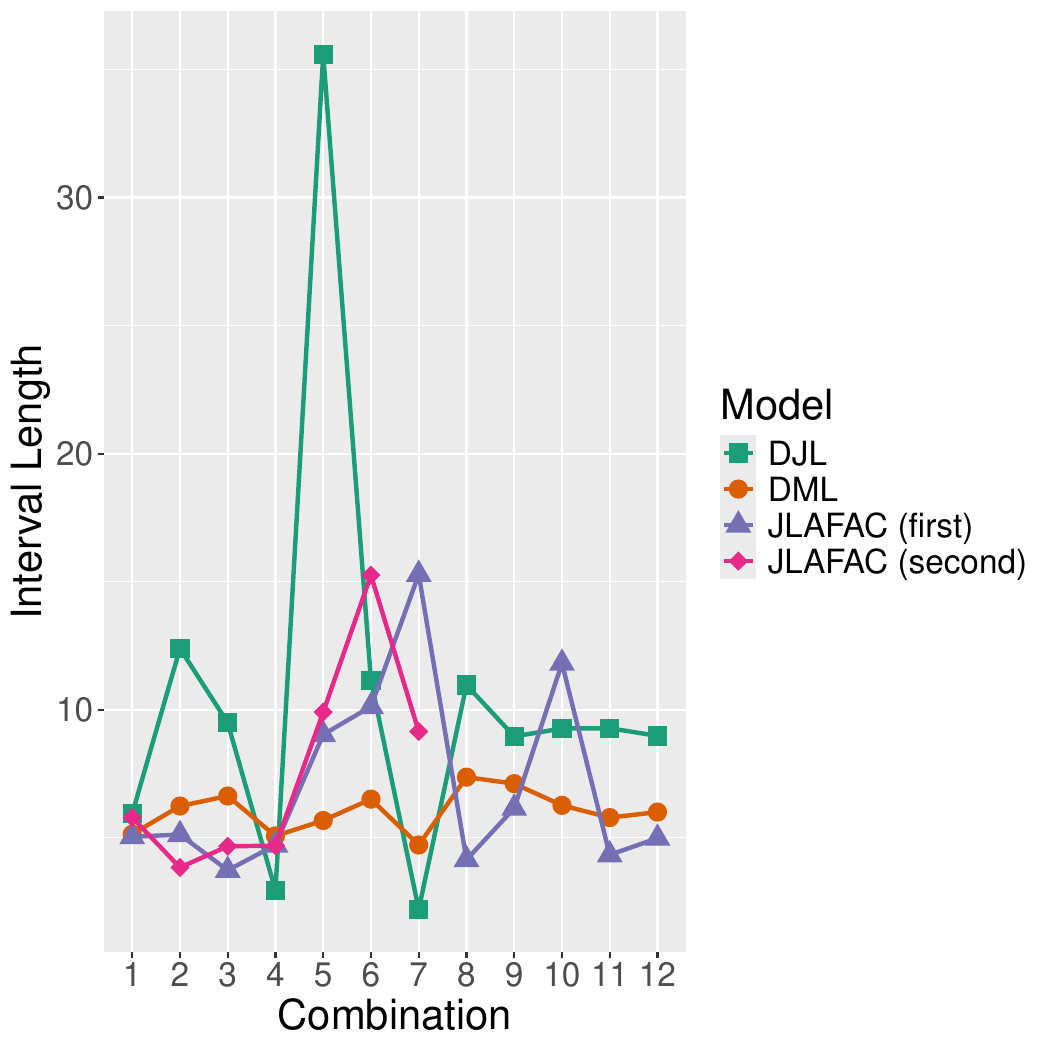}} \\
    \subfloat[Scheme 2: In-Sample]{\includegraphics[width=.33\linewidth]{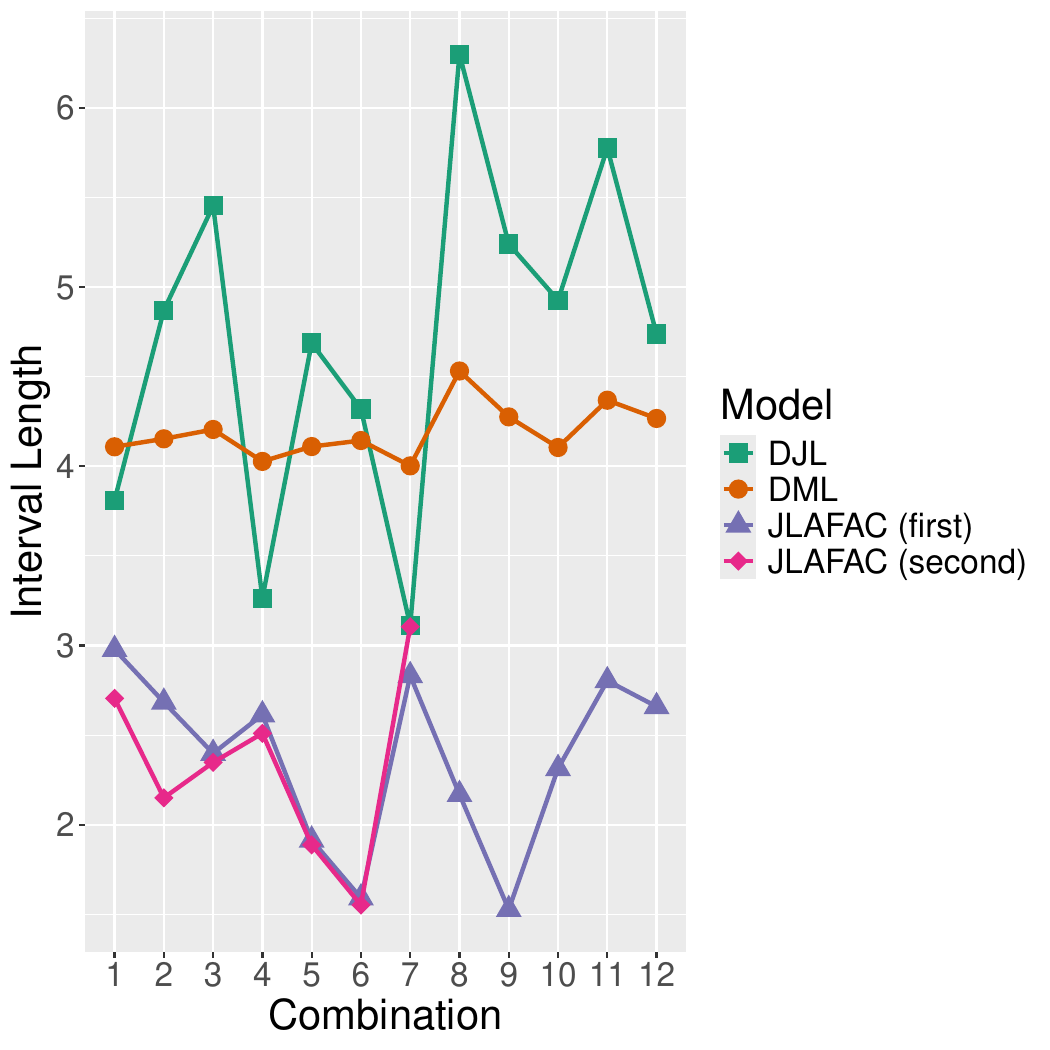}}
    \subfloat[Scheme 2: Missing] {\includegraphics[width=.33\linewidth]{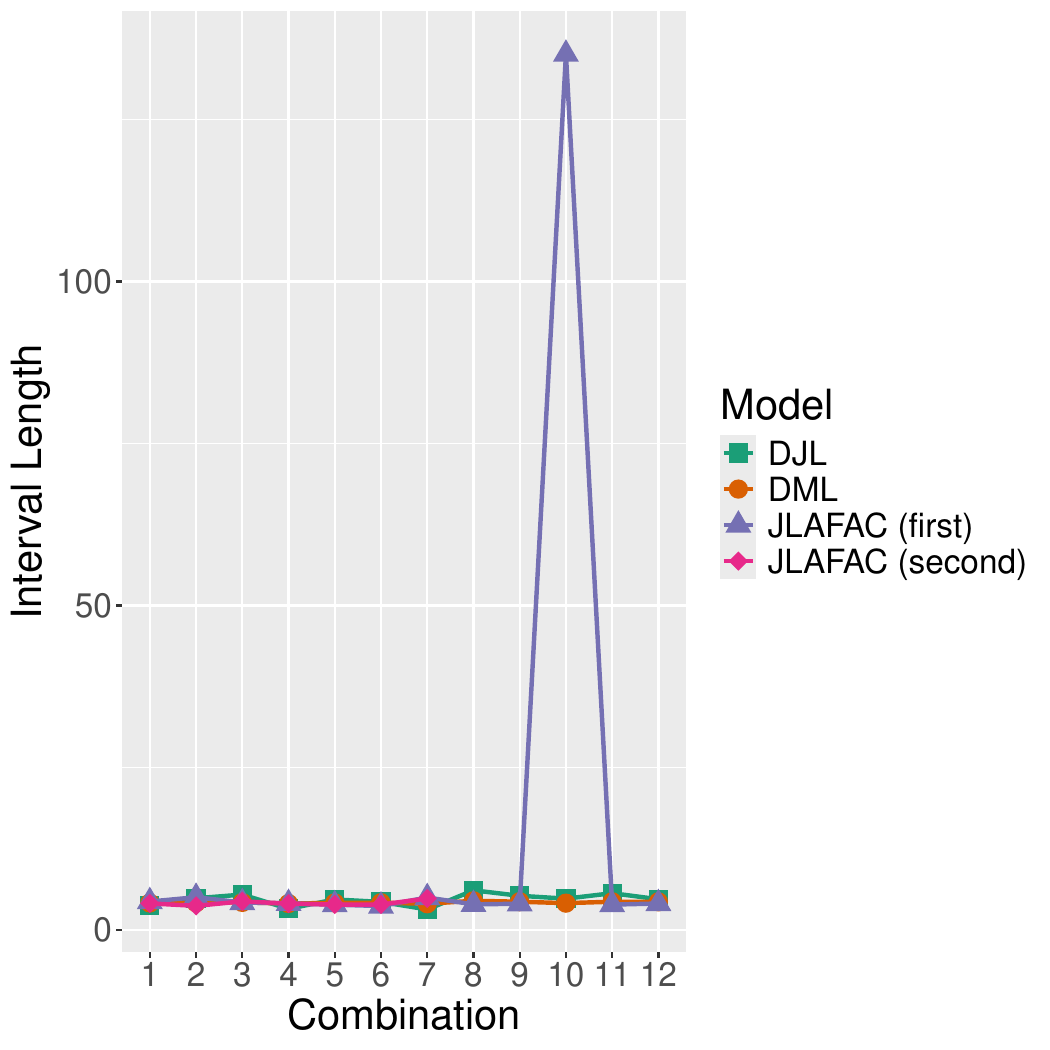}} 
    \subfloat[Scheme 2: Out-of-Sample]{\includegraphics[width=.33\linewidth]{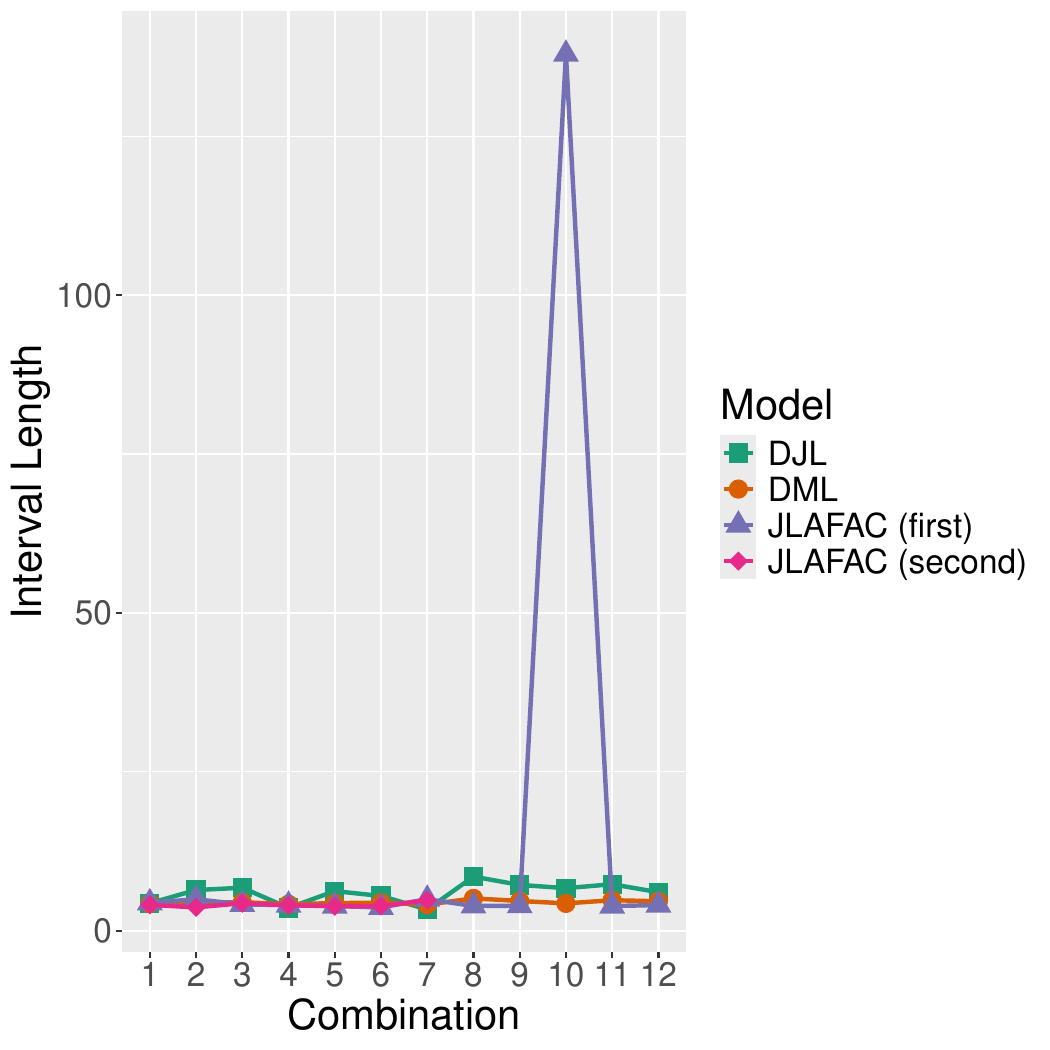}} \\
    \subfloat[Scheme 3: In-Sample]{\includegraphics[width=.33\linewidth]{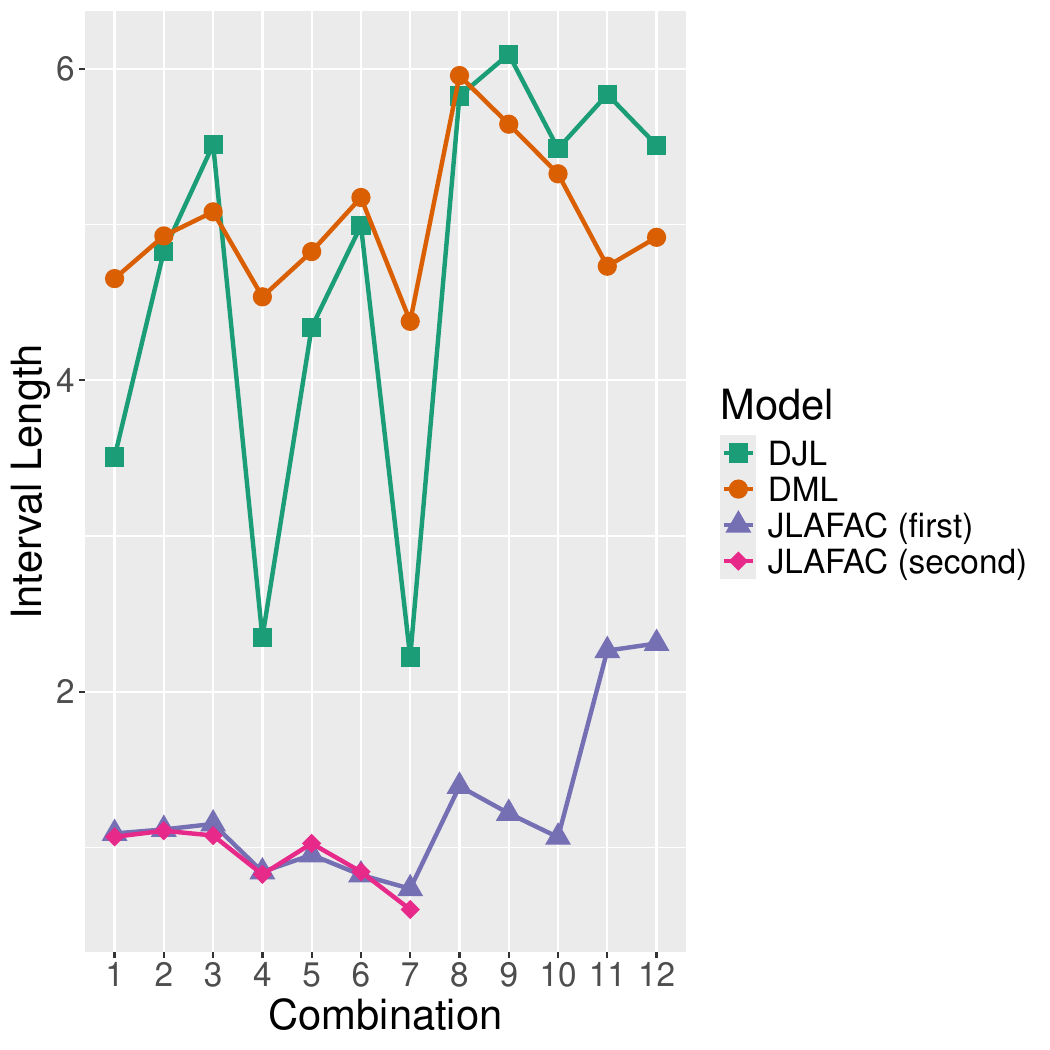}}
    \subfloat[Scheme 3: Missing] {\includegraphics[width=.33\linewidth]{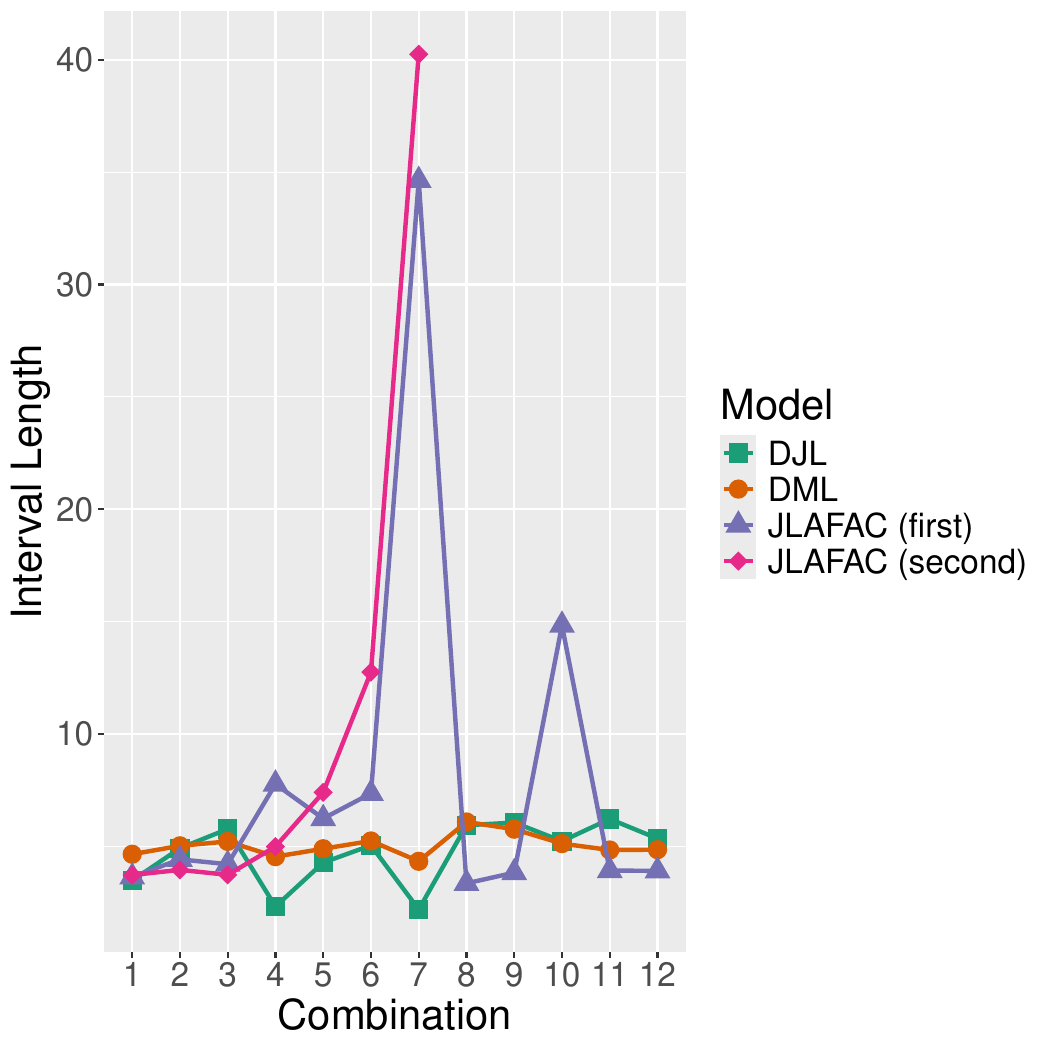}} 
    \subfloat[Scheme 3: Out-of-Sample]{\includegraphics[width=.33\linewidth]{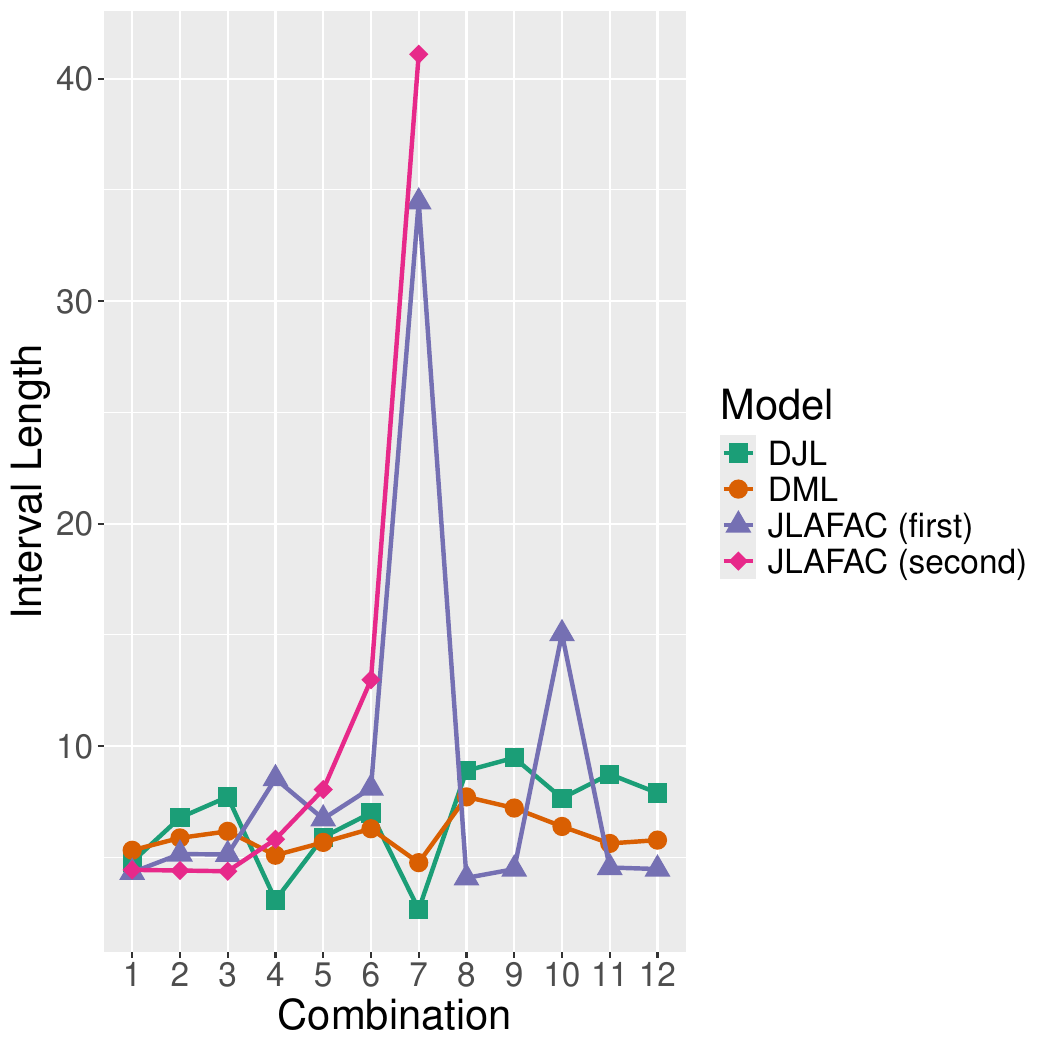}}
    \caption{Length of 95\% predictive intervals for attributes \emph{in-sample}, \emph{missing} and \emph{out-of-sample} under Simulation Schemes 1, 2, and 3.}\label{node_len_sim}
\end{figure}

All competitors achieve close to nominal coverage of 95\% predictive intervals for the \emph{in-sample} scenario across all combinations and simulation schemes, as shown in the first column of Figure~\ref{node_cov_sim}. When predicting attributes in the \emph{missing} and \emph{out-of-sample} scenarios, DJL consistently maintains near-nominal coverage under Simulation Scheme 1, while other competitors exhibit significant under-coverage for the \emph{out-of-sample} scenario, as indicated in the first row, third column of Figure~\ref{node_cov_sim}. This under-coverage is further reflected by the narrower 95\% predictive intervals of the competitors in these scenarios, as shown in the first row, third column of Figure~\ref{node_len_sim}. In Simulation Scheme 2, under the \emph{missing} scenario, all competitors achieve coverage close to nominal. In Simulation Scheme 3, which involves model mis-specification, DJL shows near-nominal coverage, while other competitors show under-coverage for the \emph{out-of-sample} scenario. Importantly, DJL achieves better coverage than its competitors with similar or smaller predictive interval widths for both the \emph{missing} and \emph{out-of-sample} scenarios.

\subsection{Sensitivity Analysis}\label{sec:sen_ana}

In our simulation study, we fit DJL with $R=R_\zeta=4$ and $F=1$. This section carefully assesses the reasonableness of these choices by examining the sensitivity of inference to different 
$(R,F)$ combinations. Specifically, we first fix $F$ and vary 
$R$, then fix $R$ and vary $F$. The results of these two analyses are presented in Sections 2.2.1 and 2.2.2, respectively. Finally, we investigate the impact of assuming $R=R_\zeta$ by fitting the model with different values of $R_\zeta$ while keeping $R$ fixed.

\subsubsection{Sensitivity to R}
To assess the sensitivity w.r.t. $R$, while fixing $F$, we fit DJL with $R=2,4,8,10,12,15,20$ when $F=1$ under three different cases of Simulation Scheme 1, with $L = 2$, $J=20$, $T=20$, and $m = 2$.
These cases, varying the true hyperparameter values, $R^{*}$ and $F^{*}$, are:
\begin{itemize}
    \item \textbf{Case 1}: $R^* = 4$, $F^* = 1$,
    \item \textbf{Case 2}: $R^* = 3$, $F^* = 2$,
    \item \textbf{Case 3}: $R^* = 2$, $F^* = 3$.
\end{itemize}
The results in Tables \ref{sens_tab_R_1_rev}, \ref{sens_tab_R_2_rev}, and \ref{sens_tab_R_3_rev}, show the AUC for edge prediction, MSPE for nodal attribute prediction, and $95\%$ Prediction Interval Coverages and Lengths for nodal attributes in the \emph{in-sample, missing}, and \emph{out-of-sample} predictive scenarios for the above three cases when setting $R=2,4,8,10,12,15,20$.
In each set of results, AUC for predicting unobserved edges remains stable across \emph{in-sample}, \emph{missing}, and \emph{out-of-sample} scenarios once $R > 2$. In each case, the MSPE for estimating nodal attributes appears to decrease as $R$ increases. However, the differences in MSPE across $R$ values are much less pronounced when $R^*$ is lower in Case 2 ($R^*=3$, $F^*=2$) and Case 3 ($R^*=2$, $F^*=3$), compared to Case 1 ($R^*=4$, $F^*=1$). For $95\%$ prediction intervals of the nodal attributes, once $R$ is at least $4$, the nominal coverage is reached. As $R$ increases further, the coverages remain similar or increase mildly, while the interval lengths increase a bit. Together, these results suggest $R=4$ strikes an effective balance between computational efficiency and statistical accuracy. The findings are in line with earlier literature, which also employs small values of $R$ to effectively capture complex graph properties via bi-linear effects of node-specific latent variables \cite{guhaniyogi2020joint,fosdick2015testing}.


\begin{table}[H]
    \centering
    \subfloat[AUC for Edge Prediction] 
     {\resizebox{0.48\textwidth}{!}{%
    \begin{tabular}{c|c c c c c c c}
         $R$ & 2 & 4 & 8 & 10 & 12 & 15 & 20 \\
         \hline
         \emph{in} & 0.9387 & 0.9810 & 0.9835 & 0.9844 & 0.9849 & 0.9852 & 0.9857 \\
         \emph{mis} & 0.9218 & 0.9571 & 0.9587 & 0.9598 & 0.9565 & 0.9538 & 0.9560  \\
         \emph{out} & 0.9735 & 0.9997 & 0.9999 & 0.9997 & 0.9998 & 0.9999 & 0.9999\\
    \end{tabular}}}
    \hfill
    \subfloat[MSPE for Nodal Attribute Prediction] 
     {\resizebox{0.48\textwidth}{!}{%
    \begin{tabular}{c|c c c c c c c}
         $R$ & 2 & 4 & 8 & 10 & 12 & 15 & 20 \\
         \hline
         \emph{in} & 0.3861 & 0.2852 & 0.1822 & 0.1571 & 0.1252 & 0.0608 & 0.0707 \\
         \emph{mis} & 0.3779 & 0.1955 & 0.1714 & 0.1559 & 0.1199 & 0.0576 & 0.0743  \\
         \emph{out} & 0.3864 & 0.2793 & 0.1725 & 0.1421 & 0.1089 & 0.0436 & 0.0376\\
    \end{tabular}}}
    \hfill
    \subfloat[$95\%$ Prediction Interval Coverage]
    {\resizebox{0.48\textwidth}{!}{%
    \begin{tabular}{c|c c c c c c c}
         $R$ & 2 & 4 & 8 & 10 & 12 & 15 & 20 \\
         \hline
         \emph{in} & 0.9764 & 0.9968 & 1.0000 & 1.0000 & 1.0000 & 1.0000 & 1.0000 \\
         \emph{mis} & 0.9375 & 1.0000 & 1.0000 & 1.0000 & 1.0000 & 1.0000 & 1.0000  \\
         \emph{out} & 0.8667 & 0.9917 & 1.0000 & 1.0000 & 1.0000 & 1.0000 & 1.0000\\
    \end{tabular}}}
    \hfill
    \subfloat[$95\%$ Prediction Interval Length]
    {\resizebox{0.48\textwidth}{!}{%
    \begin{tabular}{c|c c c c c c c}
         $R$ & 2 & 4 & 8 & 10 & 12 & 15 & 20 \\
         \hline
         \emph{in} & 2.1348 & 2.8725 & 3.7874 & 3.5490 & 4.3305 & 4.4654 & 5.1631 \\
         \emph{mis} & 2.1793 & 3.2326 & 4.0007 & 3.7774 & 4.5477 & 4.8024 & 5.4087 \\
         \emph{out} & 2.3901 & 4.4664 & 5.4613 & 4.9834 & 6.0416 & 6.3324 & 7.1336\\
    \end{tabular}}}
    \caption{AUC, MSPE, coverage and length of $95\%$ prediction intervals for different values of $R=2,4,8,10,12,15,20$ used to fit DJL in the \emph{in-sample, missing}, and \emph{out-of-sample} predictive scenarios in Case 1. This case assumes $L = 2$, $J=20$, $T=20$, $m = 2$, $R^* = 4$, and $F^* = 1$. DJL is fitted with $F = 1$ for different choices of $R$.}
    \label{sens_tab_R_1_rev}
\end{table}

\begin{table}[H]
    \centering
    \subfloat[AUC for Edge Prediction] 
     {\resizebox{0.48\textwidth}{!}{%
    \begin{tabular}{c|c c c c c c c}
         $R$ & 2 & 4 & 8 & 10 & 12 & 15 & 20 \\
         \hline
         \emph{in} & 0.8634 & 0.9087&  0.9137 & 0.9169 & 0.9188 & 0.9197 & 0.9221 \\
         \emph{mis} & 0.8939 & 0.9357 & 0.9344 & 0.9396 & 0.9373 & 0.9375 & 0.9339  \\
         \emph{out} & 0.9565 & 0.9903 & 0.9875 & 0.9902 &  0.9917 & 0.9913 & 0.9915 \\
    \end{tabular}}}
    \hfill
    \subfloat[MSPE for Nodal Attribute Prediction] 
     {\resizebox{0.48\textwidth}{!}{%
    \begin{tabular}{c|c c c c c c c}
         $R$ & 2 & 4 & 8 & 10 & 12 & 15 & 20 \\
         \hline
         \emph{in} & 0.1799 & 0.1706 & 0.2611 & 0.1529 & 0.1106 & 0.1029 & 0.0973 \\
         \emph{mis} & 0.1454 & 0.1416 & 0.2455 & 0.1407 & 0.0803 & 0.0908 & 0.0673 \\
         \emph{out} & 0.1253 & 0.1147 & 0.1575 & 0.0882 &  0.0500 & 0.0397 & 0.0380\\
    \end{tabular}}}
    \hfill
    \subfloat[$95\%$ Prediction Interval Coverage]
    {\resizebox{0.48\textwidth}{!}{%
    \begin{tabular}{c|c c c c c c c}
         $R$ & 2 & 4 & 8 & 10 & 12 & 15 & 20 \\
         \hline
         \emph{in} & 0.9903 & 0.9968 & 0.9954 & 0.9968 & 0.9949 & 0.9954 & 0.9954 \\
         \emph{mis} & 0.9750 & 0.9917 & 0.9958 & 0.9958 & 0.9958 & 0.9958 & 0.9958 \\
         \emph{out} & 0.9750 & 0.9917 & 1.0000 & 1.0000 & 1.0000 & 1.0000 & 1.0000\\
    \end{tabular}}}
    \hfill
    \subfloat[$95\%$ Prediction Interval Length]
    {\resizebox{0.48\textwidth}{!}{%
    \begin{tabular}{c|c c c c c c c}
         $R$ & 2 & 4 & 8 & 10 & 12 & 15 & 20 \\
         \hline
         \emph{in} & 2.6727 & 3.2946 & 5.2308 & 4.4391 & 4.6172 & 4.5568 & 5.0013 \\
         \emph{mis} & 2.7085 & 3.4922 & 5.4913 & 4.6005 & 4.7517 & 4.7375 & 5.1982 \\
         \emph{out} & 3.0304 & 4.2926 & 7.2253 & 5.9734 & 6.1377 & 6.1912 & 6.7165\\
    \end{tabular}}}
    \caption{AUC, MSPE, coverage and length of $95\%$ prediction intervals for different values of $R=2,4,8,10,12,15,20$ used to fit DJL in the \emph{in-sample, missing}, and \emph{out-of-sample} predictive scenarios in Case 1. This case assumes $L = 2$, $J=20$, $T=20$, $m = 2$, $R^* = 3$, and $F^* = 2$. DJL is fitted with $F = 1$ for different choices of $R$.}
    \label{sens_tab_R_2_rev}
\end{table}

\begin{table}[H]
    \centering
    \subfloat[AUC for Edge Prediction] 
     {\resizebox{0.48\textwidth}{!}{%
    \begin{tabular}{c|c c c c c c c}
         $R$ & 2 & 4 & 8 & 10 & 12 & 15 & 20 \\
         \hline
         \emph{in} & 0.7068 & 0.7311 & 0.7497 & 0.7542 & 0.7592 & 0.7662 & 0.7729 \\
         \emph{mis} & 0.7952 & 0.7842 & 0.7825 & 0.7793 & 0.7722 & 0.7638 & 0.7638  \\
         \emph{out} & 0.8547 & 0.8530 & 0.8419 & 0.8382 & 0.8342 & 0.8266 & 0.8211\\
    \end{tabular}}}
    \hfill
    \subfloat[MSPE for Nodal Attribute Prediction] 
     {\resizebox{0.48\textwidth}{!}{%
    \begin{tabular}{c|c c c c c c c}
         $R$ & 2 & 4 & 8 & 10 & 12 & 15 & 20 \\
         \hline
         \emph{in} & 0.7007 & 0.6442 & 0.5752 & 0.5627 & 0.5538 & 0.5535 & 0.5370 \\
         \emph{mis} & 0.6517 & 0.6079 & 0.5456 & 0.5354 & 0.5300 & 0.5426 & 0.5349  \\
         \emph{out} & 0.4625 & 0.3531 & 0.2536 & 0.2495 & 0.2564 & 0.2549 & 0.2710\\
    \end{tabular}}}
    \hfill
    \subfloat[$95\%$ Prediction Interval Coverage]
    {\resizebox{0.48\textwidth}{!}{%
    \begin{tabular}{c|c c c c c c c}
         $R$ & 2 & 4 & 8 & 10 & 12 & 15 & 20 \\
         \hline
         \emph{in} & 0.9787 & 0.9852 & 0.9852 & 0.9843 & 0.9861 & 0.9852 & 0.9856 \\
         \emph{mis} & 0.9792 & 0.9833 & 0.9833 & 0.9792 & 0.9792 & 0.9833 & 0.9875 \\
         \emph{out} & 0.9583 & 0.9833 & 0.9750 & 0.9750 & 0.9750 & 0.9750 & 0.9833\\
    \end{tabular}}}
    \hfill
    \subfloat[$95\%$ Prediction Interval Length]
    {\resizebox{0.48\textwidth}{!}{%
    \begin{tabular}{c|c c c c c c c}
         $R$ & 2 & 4 & 8 & 10 & 12 & 15 & 20 \\
         \hline
         \emph{in} & 3.9878 & 4.2408 & 4.4667 & 4.5862 & 4.7514 & 5.0306 & 5.2104 \\
         \emph{mis} & 4.3393 & 4.4945 & 4.6907 & 4.8298 & 4.9909 & 5.2449 & 5.4473  \\
         \emph{out} & 5.1305 & 5.4748 & 5.6805 & 5.8477 & 6.0665 & 6.4294 & 6.6833\\
    \end{tabular}}}
    \caption{AUC, MSPE, coverage and length of $95\%$ prediction intervals for different values of $R=2,4,8,10,12,15,20$ used to fit DJL in the \emph{in-sample, missing}, and \emph{out-of-sample} predictive scenarios in Case 1. This case assumes $L = 2$, $J=20$, $T=20$, $m = 2$, $R^* = 2$, and $F^* = 3$. DJL is fitted with $F = 1$ for different choices of $R$.}
    \label{sens_tab_R_3_rev}
\end{table}

\subsubsection{Sensitivity to F}
To examine sensitivity with respect to the choice of $F$, we fit DJL with $F = 1, 2, 3, 4$ (holding $R = 4$) under three simulation settings of Scheme 1, as described in Section~2.2.1 of this supplementary file, with $L = 2$, $J = 20$, $T = 20$, and $m = 2$.
Tables~\ref{sens_tab_F_1_rev}, \ref{sens_tab_F_2_rev}, and \ref{sens_tab_F_3_rev} report the AUC for edge prediction, MSPE for nodal attribute prediction, and coverage and length of 95\% predictive intervals across the \emph{in-sample}, \emph{missing}, and \emph{out-of-sample} scenarios for each choice of $F$.

For edge prediction, results remain relatively consistent across $F$, but performance is generally best when $F=1$. In Case 1 ($R^* = 4, F^* = 1$), AUC is highest at $F=1$ across all scenarios. In Case 2 ($R^* = 3, F^* = 2$), AUC peaks at $F=1$ and declines with larger $F$. In Case 3 ($R^* = 2, F^* = 3$), AUC is highest at $F=1$ in the \emph{in-sample} and \emph{out-of-sample} settings, while in the \emph{missing} scenario, the best AUC occurs at $F=2$, though values at $F=1$ are comparable.

For nodal attribute prediction, MSPE values are consistently lower at $F=1$ and $F=2$ than at $F=3$ or $F=4$. In Case 1, $F=1$ yields the lowest MSPE in the \emph{in-sample} and \emph{out-of-sample} settings. In Case 2, $F=2$ achieves the lowest MSPE across all scenarios. In Case 3, $F=1$ outperforms $F=2$ in the \emph{in-sample} and \emph{missing} settings, while they are comparable for \emph{out-of-sample} scenario.

For predictive intervals, $F=1$ is the only setting that consistently achieves at least nominal coverage across all cases and scenarios. When $F=2, 3, 4$, under-coverage occurs frequently in the \emph{out-of-sample} setting. Interval lengths vary across $F$: in Cases 1 and 3, the widest intervals appear at $F=1$, while in Case 2, $F=4$ produces the widest intervals in the \emph{in-sample} and \emph{missing} scenarios, with $F=1$ yielding the widest in the \emph{out-of-sample} scenario.

Taken together, these results demonstrate that fitting DJL with $F=1$ is sufficient to achieve accurate prediction and well-calibrated uncertainty quantification across a range of scenarios, with little or no benefit gained from larger values of $F$.

\begin{table}[H]
        \centering
    \subfloat[AUC for Edge Prediction] 
     {\resizebox{0.48\textwidth}{!}{%
    \begin{tabular}{c|c c c c c c c}
         $F$ & 1 & 2 & 3 & 4 \\
         \hline
         \emph{in} & 0.9822 & 0.9649 & 0.9409 & 0.9415 \\
         \emph{mis} & 0.9891 & 0.9864 & 0.9811 & 0.9844\\
         \emph{out} & 0.9981 & 0.9980 & 0.9951 & 0.9974\\
    \end{tabular}}}
    \hfill
    \subfloat[MSPE for Nodal Attribute Prediction] 
     {\resizebox{0.48\textwidth}{!}{%
    \begin{tabular}{c|c c c c c c c}
         $F$ & 1 & 2 & 3 & 4 \\
         \hline
         \emph{in} & 0.2870 & 0.2883 & 0.3415 & 0.5741 \\
         \emph{mis} &  0.2954 & 0.2927 & 0.3471 & 0.5499\\
         \emph{out} & 0.2831 & 0.3014 & 0.3808 & 0.6305\\
    \end{tabular}}}
    \hfill
    \subfloat[$95\%$ Prediction Interval Coverage]
    {\resizebox{0.48\textwidth}{!}{%
    \begin{tabular}{c|c c c c c c c}
         $F$ & 1 & 2 & 3 & 4 \\
         \hline
         \emph{in} & 1.0000 & 0.9838 & 0.9671 & 0.9560 \\
         \emph{mis} & 1.0000 & 0.9750 & 0.9417 & 0.9458 \\
         \emph{out} & 0.9917 & 0.9000 & 0.6583 & 0.6667 \\
    \end{tabular}}}
    \hfill
    \subfloat[$95\%$ Prediction Interval Length]
    {\resizebox{0.48\textwidth}{!}{%
    \begin{tabular}{c|c c c c c c c}
         $F$ & 1 & 2 & 3 & 4 \\
         \hline
         \emph{in} & 3.7482 & 2.1035 & 2.2223 & 3.4283 \\
         \emph{mis} & 3.9641 & 2.1071 & 2.2238 & 3.4235 \\
         \emph{out} & 5.7746 & 2.4053 & 2.2930 & 3.4561 \\
    \end{tabular}}}
    \caption{AUC, MSPE, and $95\%$ Prediction Interval Coverages and Lengths for different values of $F$ used to fit DJL in the in-sample, missing, and out-of-sample predictive scenarios in Case 1. Here, we use $L = 2$, $J=20$, $T=20$, $m = 2$, $R^* = 4$, and $F^* = 1$ for data generation. DJL is fitted with $R = 4$.}
    \label{sens_tab_F_1_rev}
\end{table}

\begin{table}[H]
\centering
    \subfloat[AUC for Edge Prediction] 
     {\resizebox{0.48\textwidth}{!}{%
    \begin{tabular}{c|c c c c c c c}
         $F$ & 1 & 2 & 3 & 4 \\
         \hline
         \emph{in} & 0.9244 & 0.9011 & 0.8659 & 0.8631 \\
         \emph{mis} & 0.9155 & 0.9084 & 0.8901 & 0.8844\\
         \emph{out} & 0.9928 & 0.9853 & 0.9826 & 0.9850\\
    \end{tabular}}}
    \hfill
    \subfloat[MSPE for Nodal Attribute Prediction] 
     {\resizebox{0.48\textwidth}{!}{%
    \begin{tabular}{c|c c c c c c c}
         $F$ & 1 & 2 & 3 & 4 \\
         \hline
         \emph{in} & 0.3042 & 0.2810 & 0.4233 & 0.5811\\
         \emph{mis} & 0.3152 & 0.2834 & 0.4018 & 0.5266\\
         \emph{out} & 0.2949 & 0.2922 &  0.4844 & 0.6333\\
    \end{tabular}}}
    \hfill
    \subfloat[$95\%$ Prediction Interval Coverage]
    {\resizebox{0.48\textwidth}{!}{%
    \begin{tabular}{c|c c c c c c c}
         $F$ & 1 & 2 & 3 & 4 \\
         \hline
         \emph{in} & 0.9866 & 0.9801 & 0.9593 & 0.9569 \\
         \emph{mis} & 0.9958 & 0.9750 & 0.9292 & 0.9292\\
         \emph{out} & 0.9417 & 0.8917 & 0.6500 & 0.6667 \\
    \end{tabular}}}
    \hfill
    \subfloat[$95\%$ Prediction Interval Length]
    {\resizebox{0.48\textwidth}{!}{%
    \begin{tabular}{c|c c c c c c c}
         $F$ & 1 & 2 & 3 & 4 \\
         \hline
         \emph{in} & 2.6458 & 2.4136 & 2.8971 & 3.4686\\
         \emph{mis} & 2.7697 & 2.4463 & 2.9068 & 3.4673\\
         \emph{out} &  3.5683 & 2.7643 & 2.9529 & 3.4491\\
    \end{tabular}}}
    \caption{AUC, MSPE, and $95\%$ Prediction Interval Coverages and Lengths for different values of $F$ used to fit DJL in the in-sample, missing, and out-of-sample predictive scenarios in Case 1. Here, we use $L = 2$, $J=20$, $T=20$, $m = 2$, $R^* = 3$, and $F^* = 2$ for data generation. DJL is fitted with $R = 4$.}
    \label{sens_tab_F_2_rev}
\end{table}

\begin{table}[H]
   \centering
    \subfloat[AUC for Edge Prediction] 
     {\resizebox{0.48\textwidth}{!}{%
    \begin{tabular}{c|c c c c c c c}
         $F$ & 1 & 2 & 3 & 4 \\
         \hline
         \emph{in} & 0.7460  & 0.7243 & 0.6987 & 0.6959\\
         \emph{mis} & 0.6714 & 0.6863 & 0.6657 & 0.6703\\
         \emph{out} &  0.8828 & 0.8792 & 0.8639 & 0.8622\\
    \end{tabular}}}
    \hfill
    \subfloat[MSPE for Nodal Attribute Prediction] 
     {\resizebox{0.48\textwidth}{!}{%
    \begin{tabular}{c|c c c c c c c}
         $F$ & 1 & 2 & 3 & 4 \\
         \hline
         \emph{in} & 0.5660 & 0.5668 & 0.7311 & 0.7663 \\
         \emph{mis} & 0.5772 & 0.6259 & 0.7755 & 0.8105 \\
         \emph{out} & 0.3617 & 0.3589 & 0.6523 & 0.7112 \\
    \end{tabular}}}
    \hfill
    \subfloat[$95\%$ Prediction Interval Coverage]
    {\resizebox{0.48\textwidth}{!}{%
    \begin{tabular}{c|c c c c c c c}
         $F$ & 1 & 2 & 3 & 4 \\
         \hline
         \emph{in} & 0.9884 & 0.9806 & 0.9657 & 0.9630\\
         \emph{mis} & 0.9833 & 0.9792 & 0.9583 & 0.9542 \\
         \emph{out} & 0.9833 & 0.9417 & 0.7833 & 0.7583 \\
    \end{tabular}}}
    \hfill
    \subfloat[$95\%$ Prediction Interval Length]
    {\resizebox{0.48\textwidth}{!}{%
    \begin{tabular}{c|c c c c c c c}
         $F$ & 1 & 2 & 3 & 4 \\
         \hline
         \emph{in} & 3.8890 & 3.5408 & 3.7512 & 3.8374\\
         \emph{mis} & 3.9475 & 3.5656 & 3.7532 & 3.8245 \\
         \emph{out} & 5.0399 & 3.8755 & 3.7613 & 3.8382\\
    \end{tabular}}}
    \caption{AUC, MSPE, and $95\%$ Prediction Interval Coverages and Lengths for different values of $F$ used to fit DJL in the in-sample, missing, and out-of-sample predictive scenarios in Case 1. Here, we use $L = 2$, $J=20$, $T=20$, $m = 2$, $R^* = 2$, and $F^* = 3$ for data generation. DJL is fitted with $R = 4$.}
    \label{sens_tab_F_3_rev}
\end{table}

\subsubsection{Sensitivity to Choosing Equal Dimensions for Layer-Specific and Shared Latent Factors}\label{sec:dim_layer_shared}
This section evaluates the sensitivity of inference to the assumption $R = R_\zeta$. 
Table~\ref{tab_R_zeta_rev} reports results obtained by varying 
$R_{\zeta} \in \{2,3,4,6,8,10,12,15\}$, while fixing $R = 4$ and $F = 1$, 
under Simulation Scheme~1 with $L = 2$, $J = 20$, $T = 20$, $m = 2$, 
$R^* = 4$, and $F^* = 1$.

For edge prediction, the AUC values remain largely stable across all choices of 
$R_\zeta$ in the \emph{in-sample}, \emph{missing}, and \emph{out-of-sample} scenarios. 
For nodal attribute prediction, MSPE shows a modest decrease when increasing 
$R_\zeta$ from 2 to 4, with $R_\zeta = 8$ also yielding slightly lower MSPE values. 
However, this improvement comes at the cost of wider intervals and over-coverage in 
the 95\% predictive intervals compared to $R_\zeta = 4$. 
Overall, these findings suggest that setting $R = R_\zeta$ does not substantially 
affect inference in simulation scenarios designed to mimic real data, and that the 
model remains robust to moderate deviations in this assumption.

\begin{table}[H]
    \centering
    \subfloat[AUC for Edge Prediction] 
     {\resizebox{0.49\textwidth}{!}{%
    \begin{tabular}{c|c c c c c c c c}
          $R_{\zeta}$ & 2 & 3 & 4 & 6 & 8 & 10 & 12 & 15 \\
         \hline
         \emph{in} &  0.9810 & 0.9824 & 0.9836 & 0.9833 & 0.9842 & 0.9837 & 0.9844 & 0.9847 \\
         \emph{mis} & 0.9620 & 0.9670 & 0.9703 & 0.9668 & 0.9676 & 0.9698 & 0.9713 & 0.9737 \\
         \emph{out} & 0.9998 & 0.9999 & 0.9998 & 0.9999 & 0.9999 & 0.9999 & 0.9999 & 1.0000 \\
    \end{tabular}}}
    \hfill
    \subfloat[MSPE for Nodal Attribute Prediction] 
     {\resizebox{0.49\textwidth}{!}{%
    \begin{tabular}{c|c c c c c c c c}
          $R_{\zeta}$ & 2 & 3 & 4 & 6 & 8 & 10 & 12 & 15 \\
         \hline
         \emph{in} & 0.2718 & 0.2717 & 0.2582 & 0.2879 & 0.2453 & 0.3020 & 0.3042 & 0.3154 \\
         \emph{mis} & 0.2741 & 0.2744 &0.2554 & 0.2872 & 0.2434 & 0.3016 & 0.3071 & 0.3146\\
         \emph{out} & 0.2759 & 0.2778 & 0.2556 & 0.2863 & 0.2444 & 0.3027 & 0.3030 & 0.3140 \\
    \end{tabular}}}
    \hfill
    \subfloat[$95\%$ Prediction Interval Coverage]
    {\resizebox{0.49\textwidth}{!}{%
    \begin{tabular}{c|c c c c c c c c}
          $R_{\zeta}$ & 2 & 3 & 4 & 6 & 8 & 10 & 12 & 15 \\
         \hline
         \emph{in} & 0.9931 & 0.9948 & 0.9962 & 0.9976 & 0.9983 & 0.9934 & 0.9972 & 0.9920 \\
         \emph{mis} & 0.9875 & 0.9906 & 0.9906 & 0.9938 & 0.9969 & 0.9875 & 0.9969 & 0.9906\\
         \emph{out} & 0.9375 & 0.9625 & 0.9750 & 0.9875 & 0.9938 & 0.9562 & 0.9813 & 0.9625 \\
    \end{tabular}}}
    \hfill
    \subfloat[$95\%$ Prediction Interval Length]
    {\resizebox{0.49\textwidth}{!}{%
    \begin{tabular}{c|c c c c c c c c}
          $R_{\zeta}$ & 2 & 3 & 4 & 6 & 8 & 10 & 12 & 15 \\
         \hline
         \emph{in} & 2.1742 & 2.5704 & 2.7038 & 3.0166 & 3.2360 & 2.6908 & 2.7022 & 2.5148 \\
         \emph{mis} & 2.2222 & 2.6501 & 2.7568 & 3.1714 & 3.4923 & 2.7291 & 2.7846 & 2.6119 \\
         \emph{out} & 2.9467 & 3.6389 & 3.7933 & 4.6257 & 5.1747 & 3.9137 & 4.0218 & 3.6005 \\
    \end{tabular}}}
    \caption{AUC, MSPE, and $95\%$ Prediction Interval Coverages and Lengths for different values of $R_{\zeta}$ used to fit DJL in the \emph{in-sample, missing}, and \emph{out-of-sample} predictive scenarios. Here, the data is generated under simulation scenario 1, with $L = 2$, $J=20$, $T=20$, $m = 2$, $R^* = 4$, and $F^* = 1$. DJL is fitted with $R = 4$ and $F = 1$.}
    \label{tab_R_zeta_rev}
\end{table}

\section{Full Conditional Distributions}
Although the full posterior distribution for model parameters does not come in closed form, full conditional distributions can be obtained in closed form to construct Gibbs sampler for estimating model parameters. These full conditional distributions are given below. 
\begin{itemize}

\item  $\bmu|-\sim N(\bS_{\bmu}\bm_{\bmu}, \bS_{\bmu})$, where \newline $\bS_{\bmu}=(\sum_{l=1}^L\sum_{1\leq j<j'\leq J} \bD_{\omega,jj',l}+ \bSigma(\bbeta_{\mu})^{-1})^{-1}$, 
$\bm_{\bmu}=\sum_{l=1}^L\sum_{1\leq j<j'\leq J} \bD_{\omega,jj',l}\tilde{\by}_{jj',l}$, with \newline 
$\tilde{y}_{jj',l}^{t(o)}=y_{jj',l}^{t(o)}-(\bzeta_{j}^t)^T \bzeta_{j'}^t-(\bxi_{j,l}^t)^T \bxi_{j',l}^t$, 
$\tilde{\by}_{jj',l}=(\tilde{y}_{jj',l}^{t_{1}(o)},...,\tilde{y}_{jj',l}^{t_{T}(o)})^T$, and $\bD_{\omega,jj',l}=diag(\omega_{jj',l}^{t_{1}(o)},...,\omega_{jj',l}^{t_{T}(o)})$.

\item $\bet_k|-\sim N(\bS_{\bet_k}\bm_{\bet_k},\bS_{\bet_k})$, where \newline
$\bS_{\bet_k}= (J/\sigma_k^2\bI_T+\bSigma(\bbeta_{\eta})^{-1})^{-1}$ and 
$\bm_{\bet_k}=\sum_{j=1}^J(\bx_{j,k}-\bs_{j, k})/\sigma_k^2$, with \newline
$\bx_{j,k} = (x_{j,k}^{t_1}, ..., x_{j,k}^{t_T})^T$ and $\bs_{j, k} = (\sum_{l=1}^L (\bxi_{j,l}^{t_1})^T\balpha_{k,l}^{t_1}, ..., \sum_{l=1}^L (\bxi_{j,l}^{t_T})^T\balpha_{k,l}^{t_T})^T$.

\item $\sigma^2_k|- \sim IG\big(a_{\sigma} + \frac{JT}{2},\; b_{\sigma} + \dfrac{1}{2}\sum_{j=1}^J \big(\bx_{j,k} - (\bet_k + \bs_{j, k})  \big)^T \big( \bx_{j,k} - (\bet_k + \bs_{j, k})\big) \big).$

\item $\omega_{jj',l}^{t(o)} | - \sim PG(1, \mu^t +(\bzeta_{j}^t)^T \bzeta_{j'}^t+ (\bxi_{j,l}^t)^T \bxi_{j',l}^t).$

\item $\bzeta_{j,r} | -  N(\bS_{\bzeta_{j,r}}\bm_{\bzeta_{j,r}},\bS_{\bzeta_{j,r}})$, where \newline
$\bS_{\bzeta_{j,r}}=(\sum_{l=1}^L\sum_{j'\neq j}\bH_{j',r}^T \bD_{\omega,jj',l} \bH_{j',r}+ \bSigma(\bbeta_{\zeta})^{-1})^{-1}$ and
$\bm_{\bzeta_{j,r}}=\sum_{l=1}^L\sum_{j'\neq j}\bH_{j',r}^T \bD_{\omega,jj',l}\tilde{\tilde{\by}}_{jj',l,r}^{(o)}$,
with \newline
$\tilde{\tilde{y}}_{jj',l,r}^{t(o)}=y_{jj',l}^{t(o)}-\mu^t-(\bxi_{j,l}^t)^T \bxi_{j',l}^t -\sum_{r'\neq r}\zeta^{t}_{j',r}\zeta^{t}_{j,r'}$, 
$\bH_{j',r}=diag(\zeta_{j',r}^{t_1},...,\zeta_{j',r}^{t_T})$, and \newline $\tilde{\tilde{\by}}_{jj',l,r}^{(o)}=(\tilde{\tilde{y}}_{jj',l,r}^{t_1(o)},...,\tilde{\tilde{y}}_{jj',l,r}^{t_T(o)})^T$.

\item $\bxi_{j,l,r} | - \sim N(\bS_{\bxi_{j,l,r}}\bm_{\bxi_{j,l,r}},\bS_{\bxi_{j,l,r}})$, where \newline
$\bS_{\bxi_{j,l,r}}=\left[ \sum_{j'\neq j}\tilde{\bH}_{j',l,r}^T\bD_{\omega,jj',l}\tilde{\bH}_{j',l,r}+\sum_{k=1}^m \tilde{\tilde{\bH}}_{k,l,r}^T\tilde{\tilde{\bH}}_{k,l,r}/\sigma_k^2 + \bSigma(\bbeta_{\xi})^{-1} \right]^{-1}$ and \newline
 $\bm_{\bxi_{j,l,r}}=\sum_{j'\neq j}\tilde{\bH}_{j',l,r}^T\bD_{\omega,jj',l}\bar{\by}_{jj',l,r}^{(o)}+\sum_{k=1}^m\tilde{\tilde{\bH}}_{k,l,r}^T\bar{\bx}_{j,k,l,r}/\sigma_k^2$
, with \newline
$\tilde{\bH}_{j',l,r}=diag(\xi_{j',l,r}^{t_1},..,\xi_{j',l,r}^{t_T})$, $\tilde{\tilde{\bH}}_{k,l,r}=diag(\alpha_{k,l,r}^{t_1},..,\alpha_{k,l,r}^{t_T})$, \newline
$\bar{y}_{jj',l,r}^{t(o)}=y_{jj',l}^{t(o)}-\mu^t -(\bzeta_{j}^t)^T \bzeta_{j'}^t- \sum_{r'\neq r}\xi_{j,l,r}^t \xi_{j',l,r}^t$, \newline
$\bar{x}_{j,k,l,r}^t=x_{j,k}^t-\eta_k^t-\sum_{l'\neq l}(\bxi_{j,l'}^t)^T\balpha_{k,l'}^t
-\sum_{r'\neq r}\xi_{j,l,r'}^t\alpha_{k,l,r'}^t$, \newline
$\bar{\by}_{jj',l,r}^{(o)}=(\bar{y}_{jj',l,r}^{t_1(o)},...,\bar{y}_{jj',l,r}^{t_T(o)})^T$ and $\bar{\bx}_{j,k,l,r}=(\bar{x}_{j,k,l,r}^{t_1},...,\bar{x}_{j,k,l,r}^{t_T})^T.$

\item $\balpha_{k,l,r} | - \sim N(\bS_{\balpha_{k,l,r}}\bm_{\balpha_{k,l,r}},\bS_{\balpha_{k,l,r}})$,
where \newline $\bS_{\balpha_{k,l,r}}=\left[\sum_{j=1}^J \bA_{j,l,r}^T\bA_{j,l,r}/\sigma_k^2+\bSigma(\bbeta_{\alpha})^{-1}\right]^{-1}$ and 
$\bm_{\balpha_{k,l,r}}=\sum_{j=1}^J\bA_{j,l,r}^T\tilde{\bx}_{j,k,l,r}/\sigma_k^2$, 
with \newline
$\bA_{j,l,r}=diag(\xi_{j,l,r}^{t_1},...,\xi_{j,l,r}^{t_T})$,
$\tilde{x}_{j,k,l,r}^t=x_{j,k}^t-\eta_k^t-\sum_{l'\neq l}(\bxi^{t}_{j,l'})^T\balpha^{t}_{k,l'}-\sum_{r'\neq r}\alpha_{k,l,r'}^t\xi_{j,l,r'}^t$, and
$\tilde{\bx}_{j,k,l,r}=(\tilde{x}_{j,k,l,r}^{t_1},...,\tilde{x}_{j,k,l,r}^{t_T})^T.$

\item To update $\bbeta_{\eta}$, $\bbeta_{\alpha}$, $\bbeta_{\zeta}$, $\bbeta_{\xi}$, and $\bbeta_{\mu}$, we use the values on discrete uniform grids that maximize the model likelihood. In this work, we use the grid, $\{0.01,0.02, ..., 0.1\}^{2}$, for each case. The updated hyperparameter values are then used to update $\bSigma(\bbeta_{\eta})$, $\bSigma(\bbeta_{\alpha})$, $\bSigma(\bbeta_{\zeta})$, $\bSigma(\bbeta_{\xi})$, and $\bSigma(\bbeta_{\mu})$, respectively.

\end{itemize}

\section{Convergence Analysis}\label{sec:convergence_analysis}
\begin{figure}[H]
\centering
\subfloat[$\mu^{(1)}$]{
\includegraphics[width=0.46\textwidth]{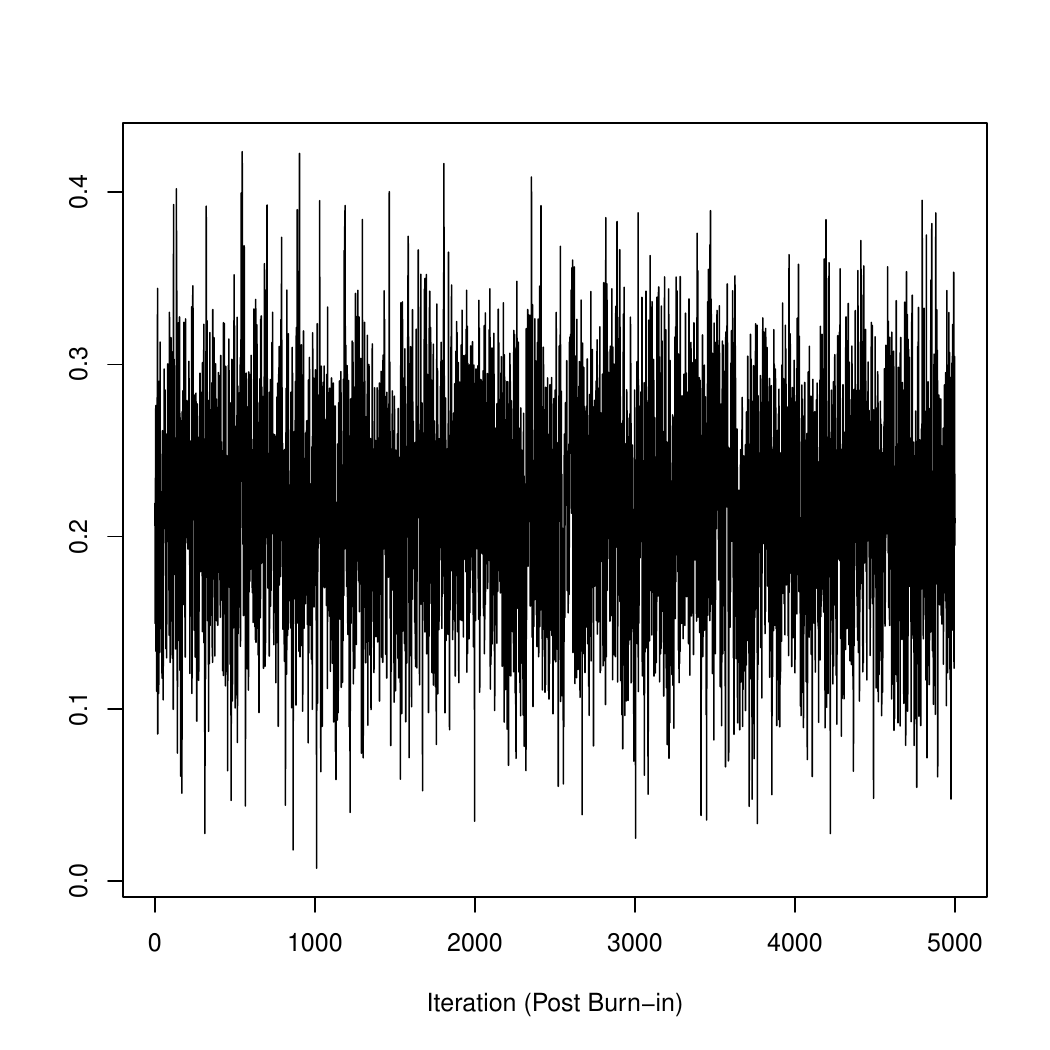}}
\subfloat[$\eta_2^{(1)}$]{
\includegraphics[width=0.46\textwidth]{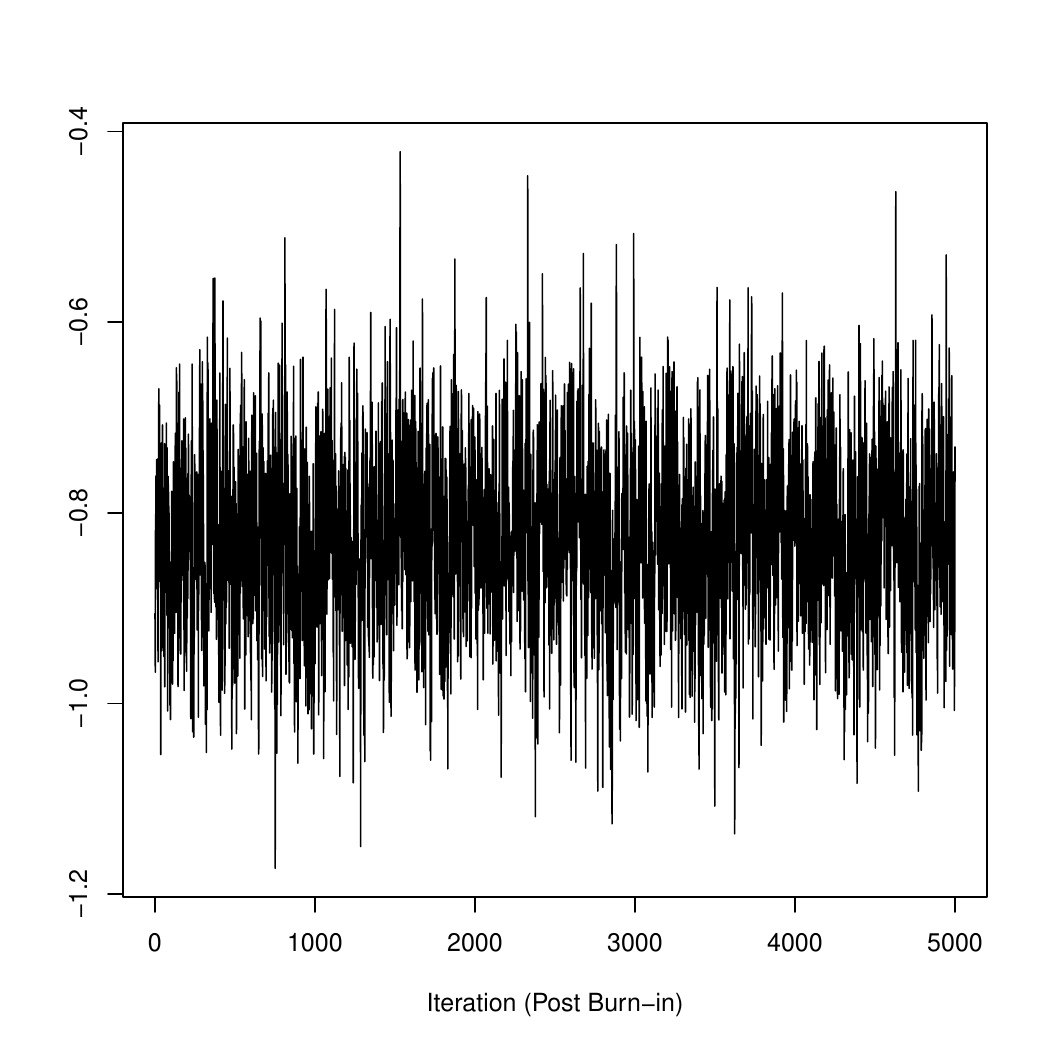}}
\hfill
\subfloat[$\alpha^{(1)}_{1,1,4} $]{
\includegraphics[width=0.46\textwidth]{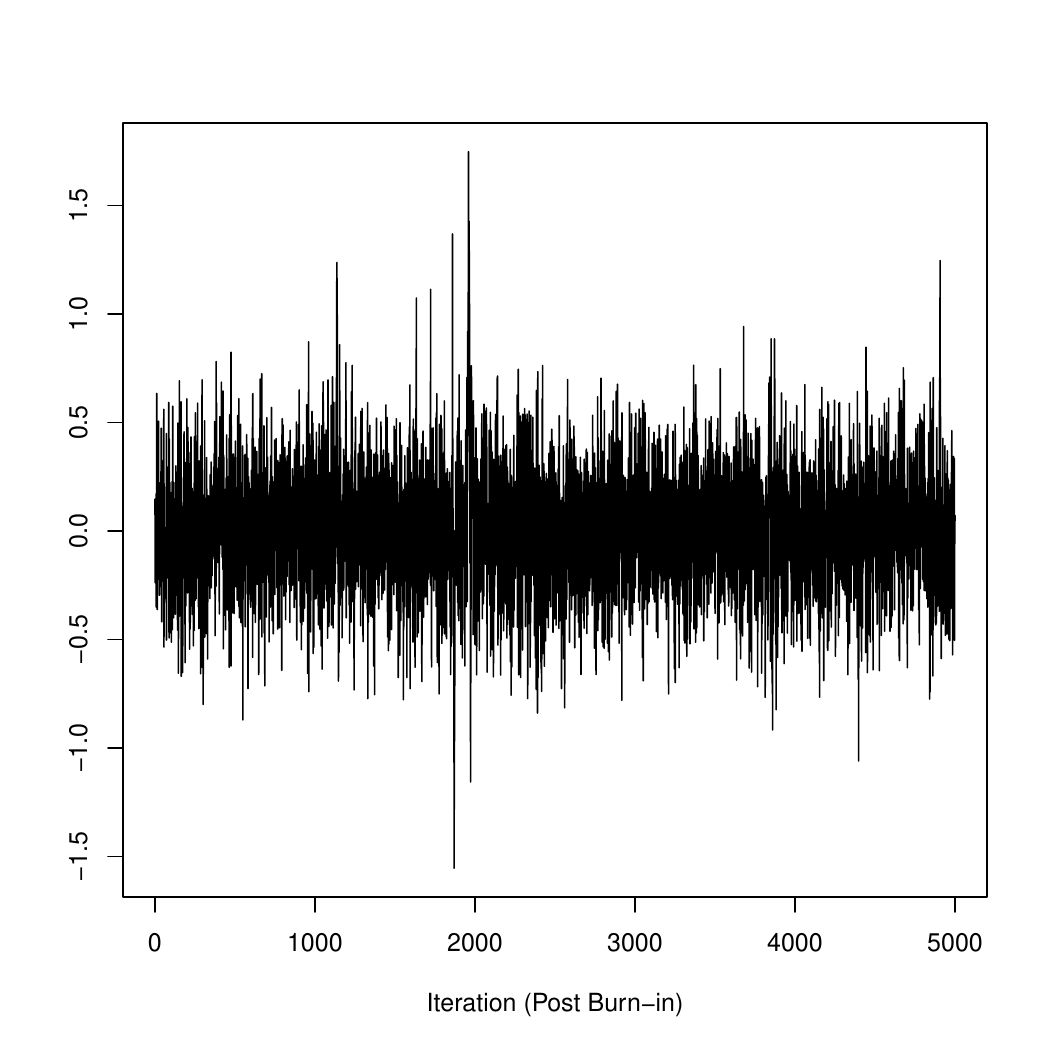}}
\subfloat[$\xi^{(1)}_{1,1,3}$]{
\includegraphics[width=0.46\textwidth]{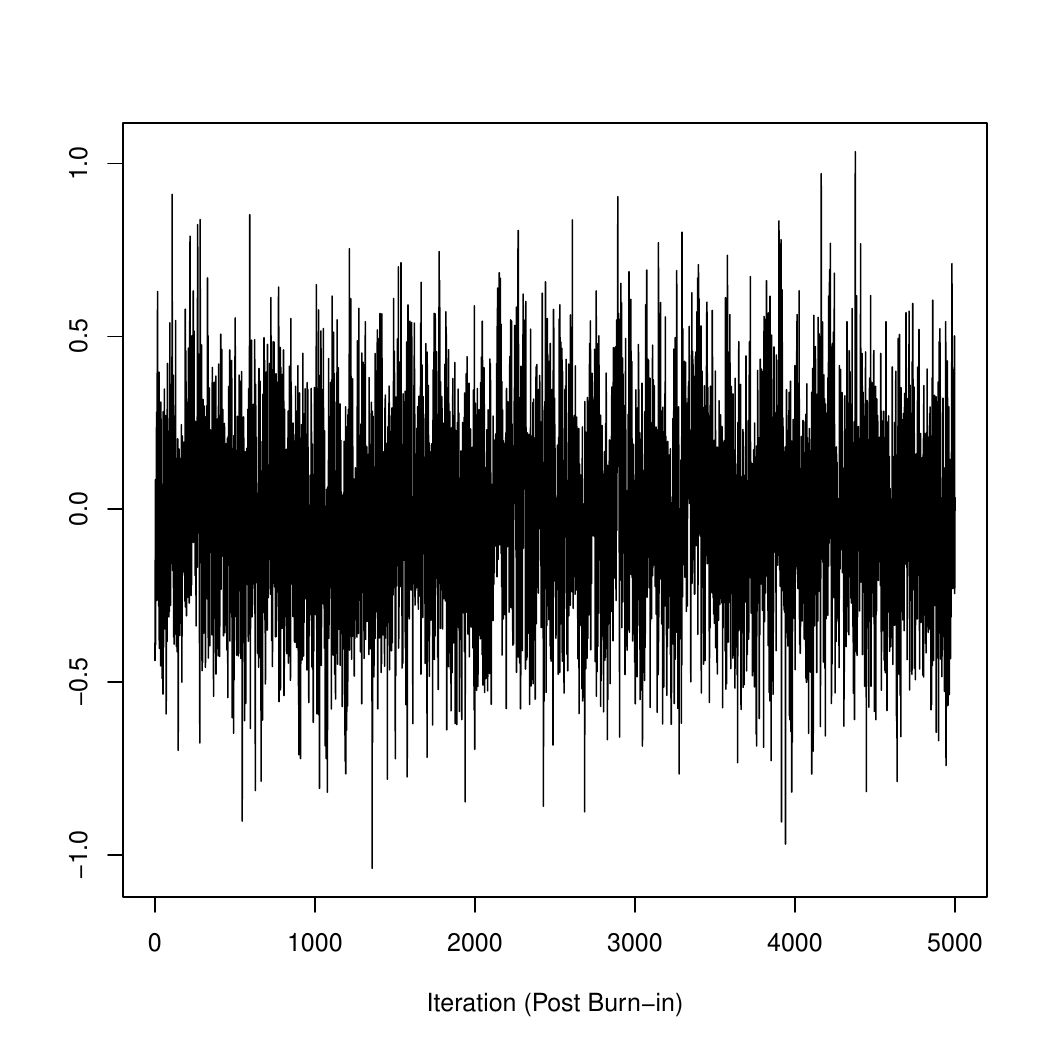}}
\caption{Traceplots for posterior distributions for a few representative model parameters.}
\label{conv-plots}
\end{figure}

The traceplots of MCMC samples show excellent convergence for the model parameters in DJL under all simulation scenarios. In particular, under all simulation schemes, the average effective sample size for $5000$ post burn-in samples always turns out to be above $3500$, indicating excellent convergence behavior. Traceplots for some representative parameters for Simulation Scheme 1 are given in Figure~\ref{conv-plots}.

\begin{singlespace}
\bibliographystyle{plain}
\bibliography{refs}
\end{singlespace}